\documentclass[12pt]{article}
\usepackage[T1]{fontenc}
\usepackage[utf8]{inputenc}
\usepackage[margin=0.8in]{geometry}
\usepackage[colorlinks,citecolor=blue,linkcolor=blue,urlcolor=blue,pagebackref]{hyperref}
\usepackage{graphicx}
\usepackage{booktabs}
\usepackage{subcaption}
\usepackage{tabularx}
\usepackage{longtable}
\usepackage{array}
\usepackage{pdflscape}
\usepackage{lscape}
\usepackage{float}
\usepackage{amsmath}
\usepackage{amssymb}
\usepackage{xcolor}
\usepackage{tcolorbox}
\tcbuselibrary{breakable}
\tcbset{
  colback=gray!2,
  colframe=gray!40,
  boxrule=0.4pt,
  arc=2pt,
  left=4pt,
  right=4pt,
  top=4pt,
  bottom=4pt
}
\usepackage{fvextra}
\usepackage{natbib}
\usepackage{ragged2e}
\usepackage{setspace}
\usepackage{mathpazo}
\usepackage{tikz}
\usetikzlibrary{arrows.meta,positioning}
\tikzset{
  cg node/.style={
    draw=gray!70,
    fill=gray!8,
    rounded corners=2pt,
    align=center,
    font=\scriptsize,
    text width=2.55cm,
    minimum height=8mm,
    inner sep=2pt
  },
  cg causal/.style={
    -{Stealth[length=3.0mm,width=2.0mm]},
    draw=blue!70!black,
    line width=0.95pt,
    shorten >=2pt,
    shorten <=2pt
  },
  cg noncausal/.style={
    -{Stealth[length=3.0mm,width=2.0mm]},
    draw=orange!85!black,
    line width=0.95pt,
    shorten >=2pt,
    shorten <=2pt
  }
}

\DefineVerbatimEnvironment{PromptBlock}{Verbatim}{breaklines,breakanywhere,fontsize=\scriptsize}
\newcolumntype{L}[1]{>{\raggedright\arraybackslash}p{#1}}

\title{Causal Claims in Economics\thanks{We thank Susan Athey, Elias Bareinboim, David Card, Matt Clancy, Scott Cunningham, Sanjeev Goyal, Marcin Kacperczyk, Samuel Kortum, Peter J. Lambert, Ralf Martin, Andrew Oswald, Jörg Ankel-Peters, Carol Propper, Chris Roth, Tommaso Valletti and seminar and conference participants at CEPR Paris Symposium, Metascience, NetSciSci, EAYE, Groningen, AYEW, ZBW, MPWZ-CEPR Text-As-Data, PolMeth, Leibniz Open Science, and Causal Data Science Meeting for helpful comments. We thank Lakshya Kaviya for excellent research assistance. We also thank Carina Neisser and Abel Brodeur for sharing hand-coded data, and Sangmin Oh for hosting the Plausibly Exogenous Galore dataset and for guidance on accessing it. The authors declare no competing interests. }

\author{Prashant Garg \and Thiemo Fetzer \footnote{Garg is based at Imperial College London. Email: prashant.garg@imperial.ac.uk. Fetzer is based at University of Warwick \& Bonn. Data, code, prompts, and workflow documentation are publicly available at \href{https://github.com/prashgarg/CausalClaimsInEconomics}{Github}. Email: \href{mailto:team@causal.claims}{team@causal.claims}}}}

\date{\today}

\begin{document}
\maketitle
\vspace{-1cm}

\begin{abstract}
\noindent
As economics scales, a key bottleneck is representing what papers claim in a comparable, aggregable form. We introduce evidence-annotated \emph{claim graphs} that map each paper into a directed network of standardized economic concepts (nodes) and stated relationships (edges), with each edge labeled by evidentiary basis, including whether it is supported by causal inference designs or by non-causal evidence. Using a structured multi-stage AI workflow, we construct claim graphs for 44{,}852 economics papers from 1980--2023. The share of causal edges rises from 7.7\% in 1990 to 31.7\% in 2020.  Measures of causal narrative structure and causal novelty are positively associated with top-five publication and long-run citations, whereas non-causal counterparts are weakly related or negative.

\vspace{0.75em}
\noindent\textbf{Keywords:} Claim Graph; Credibility Revolution; Causal Inference; Narrative Complexity; Novelty; Large Language Models
\end{abstract}

\setstretch{1.65}

\clearpage
\section{Introduction}

Economic research is produced and consumed at ever larger scale. Working-paper repositories have expanded for decades, and recent machine-learning tools, including large language models (LLMs), further reduce the marginal cost of drafting and revising manuscripts \citep{ludwig2025economistsllm, humlum2025large, feyzollahi2025adoptionllm}. As output scales, the constraint shifts from producing papers to organizing and interpreting large literatures: identifying what a paper claims, how claims connect into mechanisms, what is genuinely new, and how incentives shape the production and dissemination of evidence. Yet the core unit of scholarly communication remains narrative prose. Claims are embedded in heterogeneous language and dispersed across sections, making systematic comparison difficult precisely when manual synthesis is least feasible.

This paper takes a step toward solving that representation problem. We introduce a paper-level claim graph: a structured, machine-readable representation that maps each paper into a directed network of standardized economic concepts and their stated relationships. Nodes correspond to economic concepts mapped to a common ontology (JEL codes), and directed edges represent claims from a source concept to a sink concept as expressed by the authors. Crucially, edges are annotated with the type of evidentiary support, including whether a relationship is documented using a credible causal inference design (e.g., Difference-in-Differences, Instrumental Variables, Regression Discontinuity, randomized experiments, event studies, synthetic control), or instead reflects theoretical reasoning, descriptive evidence, or correlational analysis.\footnote{Throughout, we use ``claim'' to emphasize that we extract what papers state and how they support it; we do not adjudicate whether claims are true. This distinction is particularly important in empirical economics, where strong identification claims depend on assumptions and are subject to ongoing debates about validity, interpretation, and generalizability \citep{deaton2010instruments, heckman2001micro, sims2010but, deaton2018understanding, cartwright2010rcts}.} By stripping away rhetoric while preserving the skeleton of a paper's argument in a form that is consistent across documents, claim graphs can be compared, aggregated, and analyzed at scale.\footnote{A standardized claim representation has multiple potential uses beyond the applications we pursue here. For example, it can support claim-level evidence synthesis across literatures; enable structured search and discovery by linking questions, concepts, and evidentiary support; and provide building blocks for tracing how research claims are summarized and reframed in policy and media discourse.}

% Stripping down papers to their claims is relevant, as claims do not remain inside scholarly texts: they can become narrative units that diffuse through media and policy discourse, where framing and political language shape interpretation and hence, may actively shape societal discourse.

\paragraph{Positioning in the literature.}
Our approach connects three strands of work that are often studied separately. First, it complements the ``credibility revolution'' literature documenting the rise of quasi-experimental and experimental designs and the discipline's shift toward empirical identification \citep{angrist2010credibility, angrist2008mostly, imbens2015causal, card2013nine, hamermesh2013six}. Related large-scale accounts document how methods and fields evolve over time \citep{angrist2017economic, backhouse2017age, currie2020technology}, and recent work tracks the diffusion of quasi-experimental styles across subfields, with applied micro leading and areas such as finance and macro lagging, though the gap is narrowing \citep{goldsmith2024tracking}. Existing approaches typically classify papers by broad method tags, topics, or fields. We instead measure causality at the level of relationships: within a paper, which specific links are supported by which identification strategies, and how those links chain into mechanisms. Second, our analysis relates to work on the economics of science and the production of ideas, including evidence that novelty is harder to achieve and that incentives shape what gets pursued and rewarded \citep{bloom2020ideas, park2023papers, carnehl2025quest}. Claim graphs provide new primitives, mechanism depth, causal novelty, and gap filling, that are difficult to observe using bibliometrics alone. Third, we build on text-as-data and scientometric traditions that map knowledge using co-occurrence and network methods \citep{small1973co, waltman2012new, van2014visualizing, hidalgo2009building}, and on recent interest in knowledge graphs and LLM-assisted synthesis pipelines \citep{pugliese2024conduct, chan2024steps, buehler2024accelerating}. Our contribution is to operationalize a paper-level, evidence-annotated claim representation that is consistent across documents and supports claim-level measurement, not only paper-level classification.\footnote{We use LLMs as constrained retrieval under fixed schemas rather than open-ended generators, addressing concerns about hallucinated synthesis in AI-assisted reviewing \citep{pearson2024can}.} 

We construct claim graphs for 44{,}852 NBER and CEPR working papers from 1980--2023 using a structured multi-stage retrieval workflow. We use an LLM as an information-retrieval engine rather than as an open-ended reasoner: Stage 1 extracts a structured summary of research questions, designs, and key metadata from the first 30 pages; Stage 2 extracts candidate edges with evidence attributes; Stage 3 maps free-text entities to standardized JEL concepts via embeddings-based matching. To quantify and improve robustness, we run multiple independent extraction passes and aggregate edges using an explicit edge-overlap rule that yields a transparent precision--recall frontier. We validate key retrieval dimensions using layered checks (iteration-stability diagnostics, snippet-based support validation, and external benchmarks), described in detail in the Appendix.

We use economics as a natural flagship domain for this approach because the credibility revolution provides both a substantive transformation and a measurement challenge. It raised the bar for empirical work by emphasizing design-based identification, transparent reporting, and sensitivity to assumptions \citep{angrist2008mostly, imbens2015causal}, while also intensifying debates about whether methodological innovation crowds out novelty in questions, whether narrative complexity and ``salesmanship'' matter for dissemination, and how editorial selection relates to long-run influence \citep{falk2021whats, rawat2014publish}. Claim graphs make these debates measurable at the level of a paper's argumentative content: they allow us to separate non-causal narrative structure from causally identified structure, and to quantify novelty and ``gap filling'' conditional on evidentiary support.\footnote{Beyond this paperâ€™s applications within economics, representing claims as portable, evidence-annotated units could support future work on how research findings diffuse and are reframed outside academia. Classic work emphasizes agenda-setting and framing in media environments \citep{mccombs1972agenda,entman1993framing}, and a large empirical literature documents that media coverage and slant can affect beliefs, political behavior, and policy \citep{dellavigna2007fox,gentzkow2010drives,snyder2010press,enikolopov2011media,stromberg2004radio}. This perspective naturally connects to work on amplification and mediated effects, among others \citep{besley2024mediamultiplier,fetzer2025losinghomefront,garg2024political}.}
% More broadly, by representing claims as portable, evidence-annotated narrative units, claim graphs provide the foundation for tracing how research narratives propagate beyond papers into media systems and politically coded discourse, a natural extension given evidence on media amplification and mediated effects on societal outcomes \citep{besley2024mediamultiplier,fetzer2025losinghomefront,garg2024political}, and concerns that public self-correction may be slow or incomplete in misinformation-prone environments \citep{alabrese2022badscience,alabreseadena_capozza_selfcorrecting_wip}.

Three headline findings emerge from this first application. First, we observe a sharp rise in causally documented claims: the average share of causal edges increases from 7.7\% in 1990 to 31.7\% in 2020 (and continues rising through 2023), with substantial heterogeneity across fields. Second, causal narrative structure is strongly rewarded. Doubling causal edge volume is associated with about +1.36 percentage points in top-5 publication probability (roughly +12\% relative to the 11.35\% baseline rate) and +11.2\% in citations; doubling causal new edges is associated with +1.71 percentage points in top-5 probability (roughly +15\%) and +10.7\% in citations. By contrast, non-causal edge expansion is weakly related or negative. Third, novelty matters primarily when grounded in credible designs: introducing genuinely new causal edges or causal paths is strongly associated with top-5 publication and long-run citations, while non-causal novelty exhibits weak or even negative effects. At the same time, citation impact remains tightly linked to conceptual positioning: papers anchored in central, widely recognized concepts tend to receive more citations, highlighting a wedge between features associated with editorial gatekeeping and those associated with broad diffusion. Finally, bridging underexplored concept pairs (``gap filling'') appears to be rewarded primarily when grounded in causal methods, yet exhibits no consistent relationship with future citations.

Our contributions are therefore both methodological and substantive. Methodologically, we provide (i) a scalable representation of papers as evidence-annotated claim graphs, enabling direct comparison and aggregation of research claims across a large corpus; (ii) an open dataset and tooling that support claim-level queries and measurement; and (iii) a compact, interpretable set of paper-level graph measures capturing narrative complexity, novelty and contribution, and conceptual positioning, with robustness and validation. Substantively, we use this infrastructure to quantify how the credibility revolution manifests at the level of claims rather than papers, and to document how journals and the profession reward causal depth, causal novelty, and topic positioning.

The remainder of the paper is organized as follows. Section~\ref{sec:data_methods} describes the corpus, retrieval workflow, graph construction, and validation strategy. Section~\ref{sec:application_causal_evidence} presents our flagship application measuring causally documented claims across time, fields, and methods. Section~\ref{sec:graphical_measures} defines paper-level measures of narrative complexity, novelty and contribution, and conceptual importance and diversity. Section~\ref{sec:publication_citation_outcomes} links these measures to publication outcomes and citations. Section~\ref{sec:discussion} discusses broader implications and future modules, and Section~\ref{sec:discussion_conclusion} concludes with implications for research incentives, evidence synthesis, and the development of claim-level infrastructures for cumulative science.

\section{Claim-Graph Dataset and Construction} \label{sec:data_methods}
This section describes the corpus, the claim-graph representation, the extraction/standardization pipeline, and validation. It documents the infrastructure; subsequent sections use it for our flagship application to causal claims. Further methodological details and technical specifications are provided in the Appendix.\footnote{Data, code, prompts, and workflow documentation are publicly available at \url{https://github.com/prashgarg/CausalClaimsInEconomics}.}

\subsection{Working Paper Corpus}
Our analysis is based on a comprehensive corpus of working papers from two primary sources: the National Bureau of Economic Research (NBER) and the Centre for Economic Policy Research (CEPR). The NBER dataset comprises 28,186 working papers, while the CEPR dataset includes 16,666 papers, resulting in a total sample of 44,852 papers. These papers span several decades and encompass various subfields of economics, providing a broad view of the research landscape.

Because NBER and CEPR can host overlapping versions of the same working paper, we run an explicit deduplication audit. We first preserve repository-native IDs, then form normalized title-year keys (lower-cased, punctuation-normalized titles with publication year). Cross-repository matches are collapsed to a canonical paper identifier before paper-level aggregation. The Appendix reports these audit counts in a dedicated Corpus Deduplication Audit Summary table.

All publication outcomes and citation counts are measured as of September 2024 and treated as fixed snapshots to avoid repeated re-baselining of the results.

To refine the sample and focus on relevant content, we applied specific filtering criteria. We included only papers containing more than 1,000 characters to exclude incomplete documents. Additionally, we limited the analysis to the first 30 pages of each paper, ensuring that we captured the sections most likely to contain causal claims, such as abstracts, introductions, identification strategy discussions, and core empirical results.\footnote{The 30-page window balances coverage, compute cost and the context attention of a model. In economics papers, central claims and identification logic are typically stated early. We also report robustness using nine-run edge-overlap aggregation to reduce sensitivity to occasional truncation.}

The corpus covers a wide range of economics subfields, including Labour Economics, Public Economics, Macroeconomics, Development Economics, and Finance. The papers employ diverse empirical strategies, such as Randomized Controlled Trials (RCTs), Instrumental Variables (IV), Difference-in-Differences (DiD), and Regression Discontinuity Designs (RDD), allowing us to examine methodological trends across the discipline.

\paragraph{Pre-processing.}
We convert PDFs to text and apply standard normalization steps to remove artifacts from PDF extraction (e.g., errant whitespace and encoding noise). We preserve the substantive content of the text while ensuring a consistent input format for the extraction prompts.
% \paragraph{Pre-processing} The preprocessing of the text data followed a structured pipeline aimed at cleaning and normalizing the text for analysis. The preprocessing steps included removing excessive whitespace, converting all characters to lowercase, and filtering out non-alphanumeric characters, keeping only spaces for readability. Additionally, we stripped leading and trailing whitespace to ensure uniformity in the text. These steps were helpful for the large language model to efficiently and accurately process the text, especially considering the large volume of data involved.

\subsection{Claim-Graph Representation and Evidence Annotation}\label{sec:claim_graph_representation}

For each paper $p$, we define a directed claim graph $G_p=(V_p,E_p)$, where $V_p$ is the set of nodes representing economic concepts mapped to JEL codes, and $E_p$ is the set of directed edges representing claims from a source node to a sink node as stated by the authors. We use the terms ``source'' and ``sink'' to denote narrative direction without presupposing causality; for intuition, these can be read as ``driver/predictor'' and ``outcome'' labels. Graph-based representations are increasingly used to synthesize evidence across scientific domains \citep{buehler2024accelerating}.

Each edge carries attributes describing the relationship type and the evidentiary basis. We classify an edge as causal if the associated claim is supported by canonical identification designs: Difference-in-Differences (DiD, including TWFE/event-study implementations), Instrumental Variables (IV/2SLS), Randomized Controlled Trials (RCTs/Experiments), Regression Discontinuity Design (RDD), or Synthetic Control. Structural estimation and causal language without an identified design are recorded as methods but are not classified as causal edges. The causal subgraph $G_p^{\text{causal}}$ is the subgraph induced by these edges.

This representation allows us to compare narrative structure across papers and to distinguish general relationships from those supported by identified empirical designs.

\subsection{LLM based retrieval} 
\begin{figure}[htp]
    \centering
    \caption{Master Workflow for Claim-Graph Extraction and Aggregation}
    \label{fig:pipeline_stages1_3}
    
    \includegraphics[width=\textwidth, trim=0 8cm 0 0, clip]{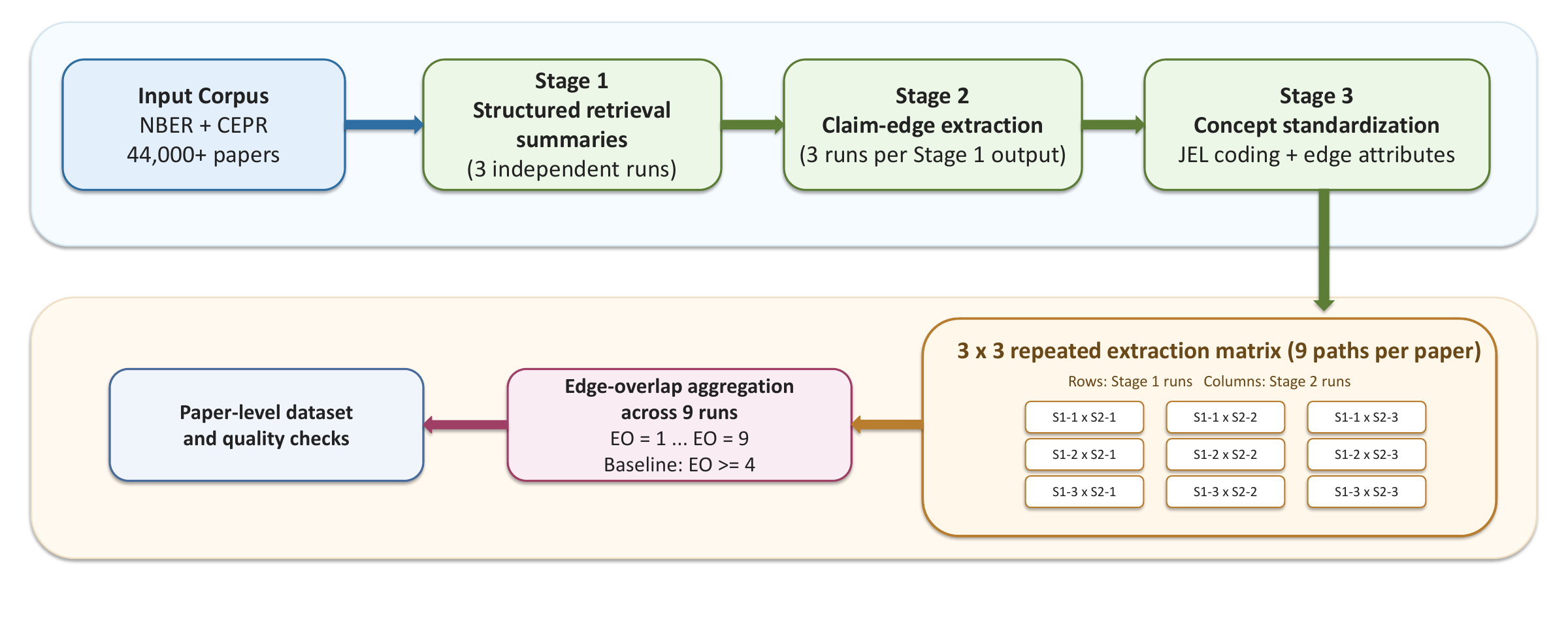}

    \captionsetup{labelformat=empty,labelsep=none}
    \caption*{\scriptsize\justifying \textbf{Note:} This figure summarizes the three-stage retrieval architecture. Stage 1 (Qualitative Curation) is run three times per paper and feeds Stage 2 (Graph Extraction), which is run three times per Stage 1 output, yielding a 3x3 design (nine edge lists). Stage 3 standardizes concepts via JEL mapping, and edge-overlap (EO) aggregation is applied with EO \(\geq 4\) as the baseline; outputs feed paper-level measures, trends, regressions, and robustness/validation checks.}
\end{figure}

We employed a multi-stage process using GPT-4o-mini, a large language model (LLM) developed by OpenAI, to extract and analyze information from the working papers in our corpus.\footnote{GPT-4o-mini is used here as an information-retrieval engine under tightly constrained prompts and schemas.} We interacted with the LLM using carefully designed prompts that guided the model to extract the required information while adhering to a predefined JSON schema. The overall process is visually summarized in Figure \ref{fig:pipeline_stages1_3}, which illustrates the flow from input text to structured data extraction and subsequent analysis. This approach allowed us to efficiently process the text and extract detailed structured data necessary for our analysis while minimizing computational and human resources.\footnote{Some scholars have called for a discourse-centric infrastructure to improve local and collaborative synthesis of research findings \citep{chan2024steps}, emphasizing how new frameworks could better integrate knowledge across multiple studies.}

Our LLM-based retrieval process consists of the following stages:

% - useful reference maybe, but not academic: https://x.com/rohanpaul_ai/status/1842734796658618670?s=43
\paragraph{Stage 1: Qualitative Summary Extraction}

In the first stage, we prompted the LLM to analyze the first 30 pages of each paper and extract a curated summary of key elements.\footnote{Recent discussions highlight both the promise and pitfalls of AI-assisted literature reviews, cautioning that while tools can speed up narrative synthesis, they can also produce hallucinated references or incomplete coverage \citep{pearson2024can}. For this reason, we do not ask the model to conduct literature review, but to retrieve explicitly stated information within a given input (i.e., the first 30 pages of the paper).} This included the research questions as presented in the abstract, introduction, and full text; information on causal identification strategies used in the paper; details on data usage, accessibility, and acknowledgements; and metadata such as authors' names, institutional affiliations, fields of study, and methods used. 

To reduce sensitivity to stochastic LLM outputs and to quantify the stability of our retrieval, we run three independent iterations of Stage 1 for each paper. Each iteration produces a structured summary that then feeds into the downstream stages. This initial extraction provided a structured overview of each paper, which was used in subsequent stages to extract more detailed information. We validate key retrieval dimensions in the Appendix using stability diagnostics, snippet-based self-consistency checks, and external benchmarks. Performance is strong for several high-frequency method labels (e.g., DiD, IV, RDD) in matched benchmark samples, while rarer labels exhibit the expected sensitivity to class imbalance and coverage in the matched subsets.

\paragraph{Stage 2: Extraction of Claim-Graph Edges}

Using the curated summaries from Stage 1, particularly the sections on identification language and claims, we prompted the LLM to extract detailed claim-graph edges between two concepts. Each edge is labeled with its evidentiary basis (causal vs non-causal) based on the methods used to support the claim. For each link, the model identified source and sink variables as described by the authors, classified the relationship type (e.g., direct causal effect, indirect causal effect, mediation, confounding, theorized relationship, correlation), and recorded the empirical method(s) used to support the link (e.g., RCT, IV, DiD, OLS, simulations). 

We execute Stage 2 three times for each of the three Stage 1 summaries, yielding nine candidate edge lists per paper. This design allows us to assess the robustness of extracted causal structures across repeated model passes. The result is an edge list per paper, where each row represents an edge with a source node and a sink node, alongside key edge attributes such as the method used to evidence the relationship. While we also collected additional attributes such as the direction of effect, magnitude, and statistical significance, these features were experimental and are not used in the main analysis due to variation in reporting standards.\footnote{These additional attributes were collected for exploratory purposes but were not included in the primary analysis.} The extracted source and sink descriptions are subsequently mapped to standardized JEL codes in Stage 3.

\paragraph{Stage 3: Mapping Variables to Standardized Economic Concepts}

To facilitate systematic network analysis and cross-paper aggregation, we standardized the free-text descriptions of the source and sink variables by mapping them to official Journal of Economic Literature (JEL) codes. We created semantic embeddings for each JEL code's overall description, which concatenates the JEL description, guidelines, and keywords.\footnote{The JEL guidelines (available at \url{https://www.aeaweb.org/jel/guide/jel.php}) provide detailed descriptions of each code and are typically a paragraph long. Including keywords enhances the semantic specificity of the JEL codes.} By generating vector embeddings of the extracted variable descriptions and comparing them to the embeddings of JEL code descriptions using cosine similarity, we identified the most relevant code for each variable.\footnote{\cite{aipnet} uses a similar embeddings-based matching approach to map product descriptions to Harmonized System (HS) codes.} 

This mapping situates each causal claim within a standardized conceptual space and enables construction of a paper-level claim graph defined over JEL nodes. Figure \ref{fig:pipeline_stages1_3} summarizes the extraction-to-standardization workflow.

\paragraph{Edge Aggregation and Network Stability}

To convert the nine candidate edge lists into a stable paper-level claim graph, we aggregate edges using an edge-overlap criterion. For each paper, we count how often a given (source $\rightarrow$ sink) edge re-appears across the nine independent extractions. We define edge-overlap as $\mathrm{EO} \in \{1,\ldots,9\}$, where higher values indicate greater stability across model passes.

This aggregation rule induces an explicit precision--recall trade-off. Low thresholds (e.g., $\mathrm{EO} \geq 1$) are inclusive and maximize recall but may admit fragile edges, while high thresholds (e.g., $\mathrm{EO} \geq 9$) isolate only the most persistent relationships, increasing precision at the cost of coverage. Empirically, we observe an an inflection point in the stability profile around $\mathrm{EO}=4$, suggesting that $\mathrm{EO} \geq 4$ captures a robust yet still inclusive set of claims. We therefore use this threshold as our primary baseline and report sensitivity checks using more permissive and more conservative thresholds.

% \paragraph{Stage 4: Data Usage and Accessibility Extraction}

% From the data-related summaries in Stage 1, we prompted the LLM to extract structured information regarding data sources and accessibility. Key elements included the ownership of the data (e.g., private company, public sector entity, researchers), data accessibility (e.g., freely accessible, restricted), and details on data granularity, units of analysis, temporal and geographical context. 

% Stage 4 is applied to each of the three Stage 1 iterations and linked to the corresponding Stage 2 outputs, ensuring that data and accessibility classifications are aligned with the same underlying summary-based interpretation. This information is crucial for assessing trends in data usage and the implications for transparency and replicability in economic research.

% \paragraph{Metadata Extraction}

% As part of Stage 1, the LLM extracted key metadata for each paper, including the authors' names, primary institutional affiliations, paper title, year of release, JEL codes, and any keywords mentioned. The model also identified the relevant fields of study (e.g., Labour Economics, Finance, Macroeconomics) using predefined categories. Additionally, papers were classified according to their empirical or theoretical nature, ranging from entirely empirical to entirely theoretical. This extraction ensured a structured and standardized representation of core metadata, facilitating subsequent analysis of causal claims and research methodologies.

\subsection{Validation} 

We use a layered validation strategy aligned with the main empirical object in this paper: paper-level claim graphs.

\paragraph{Iteration-stability validation}
For each paper we run nine extraction passes (three Stage 1 summaries crossed with three Stage 2 graph extractions). We then aggregate edges by edge-overlap (EO), defined as the number of runs in which an edge reappears. This generates a precision--recall frontier: EO \(\geq 1\) maximizes coverage, while higher EO thresholds retain only more persistent edges. In the Appendix, we show that headline trends and regression signs are stable from EO \(\geq 1\) to EO \(\geq 4\), with expected sample-thinning at stricter thresholds.

\paragraph{Machine-assisted snippet validation}
The claim snippets generated in Stage 1 are fed into a separate Stage 2 prompt that only sees verbatim snippet text, and it extracts an edge list from those snippets alone. We then compare candidate edge sets against this snippet-only edge list, classifying edges as supported (true positives), spurious (false positives), or missing (false negatives). This is a self-consistency check rather than ground-truth validation, designed to reveal how sensitive extraction is to summary-based inputs versus snippet-only evidence and to benchmark precision--recall trade-offs across EO thresholds.

\paragraph{External dataset validation}
To validate retrieval dimensions that can be benchmarked externally, we conduct two exercises (details in the Appendix). First, we compare method and field extraction against \citet{brodeur2024p}. Second, we compare causes, effects, and sources of exogenous variation against a curated expert-labeled benchmark of plausibly exogenous variation in economics. In the EO \(\geq 4\) baseline, these validations match 285 papers in Brodeur and 491 papers in the expert benchmark; we also report EO \(\geq 7\) and EO \(= 9\) sensitivity tables with explicit precision/recall and random-baseline lift metrics, and a nine-run dispersion summary for the Brodeur metrics. These checks validate core extraction components while recognizing that our target object (full paper-level claim graph) is richer than single-label benchmarks.

\paragraph{Measure-level robustness}
Finally, we assess robustness of derived measures through EO sensitivity and controlled perturbation exercises, ensuring that key relationships are not driven by a single aggregation choice.

\subsection{Citations and Publication Data}

\paragraph{Matching Publication Outcomes to Working Papers}
To analyze the publication trajectories of the working papers, we matched each paper to its eventual publication outcome using multiple data sources. Our primary goal was to determine whether a working paper was published in a peer-reviewed journal and, if so, identify the journal and publication date. This information is essential for understanding the dissemination and impact of research within the economics discipline.

We utilized four data sources to obtain publication information. First, we used official metadata from the NBER, which provides publication data collected via author submissions and automated scraping from RePEc. While comprehensive, the dataset includes duplicates and primarily covers NBER papers. Second, for residual unmatched papers, we use a model-assisted matching step that proposes candidate publication matches from bibliographic metadata (title, authors, year), which we then verify against structured records when available.\footnote{We use the model to propose candidates, not as an authoritative source of publication outcomes. The match is retained only when bibliographic fields align under our deterministic rules.} Third, we used the OpenAlex repository, matching titles of working papers and prioritizing published versions when multiple matches existed. Finally, we incorporated data from \cite{baumann2020have}, which provides manually verified publication outcomes for NBER and CEPR papers between 2000 and 2012, matched up to mid-2019. All journal placements reported in this paper are measured as of September 2024.

To ensure consistency, we standardized journal names across these sources using the SCImago Journal Rank (SJR) lists for the fields of "Economics, Econometrics and Finance" and "Business, Management and Accounting." After removing generic journal names, this list included 2,367 unique journals.

Our matching process followed a hierarchical approach, prioritizing verified data. We first checked for a match in the dataset from \cite{baumann2020have}, followed by a search in OpenAlex for exact title matches. If no match was found, we consulted the NBER metadata, and finally used the LLM retrieval method for remaining papers. This approach ensured a comprehensive and reliable matching of publication outcomes. In total, the dataset from \cite{baumann2020have} provided publication outcomes for 9,139 papers, OpenAlex matched 10,840 papers, NBER metadata contributed 15,872 matches, and the LLM retrieval identified outcomes for 1,707 papers. By consolidating these sources, we obtained a detailed picture of the publication trajectories of a substantial number of working papers.

% \subsubsection{Publication Matching Details}
% \label{sec:appendix_publication_matching}
The matching pipeline is deterministic except for the final residual LLM-assisted step. We prioritize externally validated sources first (\cite{baumann2020have}, then OpenAlex exact title matches), then repository metadata (NBER), and only then model-assisted retrieval. This ordering is designed to minimize classification error in journal-rank assignments used in downstream regressions.

Despite this extensive coverage, certain limitations remain. The NBER metadata may be incomplete due to reliance on author submissions, and the \cite{baumann2020have} dataset only covers papers up to 2012, matched to 2019. Additionally, errors may arise due to title similarities or data entry issues. However, by leveraging multiple sources, we minimized these limitations, resulting in a robust dataset for further analysis.

\paragraph{Citations Data}

To extend our analysis, we collected citations data for the working papers using three primary sources: RePEc's CiteEc service (\url{https://citec.repec.org/}), \cite{baumann2020have}, and the OpenAlex repository. We prioritized the citations data from RePEc's CiteEc service, which provides up-to-date (as of September 2024) citation counts for a large number of economics papers. For papers not included in CiteEc, we used the manually verified citations from \cite{baumann2020have}, which provides citation counts for NBER and CEPR papers published between 2000 and mid-2019. For any remaining papers, we obtained citation counts from OpenAlex, matched by exact paper titles. By merging these sources and prioritizing in this order, we assembled citations data for approximately 94.6\% of our total sample, and 97.7\% of the pre-2020 sample used in our analysis. This extensive coverage enables us to incorporate citations as a measure of research impact.

% \section{The Claim Graph of Economics} \label{sec:causal_graph_economics}
\section{Causal Evidence in Economics Claim Graphs}\label{sec:application_causal_evidence}

Using claim graphs, we measure causality at the relationship level. In our baseline EO \(\geq 4\) dataset, 23.8\% of edges are classified as causal. We begin by documenting how the share of causal edges evolves over time and across fields, and then describe method adoption patterns within the same framework.

\begin{figure}[htp]
\centering
\caption{Trends in the Proportion of Causal Edges Over Time and by Field}
\begin{subfigure}[t]{0.575\textwidth}
    \centering
    \caption{Over Time}
    \includegraphics[width=\textwidth]{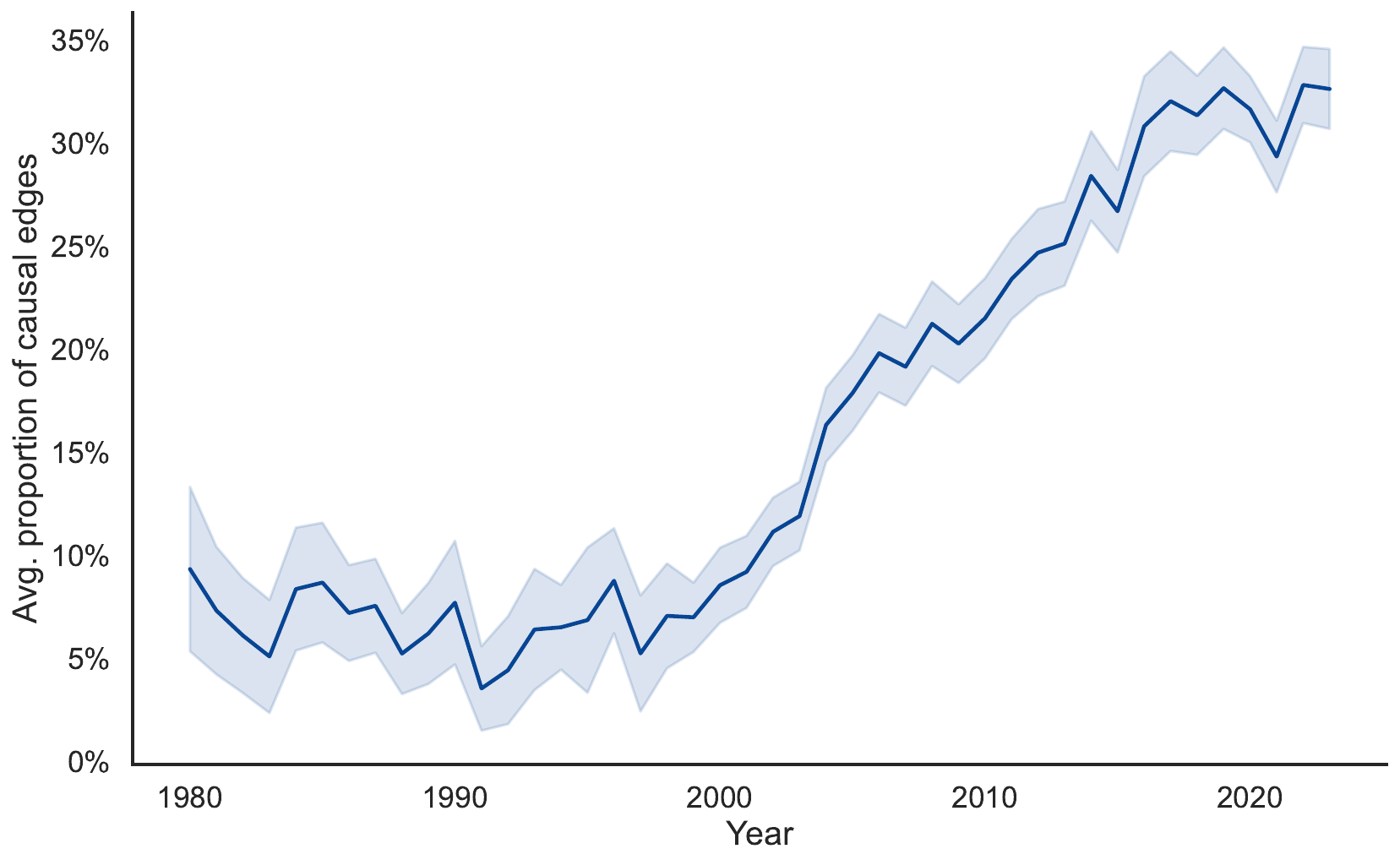}
    \label{fig:prop_causal_edges_over_time}
\end{subfigure}%
\hfill
\begin{subfigure}[t]{0.575\textwidth}
    \centering
    \caption{By Field}
    \includegraphics[width=\textwidth]{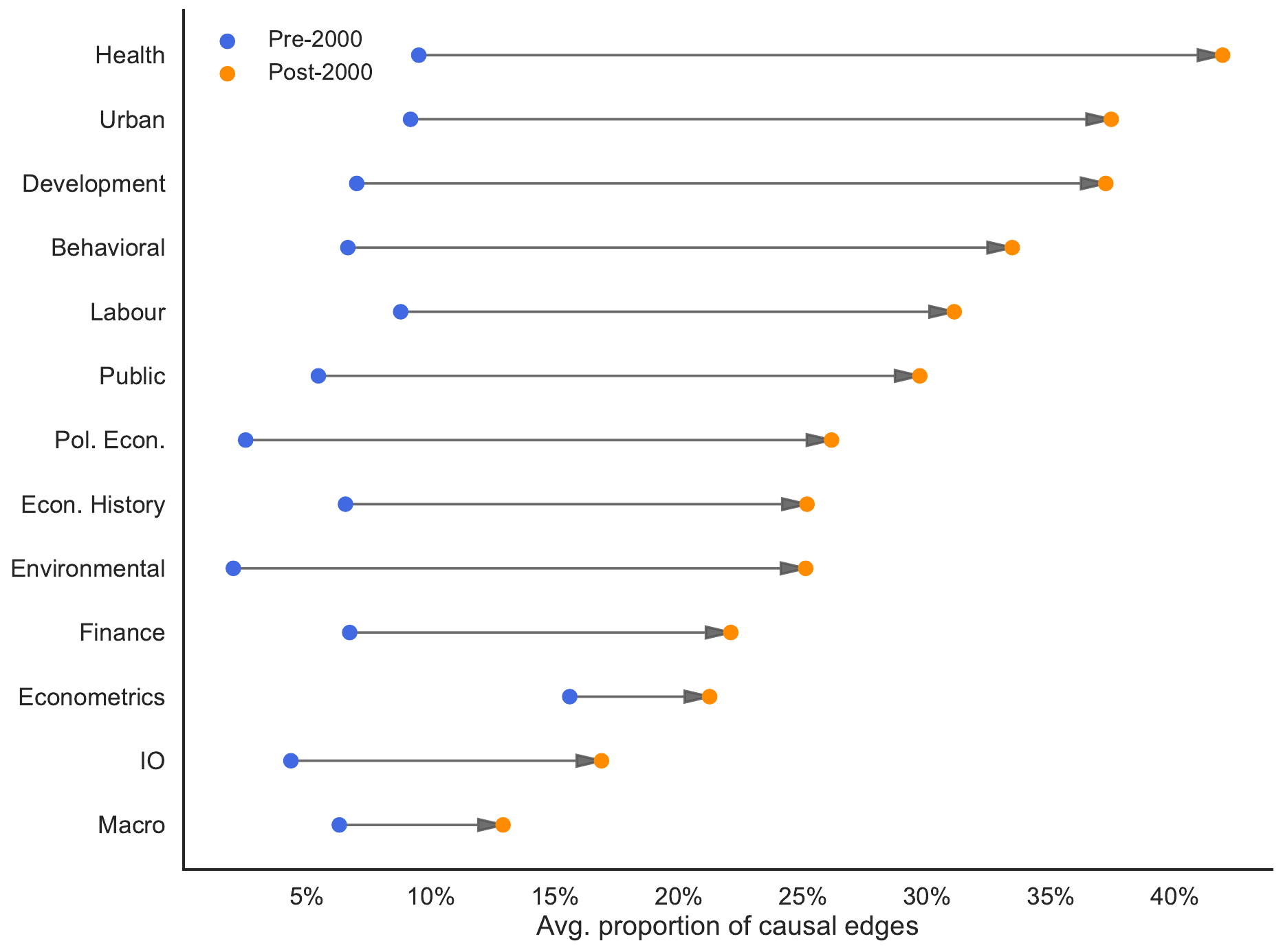}
    \label{fig:prop_causal_edges_by_field_period}
\end{subfigure}
\captionsetup{singlelinecheck=off,font=small,labelformat=empty,labelsep=none}
\caption*{\scriptsize\justifying \textbf{Note:} This figure presents the trends and distribution of the average proportion of causal edges per paper in NBER and CEPR working papers across different dimensions. Panel (a) displays the average proportion of causal edges per paper from 1980 to 2023, showing a significant increase from 7.7\% in 1990 to 31.7\% in 2020 (32.6\% in 2023). The solid blue line represents the average, and the shaded area indicates the 95\% confidence interval. Panel (b) shows the average proportion of causal edges by field, comparing the pre-2000 (royal blue) and post-2000 (orange) periods. Most fields exhibit substantial increases over time; for example, Health rises from 9.7\% to 41.9\%, Urban from 9.0\% to 37.4\%, Development from 6.9\% to 37.2\%, and Behavioral from 6.5\% to 33.4\%. These patterns suggest that the adoption of causal inference methods has become more widespread across fields in economic research.}
\label{fig:prop_causal_edges_trends}
\end{figure}

Figure \ref{fig:prop_causal_edges_trends} (a) displays the average proportion of causal edges per paper from 1980 to 2023. The data show a significant increase over time. In 1990, the average proportion of causal edges was 7.7\%. By 2000, it had risen to 8.6\%. The increase became more pronounced in subsequent decades: by 2010, the average proportion reached 21.5\%, and by 2020, it climbed to 31.7\% (reaching 32.6\% by 2023). This upward trend indicates that economic papers have increasingly incorporated causal inference methods to substantiate their claims over the past three decades.

This substantial increase suggests a growing emphasis on establishing credible causal relationships in economic research. The proliferation of empirical methods and a heightened focus on causal identification strategies have contributed to this trend, reflecting the impact of the credibility revolution in economics.

We also examine how this proportion varies across different fields within economics. Figure \ref{fig:prop_causal_edges_trends}\textbf{(b)} presents the average proportion of causal edges by field, comparing the pre-2000 and post-2000 periods. Recent evidence corroborates these cross-field patterns, showing that finance and macroeconomics have lagged behind applied micro in adopting quasi-experimental methods, though the gap is narrowing \citep{goldsmith2024tracking}. Our findings add that most fields have experienced substantial increases in the average proportion of causal edges in the post-2000 period.

Fields such as Urban, Health, Development, and Behavioral exhibit the highest increases. Urban rises from 9.0\% pre-2000 to 37.4\% post-2000. Health increases from 9.7\% to 41.9\%, achieving the highest post-2000 level among all fields. Development rises from 6.9\% to 37.2\%, and Behavioral increases from 6.5\% to 33.4\%. These fields, which often address policy-relevant questions and benefit from natural experiments or data conducive to causal analysis, have embraced causal inference methods more extensively.

Conversely, some fields experienced smaller increases or maintained lower levels. Macroeconomics increases from 6.3\% to 12.9\%, while Econometrics rises from 15.3\% to 21.2\%. Finance increases from 6.6\% to 22.1\%, but remains below leading applied-micro fields in post-2000 causal-share levels. These patterns suggest that the adoption of causal inference methods has varied across fields, influenced by the nature of the research questions, data availability, and methodological traditions within each field.

A composition-control robustness check is reported in the Appendix in a dedicated trend-robustness table. In the EO \(\geq 4\) snapshot, the unadjusted rise in share-causal-claims is \(+0.0187\) per decade; controlling for field composition yields \(+0.0174\) per decade; and adding conservative controls for method mix and graph volume yields \(+0.0011\) per decade. This confirms that compositional shifts explain part, but not all, of the observed trend.

\subsection{Evolution of Empirical Methods in Economic Research}

\begin{figure}[htp]
\centering
\caption{Proliferation of Empirical Methods Over Time in NBER and CEPR Working Papers}
\includegraphics[width=\textwidth]{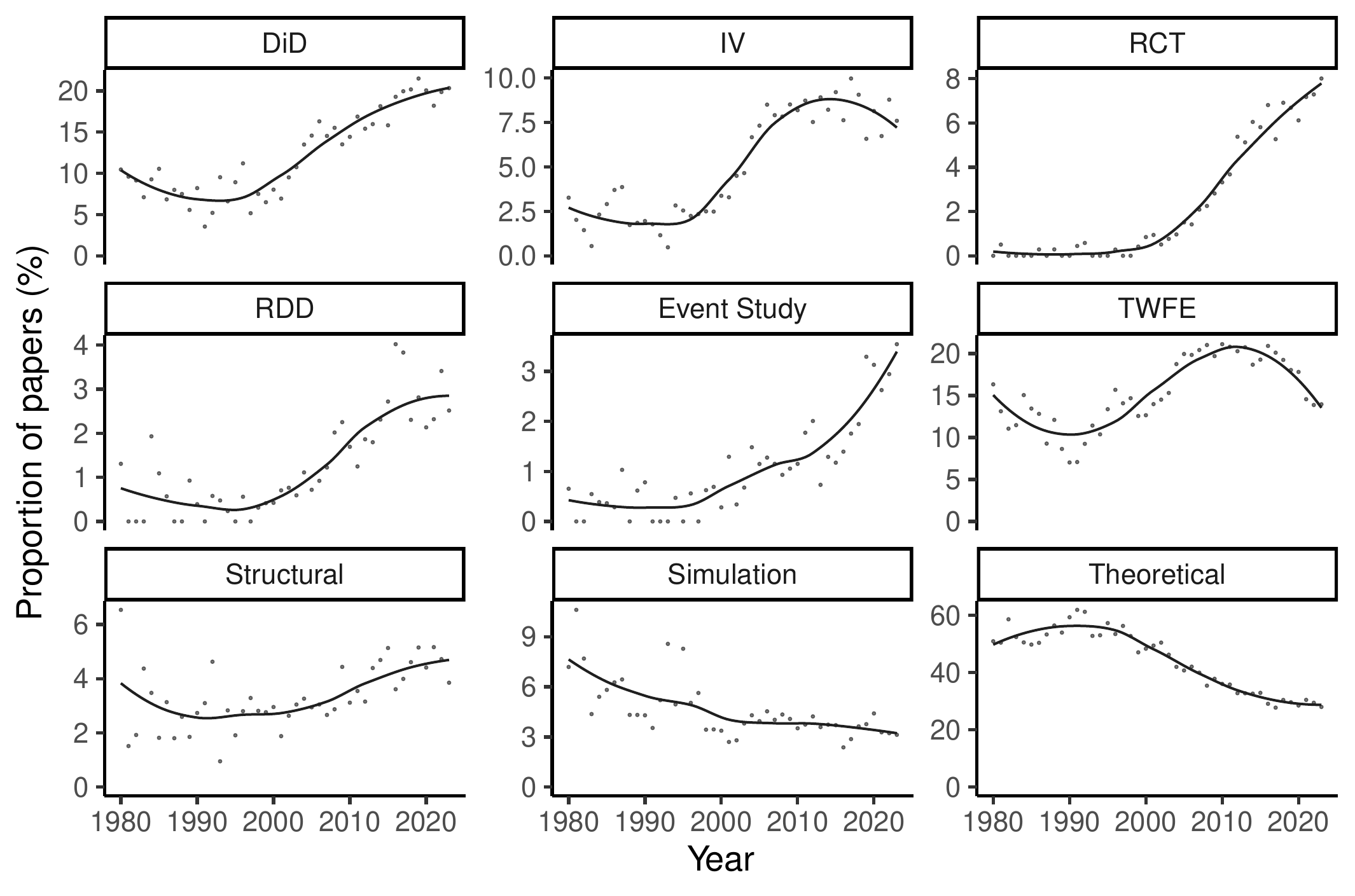}
\captionsetup{singlelinecheck=off,font=small,labelformat=empty,labelsep=none}
\caption*{\scriptsize\justifying \textbf{Note:} This figure shows the proliferation of key empirical methods used in NBER and CEPR working papers over time: Difference-in-Differences (DiD), Instrumental Variables (IV), Randomized Controlled Trials (RCTs), Regression Discontinuity Design (RDD), Two-Way Fixed Effects (TWFE), Structural Estimation, Event Studies, Simulations, and Theoretical/Non-Empirical research. Each panel represents the proportion of papers utilizing one of these methods per year, with the y-axis showing the proportion of total papers and the x-axis indicating year. In the EO \(\geq 4\) baseline series, DiD rises from 10.5\% (1980) to 20.3\% (2023), IV from 3.3\% to 7.6\%, RCT from 0.8\% (2000) to 8.0\%, and RDD from 1.3\% to 2.5\%. Theoretical/non-empirical usage declines from 51.0\% (1980) to 28.0\% (2023), while simulation declines from 7.2\% to 3.1\%. These trends highlight the increasing emphasis on credible identification strategies and the evolution of empirical methods in economics.}
\label{fig:method_proliferation}
\end{figure}

To explore the increasing focus towards causal inference, we show time trends across methods and fields. Figure \ref{fig:method_proliferation} illustrates the adoption of prominent empirical methods in NBER and CEPR working papers from 1980 to 2023. Methods such as DiD, IV, RCTs, and RDDs have seen substantial growth, reflecting the discipline's shift towards more rigorous identification strategies.\footnote{Note that a single paper may employ multiple methods and span multiple fields, as these classifications are not mutually exclusive.}

DiD has become increasingly prevalent, rising from 10.5\% of papers in 1980 to 20.3\% by 2023. This growth underscores DiD's utility in exploiting policy changes and natural experiments to identify causal effects. IV methods have also seen a steady increase, from 3.3\% in 1980 to 7.6\% in 2023, highlighting their role in addressing endogeneity through exogenous instruments \citep{angrist1994identification}. The adoption of RCTs has accelerated since the early 2000s, increasing from 0.8\% in 2000 to 8.0\% in 2023, signaling the increasing feasibility and acceptance of experimental designs in economics. RDD usage has also risen, from 1.3\% in 1980 to 2.5\% in 2023.

Conversely, the proportion of theoretical and non-empirical work has declined from 51.0\% of papers in 1980 to 28.0\% in 2023, indicating a broader emphasis on empirical analysis. The use of simulations also declines overall, from 7.2\% in 1980 to 3.1\% in 2023.

\begin{figure}[htp]
\centering
\caption{Cross-Sectional Breakdown of Empirical Methods by Field in NBER and CEPR Working Papers}
\includegraphics[width=0.95\textwidth]{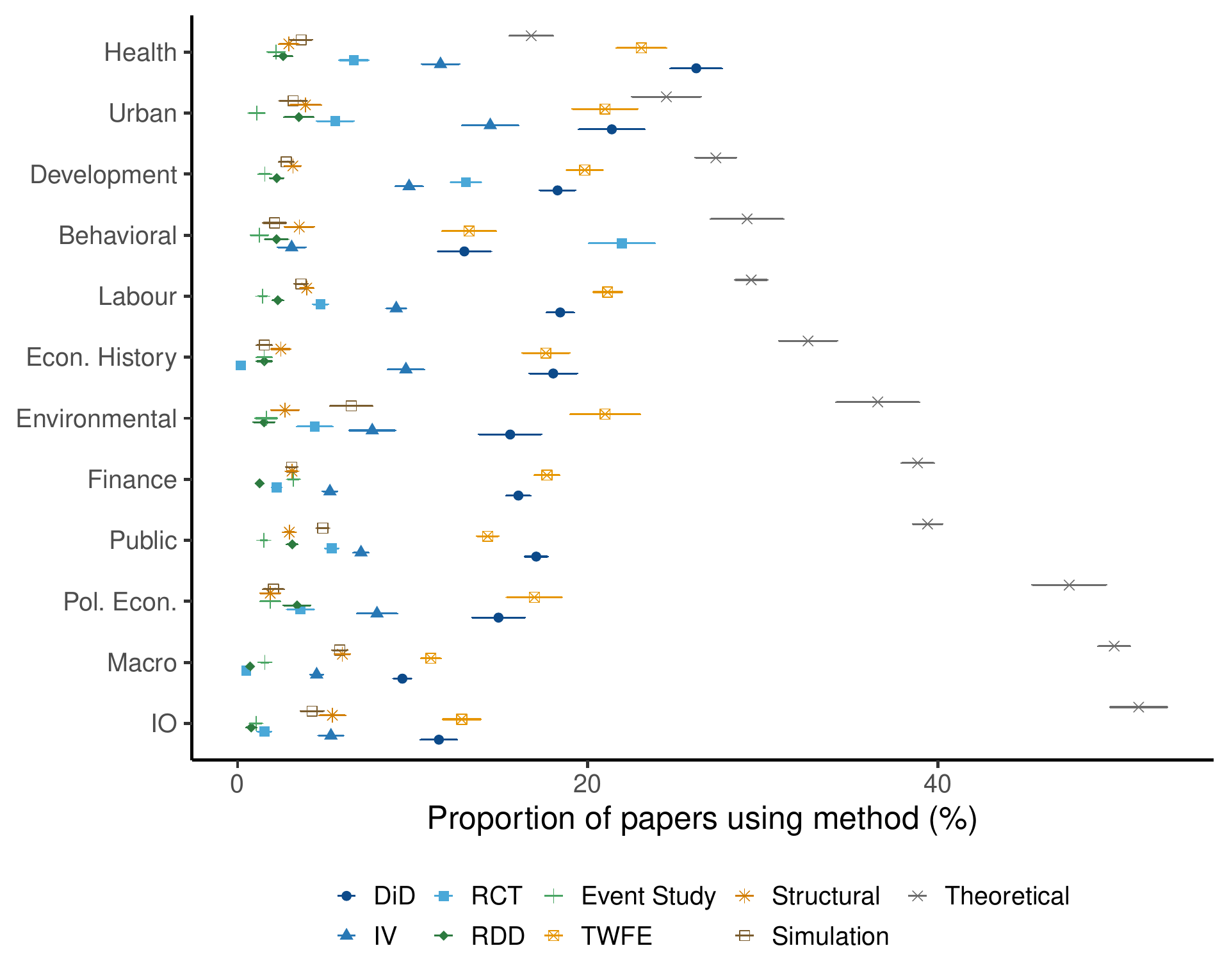}
\captionsetup{singlelinecheck=off,font=small,labelformat=empty,labelsep=none}
\caption*{\scriptsize\justifying \textbf{Note:} This figure displays the cross-sectional distribution of nine empirical methods, Difference-in-Differences (DiD), Instrumental Variables (IV), Randomized Controlled Trials (RCTs), Regression Discontinuity Design (RDD), Event Studies, Simulations, Structural Estimation, Two-Way Fixed Effects (TWFE), and Theoretical/Non-Empirical research, across twelve fields in NBER and CEPR working papers. Each point represents the proportion of papers within a specific field that utilize a given method, with 95\% confidence intervals depicted by error bars. The plot highlights substantial field heterogeneity in method adoption. In EO \(\geq 4\), DiD is especially common in Urban (21.4\%), Labour (18.4\%), and Public (17.1\%), while RCTs are highest in Behavioral (22.0\%) and Development (13.0\%). IV is comparatively high in Urban (14.4\%), and theoretical/non-empirical usage remains concentrated in Macro (50.1\%), IO (51.5\%), and Finance (38.8\%). These cross-sectional patterns reflect field-specific identification constraints, data environments, and methodological traditions.}
\label{fig:cross_section_methods_fields}
\end{figure}

These trends are not uniform across subfields. Figure \ref{fig:cross_section_methods_fields} presents the distribution of empirical methods by field. Fields such as Labour, Public, and Urban heavily utilize DiD, with 18.4\%, 17.1\%, and 21.4\% of papers using DiD, respectively. RCTs are particularly prominent in Behavioral and Development, where they are used in 22.0\% and 13.0\% of papers, respectively, reflecting the feasibility and policy relevance of experimental interventions in these areas. This trend aligns with the findings of \citet{currie2020technology}, who documented a broad increase in empirical approaches due to the availability of big data and advanced computing.

In contrast, fields like Macroeconomics, IO, and Finance rely more on theoretical modeling to infer causal relationships from observational data. In the baseline sample, theoretical/non-empirical labeling remains common in Macro (50.1\%), IO (51.5\%), and Finance (38.8\%), while simulation usage is materially lower (5.8\% in Macro and 4.3\% in IO). These patterns suggest that methodological adoption remains shaped by the structure of field-specific research questions.

\section{Measuring Narrative Structure, Novelty, and Position in Claim Graphs}\label{sec:graphical_measures}

Because claim graphs are standardized across papers, we can compute comparable measures at scale. In this section, we define a compact set of paper-level measures that characterize the structure and content of each paper's claim graph. To address concerns about measure proliferation, we organize measures into three pre-specified families that map directly to the outcomes of interest. Narrative complexity captures how broad and deep a paper's argument is; novelty and contribution capture whether a paper introduces new links or bridges underexplored combinations; and conceptual importance/diversity capture where a paper sits relative to the existing frontier of topics. We report both non-causal and causal variants, so each result compares general narrative structure with structure supported by identified empirical designs.\footnote{These categories are not exhaustive. Future work could add dimensions such as data novelty or mechanism granularity. Our focus here is on interpretable measures with clear links to editorial and citation outcomes.} This structure sets up Section~\ref{sec:publication_citation_outcomes}, where we estimate how these families are associated with publication placement and citations.

\subsection{Measures of Narrative Complexity}
We use three interpretable complexity objects. First, the number of edges captures breadth: papers with more edges make more distinct claims. Second, the number of unique directed paths captures internal interconnection: papers with more paths connect claims into richer mechanism chains rather than isolated pairwise links. Third, the longest path captures depth: longer paths indicate longer sequences of linked reasoning. We compute each quantity for the full graph and for the causal-only subgraph. This allows us to ask whether publication and citation outcomes are related to complexity in general, or specifically to complexity backed by identified empirical strategies.

The formal notation is reported in the Appendix. In the main text, we focus on interpretation and empirical relevance.

\paragraph{Illustrative examples}

To concretely illustrate our graphical framework and the measures derived from it, we examine four landmark economic papers: \citet{chetty2014land}, \citet{banerjee2015miracle}, \citet{gabaix2011granular}, and \citet{goldberg2010imported}. These papers cover a diverse range of topics and methodologies, showcasing the versatility of our approach. The visual representations of these claim graphs are provided in Figure \ref{fig:causal_graph_examples_part1} and \ref{fig:causal_graph_examples_part2}, which highlight the varying structures and complexities of the narratives in these influential papers. Arrows indicate the direction of claims from source to sink; for non-causal edges, the arrows reflect narrative direction rather than causal identification. The displayed examples use a single extraction view (Stage 1 iteration 1 and Stage 2 iteration 1) for readability, while baseline results in the paper use EO \(\geq 4\) aggregated graphs.

In \citet{chetty2014land}, the authors investigate the geography of intergenerational mobility in the United States using administrative records. The paper presents a comprehensive analysis of how various factors correlate with upward mobility. The claim graph for this paper (Figure \ref{fig:causal_graph_chetty}) includes non-causal relationships such as \textit{parent income} (JEL code D31) to \textit{child income rank} (J13), along with links from \textit{residential segregation} (R23), \textit{income inequality} (D31), \textit{primary-school quality} (I21), \textit{social capital} (Z13), and \textit{family stability} (J12) to \textit{upward mobility} (J62). The resulting structure is source-heavy, with multiple predictors feeding into mobility outcomes.

\citet{banerjee2015miracle} report on a randomized evaluation of the impact of introducing microfinance in a new market. Utilizing Randomized Controlled Trials (RCTs), the authors assess how access to microfinance affects various economic outcomes for households in Hyderabad, India. The claim graph for this paper (Figure \ref{fig:causal_graph_banerjee}) shows a design-based structure in which a common intervention node branches into multiple downstream outcomes. The mapped relationships include how \textit{the introduction of microfinance} (G21) links to \textit{new business creation} (L26), \textit{investment in existing businesses} (G31), and spending/outcome dimensions, with downstream connections to firm-level performance outcomes.

\begin{figure}[htp]
\centering
\caption{Claim Graphs of Two Landmark Economic Papers (Part 1)}
\begin{subfigure}[t]{0.49\linewidth}
    \centering
    \caption{\citet{chetty2014land}}
    \resizebox{\linewidth}{!}{
    \begin{tikzpicture}
  \node[cg node, text width=2.20cm, minimum height=7mm, inner sep=1.5pt] (par) at (-3.0,1.7) {Parent income\\(D31)};
  \node[cg node, text width=2.20cm, minimum height=7mm, inner sep=1.5pt] (ineq) at (-2.8,0.3) {Income inequality\\(D31)};
  \node[cg node, text width=2.20cm, minimum height=7mm, inner sep=1.5pt] (school) at (-2.7,-0.8) {School quality\\(I21)};
  \node[cg node, text width=2.20cm, minimum height=7mm, inner sep=1.5pt] (fam) at (-2.4,-1.9) {Family stability\\(J12)};
  \node[cg node, text width=2.20cm, minimum height=7mm, inner sep=1.5pt] (seg) at (0.0,-2.1) {Residential\\segregation (R23)};
  \node[cg node, text width=2.20cm, minimum height=7mm, inner sep=1.5pt] (soc) at (2.4,-1.6) {Social capital\\(Z13)};
  \node[cg node, text width=2.20cm, minimum height=7mm, inner sep=1.5pt] (mob) at (0.2,-0.2) {Upward mobility\\(J62)};
  \node[cg node, text width=2.20cm, minimum height=7mm, inner sep=1.5pt] (child) at (2.8,1.7) {Child income\\(J13)};

  \draw[cg noncausal] (par) -- (child);
  \draw[cg noncausal] (ineq) -- (mob);
  \draw[cg noncausal] (school) -- (mob);
  \draw[cg noncausal] (fam) -- (mob);
  \draw[cg noncausal] (seg) -- (mob);
  \draw[cg noncausal] (soc) -- (mob);
  \draw[cg noncausal] (mob) to[out=30,in=120,looseness=8] (mob);
\end{tikzpicture}
    }
    \label{fig:causal_graph_chetty}
\end{subfigure}%
\hfill
\begin{subfigure}[t]{0.49\linewidth}
    \centering
    \caption{\citet{banerjee2015miracle}}
    \resizebox{\linewidth}{!}{
    \begin{tikzpicture}
  \node[cg node, text width=2.20cm, minimum height=7mm, inner sep=1.5pt] (micro) at (-3.0,0.2) {Microfinance access\\(G21)};
  \node[cg node, text width=2.20cm, minimum height=7mm, inner sep=1.5pt] (social) at (-0.4,1.9) {Social outcomes\\(I24)};
  \node[cg node, text width=2.20cm, minimum height=7mm, inner sep=1.5pt] (borrow) at (1.2, 1.0) {Borrowing rates\\(E43)};
  \node[cg node, text width=2.20cm, minimum height=7mm, inner sep=1.5pt] (newbiz) at (1.4,0.0) {Business creation\\(L26)};
  \node[cg node, text width=2.20cm, minimum height=7mm, inner sep=1.5pt] (cons) at (1.0,-0.8) {Consumption\\patterns (D10)};
  \node[cg node, text width=2.20cm, minimum height=7mm, inner sep=1.5pt] (invest) at (-3.0,-1.7) {Investment in\\existing firms (G31)};
  \node[cg node, text width=2.20cm, minimum height=7mm, inner sep=1.5pt] (profits) at (0.3,-1.7) {Average profits\\(D33)};

  \draw[cg causal] (micro) -- (social);
  \draw[cg causal] (micro) -- (borrow);
  \draw[cg causal] (micro) -- (newbiz);
  \draw[cg causal] (micro) -- (cons);
  \draw[cg causal] (micro) -- (invest);
  \draw[cg causal] (invest) -- (profits);
\end{tikzpicture}
    }
    \label{fig:causal_graph_banerjee}
\end{subfigure}
\vspace{-0.3cm}
\captionsetup{singlelinecheck=off,font=small,labelformat=empty,labelsep=none}
\caption*{\scriptsize\justifying \textbf{Note:} This figure presents two illustrative claim graphs from a single extraction view (Stage 1 iteration 1 and Stage 2 iteration 1). Causal relationships are shown in blue, non-causal relationships in orange. Nodes are standardized economic concepts (JEL-mapped), and arrows indicate directed claim statements from source to sink. Panel (a) shows a broad association structure for \citet{chetty2014land}, with several predictors feeding into mobility outcomes. Panel (b) shows a design-based structure for \citet{banerjee2015miracle}, where a common intervention node branches into multiple downstream outcomes. Baseline paper results use EO \(\geq 4\) aggregated graphs.}
\label{fig:causal_graph_examples_part1}
\end{figure}

In \citet{gabaix2011granular}, the author proposes that idiosyncratic firm-level fluctuations can explain a significant part of aggregate economic shocks, introducing the "granular" hypothesis. The paper develops a theoretical framework and provides empirical evidence supporting the idea that shocks to large firms contribute to aggregate volatility. The claim graph (Figure \ref{fig:causal_graph_gabaix}) summarizes a compact non-causal mechanism in which \textit{idiosyncratic shocks} feed into macro-level outcomes. This highlights how claim graphs can represent theoretical mechanism structure even when edges are not identified through canonical causal designs.

Finally, \citet{goldberg2010imported} examine the impact of imported intermediate inputs on domestic product growth in India. The authors use empirical methods, including Difference-in-Differences (DiD), to establish causal relationships between declines in input tariffs and firm performance. The claim graph (Figure \ref{fig:causal_graph_goldberg}) shows a mixed structure: causal links from \textit{input-tariff declines} (F14) to firm outcomes, together with a non-causal link from \textit{imported intermediates} (Y20) to domestic product-introduction outcomes. This illustrates how the framework separates identified and non-identified components within the same paper narrative.

\begin{figure}[htp]
\centering
\caption{Claim Graphs of Two Landmark Economic Papers (Part 2)}
% \vspace{1cm}
\begin{subfigure}[t]{0.49\linewidth}
    \centering
    \caption{\citet{gabaix2011granular}}
    \resizebox{\linewidth}{!}{
    \begin{tikzpicture}
  \node[cg node, text width=2.30cm, minimum height=7.5mm, inner sep=1.6pt] (shock) at (-2.9,0.0) {Idiosyncratic shocks\\(D89)};
  \node[cg node, text width=2.30cm, minimum height=7.5mm, inner sep=1.6pt] (prop) at (0.7,1.1) {Propagation through\\the economy (E10)};
  \node[cg node, text width=2.30cm, minimum height=7.5mm, inner sep=1.6pt] (macro) at (0.7,-1.1) {Macroeconomic\\dynamics (E19)};

  \draw[cg noncausal] (shock) -- (prop);
  \draw[cg noncausal] (shock) -- (macro);
\end{tikzpicture}
    }
    \label{fig:causal_graph_gabaix}
\end{subfigure}%
\hfill
\begin{subfigure}[t]{0.49\linewidth}
    \centering
    \caption{\citet{goldberg2010imported}}
    \resizebox{\linewidth}{!}{
    \begin{tikzpicture}
  \node[cg node, text width=2.30cm, minimum height=7.5mm, inner sep=1.6pt] (tariff) at (-2.8,1.0) {Lower input\\tariffs (F14)};
  \node[cg node, text width=2.30cm, minimum height=7.5mm, inner sep=1.6pt] (imports) at (-2.8,-1.0) {New imported\\intermediates (Y20)};
  \node[cg node, text width=2.30cm, minimum height=7.5mm, inner sep=1.6pt] (prod) at (0.8,1.5) {Higher firm\\productivity (D21)};
  \node[cg node, text width=2.30cm, minimum height=7.5mm, inner sep=1.6pt] (scope) at (0.8,0.2) {Expanded product\\scope (L25)};
  \node[cg node, text width=2.30cm, minimum height=7.5mm, inner sep=1.6pt] (domestic) at (0.8,-1.2) {New domestic\\products (L68)};

  \draw[cg causal] (tariff) -- (prod);
  \draw[cg causal] (tariff) -- (scope);
  \draw[cg noncausal] (imports) -- (domestic);
\end{tikzpicture}
    }
    \label{fig:causal_graph_goldberg}
\end{subfigure}
\vspace{-0.3cm}
\captionsetup{singlelinecheck=off,font=small,labelformat=empty,labelsep=none}
\caption*{\scriptsize\justifying \textbf{Note:} This figure presents two additional illustrative claim graphs from a single extraction view (Stage 1 iteration 1 and Stage 2 iteration 1). Causal relationships are shown in blue, non-causal relationships in orange. Panel (a) summarizes the non-causal mechanism structure in \citet{gabaix2011granular}. Panel (b) summarizes the mixed causal and non-causal structure in \citet{goldberg2010imported}. Baseline paper results use EO \(\geq 4\) aggregated graphs.}
\label{fig:causal_graph_examples_part2}
\end{figure}

These examples demonstrate the diversity in the structure and complexity of claim graphs across different types of economic research. They illustrate how our measures capture key aspects of the narratives, such as the breadth of topics covered, the depth of causal analysis, and the interconnectedness of concepts.
% \paragraph{Relation to Previous Findings}

\subsection{Measures of Novelty and Contribution to Literature}
\label{sec:novelty_contribution}

% validate it with https://desci.com/publish (see https://www.nature.com/articles/d41586-024-04021-w)

Thus far, our measures of narrative complexity focus on internal graph structure. We now add novelty and contribution measures that ask whether a paper introduces new links or connects underexplored concept pairs. We compute these for both the full claim graph and the causal subgraph, so we can distinguish general novelty from causally documented novelty. Formal definitions are in the Appendix.

\subsubsection{Novel Edges}
\label{sec:novel_edges}

An edge is novel if it has not appeared in earlier papers. We track both the number of novel edges and their share within a paper. This captures direct frontier expansion: adding links that were previously absent from the cumulative literature graph.

\subsubsection{Path-Based Novelty}
\label{sec:path_novelty}

Path-based novelty captures mechanism recombination. Even if each edge has appeared before, a paper can still be novel by assembling those edges into a new directed chain. We therefore track whether a paper contributes new paths, not only new individual links.

\subsubsection{Gap Filling (Co-occurrence Analysis)}
\label{sec:gap_filling}

Gap filling asks whether a paper bridges concept pairs that were historically rare in prior work. This provides an interpretable measure of whether the paper links previously disconnected parts of the literature. Intuitively, it complements edge and path novelty by focusing on underexplored concept intersections rather than exact directed-claim strings.

\paragraph{Which Journals Fill Gaps in Literature?}
\label{sec:gap_filling_journal_comparison}

To further illustrate our gap-filling measure, we compare how papers accepted in different journal tiers (and specifically among the top 5 journals) bridge underexplored connections. The Appendix reports these gap-filling comparisons by tier and by top-5 journal, separately for the non-causal subgraph (orange) and the causal subgraph (blue), with 95\% confidence intervals.\footnote{The first 10 years are excluded from the figure to allow the cumulative claim graph to mature, as meaningful measures of gap filling require sufficient papers and stabilized concept co-occurrence frequencies. These years are included in the calculations but omitted from visualization for clarity.}

In panel comparisons of gap filling by journal tier and within top-5 journals, the Top 5 group shows a higher mean of standard gap filling for the non-causal subgraph, but we observe very little difference across tiers for the causal subgraph. This suggests that top-tier outlets more readily reward bridging underexplored concept pairs in general, although the strictly causal bridging does not differ much between tiers. Disaggregating the five elite journals highlights especially strong causal gap filling in the \emph{QJE} and \emph{AER}.

Overall, these patterns suggest that top-tier outlets reward papers forging novel or underexplored concept connections, especially when supported by credible identification strategies.

Detailed journal-tier and top-5-journal estimates are reported in the Appendix.

\bigskip

Taken together, these measures of novelty and gap filling enrich our understanding of how each paper pushes the frontier of economic knowledge. By systematically tracking new edges, new paths, and underexplored concept pairs, we capture distinct facets of a paper's originality and the extent to which it addresses previously overlooked topics or mechanisms. As shown in Section~\ref{sec:publication_citation_outcomes}, these indicators are significantly associated with both publication outcomes and the long-run citation impact of research.

% \paragraph{Gaps in political economy (or labour economics) over decade}
% - e.g. find the gaps within the JEL code (perhaps 2 digit) in 1990, 2000, 2010, 2020.

\subsection{Measures of Topic Centrality and Diversity}
\label{sec:conceptual_importance_and_diversity}

We now turn to assessing how centrally each paper's concepts lie within the broader economic literature and whether the paper balances multiple sources and targets in its argumentation. These attributes may shape both the paper's perceived significance and its eventual scholarly influence. We consider two sets of measures: (i) the source--sink ratio, which captures the balance of causal flows in the paper, and (ii) centrality-based indicators (e.g., eigenvector or PageRank scores) computed for the nodes in a cumulative claim graph of prior literature. We then aggregate these centrality values (mean, variance) over the specific concepts a paper employs, providing measures of conceptual importance and conceptual diversity.

\subsubsection{Source--Sink Ratio}
The source--sink ratio summarizes whether a paper's claim graph has a ``fan-in'' structure (many distinct drivers feeding into a smaller set of outcome) or a ``fan-out'' structure, where a small set of drivers generate claims about many downstream outcomes. Values above one indicate a source-heavy architecture (more distinct drivers than outcomes), while values below one indicate an outcome-heavy architecture (fewer drivers linked to more downstream outcomes). We compute the ratio for both the full claim graph and the causal-only subgraph.

This measure is included because it captures an aspect of argument architecture that simple count-based complexity measures miss. Two papers can have the same number of edges and paths but tell qualitatively different stories: one may aim to explain a focal outcome by adjudicating among multiple candidate determinants (fan-in), while another may emphasize the consequences of a focal shock, policy, or intervention across multiple outcomes (fan-out). These architectures plausibly matter for evaluation and diffusion: fan-in structures speak to mechanism and channel tests around a central outcome, whereas fan-out structures speak to external relevance and the breadth of implications. Distinguishing causal from non-causal versions allows us to ask whether any shift in architecture is driven by identified relationships rather than by narrative expansion per se. The Appendix reports notation and implementation details.

\begin{figure}[htp]
\centering
\caption{Source--Sink Ratio Over Time and by Field}
\begin{subfigure}[t]{0.7\textwidth}
    \centering
    \caption{Time Trends (All vs.\ Causal)}
    \includegraphics[width=\textwidth]{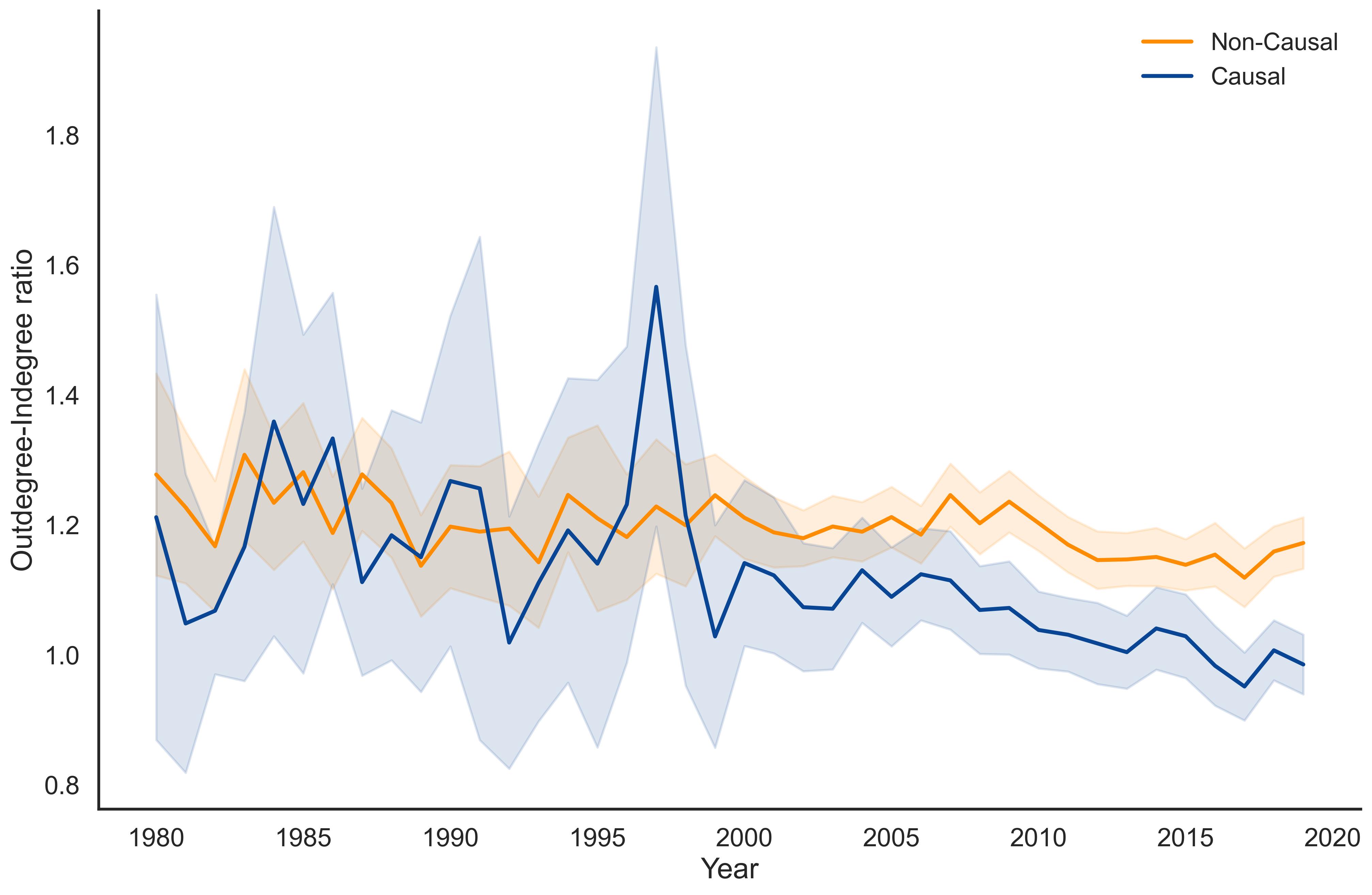}
    \label{fig:source_sink_ratio_trends_a}
\end{subfigure}
\hfill
\begin{subfigure}[t]{0.8\textwidth}
    \centering
    \caption{Cross-Field Comparisons}
    \includegraphics[width=\textwidth]{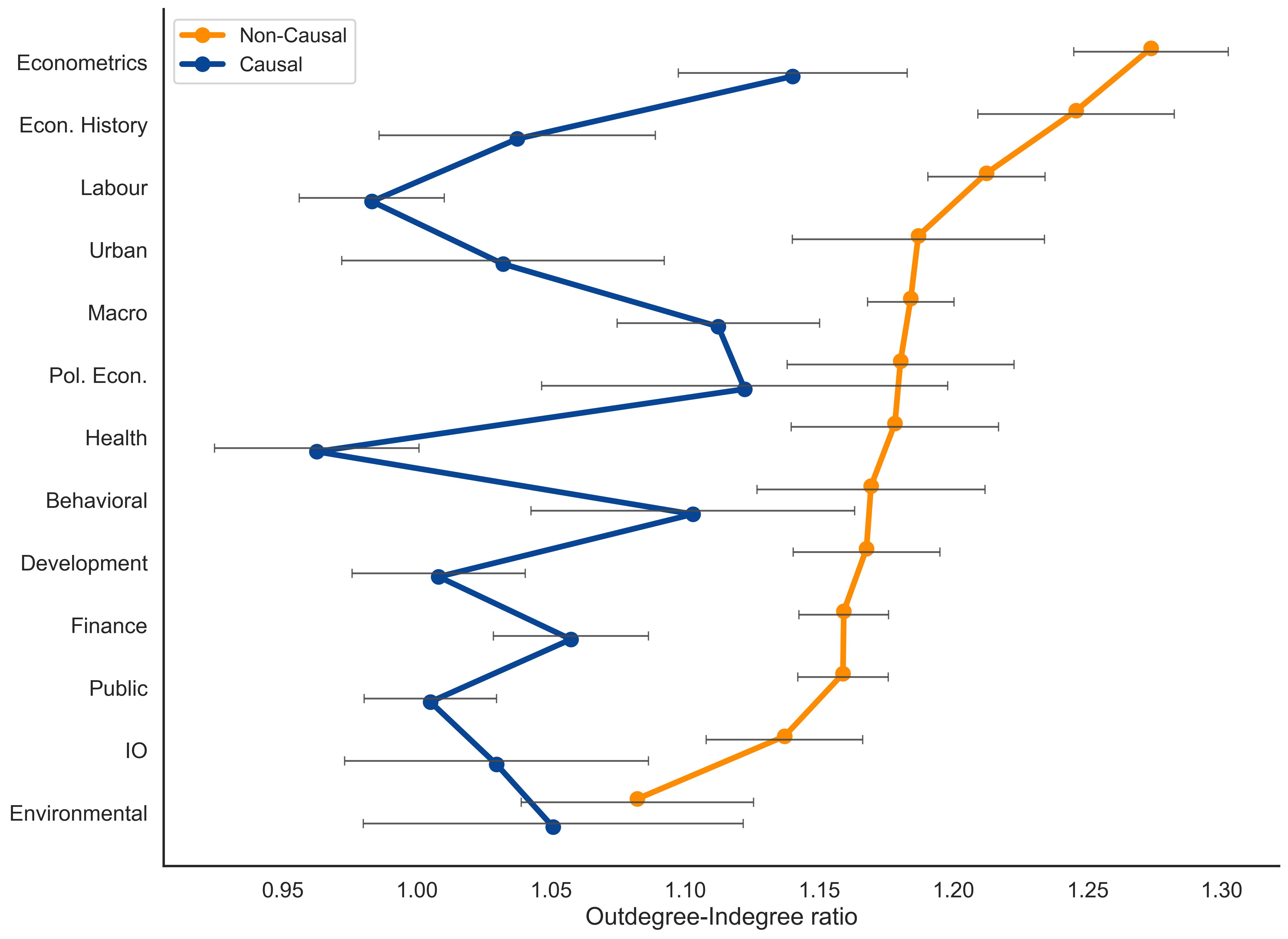}
    \label{fig:source_sink_ratio_trends_b}
\end{subfigure}
\vspace{-0.4cm}
\captionsetup{singlelinecheck=off,font=small,labelformat=empty,labelsep=none}
\caption*{\scriptsize\justifying \textbf{Note:} Panel (a) plots mean source--sink ratios (also labelled outdegree--indegree ratio) over time (1980--2023) for both the non-causal subgraph (orange) and the causal subgraph (blue), with shaded areas representing 95\% confidence intervals. The Causal series rises steadily, indicating that recent papers tend to incorporate more distinct causal drivers than outcomes. Panel (b) shows field-level average outdegree--indegree ratios, again splitting Non-Causal vs.\ Causal. Fields such as Behavioral and Health rank highly in the Causal version of the ratio, suggesting an emphasis on multiple identified factors leading to a smaller set of outcomes, whereas Econ History and Econometrics display relatively higher Non-Causal version of the ratios.}
\label{fig:source_sink_ratio_trends}
\end{figure}

To examine how the source--sink ratio evolves, Figure~\ref{fig:source_sink_ratio_trends}\textbf{(a)} plots annual averages with 95\% confidence bands. The non-causal series is comparatively flat through 2000 and then declines slightly, while the causal series rises, consistent with an increasing emphasis on designs that support multiple identified drivers or channels for a smaller set of focal outcomes.

Figure~\ref{fig:source_sink_ratio_trends}\textbf{(b)} compares fields cross-sectionally. Economic History exhibits relatively high non-causal ratios, while Behavioral and Health show higher causal ratios, consistent with greater use of designs that estimate multiple identified margins around policy-relevant outcomes.

\subsubsection{Centrality-Based Measures of Conceptual Importance and Diversity}

Centrality measures capture where a paper's concepts sit in the cumulative economics claim graph built from prior papers. We focus on two intuitive summaries at the paper level: average centrality (conceptual importance) and dispersion of centrality (conceptual diversity). Average centrality is high when a paper is anchored in already influential topics; dispersion is high when a paper mixes mainstream and peripheral concepts. We compute these summaries using both eigenvector centrality and PageRank, and we report non-causal and causal variants.

These measures are included to separate two empirically relevant mechanisms. First, papers tied to central topics may diffuse more easily and attract more citations. Second, papers that combine central and peripheral ideas may be harder to place ex ante but can create broader long-run influence.

\subsubsection{Illustrative examples}

To make the architecture and positioning measures concrete, we report the source--sink ratio and centrality summaries for the four papers whose claim graphs are shown in Figures~\ref{fig:causal_graph_examples_part1}--\ref{fig:causal_graph_examples_part2}. Centrality is computed on the cumulative claim graph formed by earlier papers, separately for the non-causal and causal subgraphs. For comparability, we report standardized (z-scored) paper-level summaries.\footnote{For each paper we compute (i) the mean eigenvector centrality of the concepts appearing in its claim graph and (ii) the dispersion of those centralities within the paper (the within-paper variance), and then standardize these summaries so that 0 is the sample mean and 1 is one standard deviation. Negative values therefore indicate below-average centrality (or below-average dispersion), not negative raw variance.}

\citet{chetty2014land}. In our extraction, retained edges for this paper are non-causal, so the reported ratio uses the non-causal graph. The source--sink ratio \(R_p \approx 2.5\) indicates a fan-in architecture: multiple determinants (e.g., residential segregation, inequality, school quality, social capital, and family stability) point to a focal outcome (upward mobility). The standardized mean eigenvector centrality is \(-0.39\), and the standardized dispersion is \(-0.46\), implying that the paper draws on concepts that are somewhat less central than average in the prior non-causal literature graph, and that those concepts are relatively similar in centrality.

\citet{banerjee2015miracle}. The causal source--sink ratio \(R_p \approx 0.71\) indicates a fan-out architecture: a focal intervention (microfinance introduction) branches into multiple downstream outcomes (e.g., borrowing behavior, business creation, and expenditures). The standardized mean eigenvector centrality is \(-0.35\) and the standardized dispersion is \(-0.22\), suggesting that the paper's causal concepts are modestly below-average in centrality in the prior causal literature graph.

\citet{gabaix2011granular}. In the non-causal graph, \(R_p \approx 0.8\) reflects a mechanism in which firm-level shocks propagate to macro outcomes (e.g., aggregate volatility). The standardized mean eigenvector centrality is \(-0.15\) (dispersion: \(-0.14\)), indicating concepts that are closer to the average position in the prior non-causal graph than in the previous two examples, with modest within-paper dispersion.

\citet{goldberg2010imported}. This paper contains both causal and non-causal components in our representation. In the non-causal subgraph, the source--sink ratio is about \(1.33\), while in the causal subgraph it is about \(2.0\), consistent with a fan-in causal structure around a narrower set of firm outcomes. The standardized mean eigenvector centrality is \(-0.60\) (dispersion: \(-0.26\)) in the non-causal graph and \(-0.57\) (dispersion: \(-0.24\)) in the causal graph, placing the paper's concepts below the average centrality level in both representations.

Taken together, these examples illustrate that the source--sink ratio and centrality-based summaries capture distinct features: the architecture of the argument (fan-in versus fan-out) and the positioning of a paper's concepts in the evolving literature graph (average centrality and dispersion). We use the same measures below in the systematic trend and outcome analyses.

\subsubsection{Evolution of Concept Centrality Over Time in Literature}
\label{subsubsec:concept_centrality_evolution}

We next investigate how the centrality of specific economic concepts has changed over time, comparing both the non-causal and the causal subgraph. First, we highlight which JEL codes rank most highly in eigenvector centrality, noting differences between non-causal edges and those established through causal inference methods. We then track how certain concepts have risen or fallen in importance since 1990.

\paragraph{Most central concepts in economics}

\begin{figure}[htp]
\centering
\caption{Largest and Smallest Causal-vs-Non-Causal Centrality Gaps}
\includegraphics[width=0.9\textwidth]{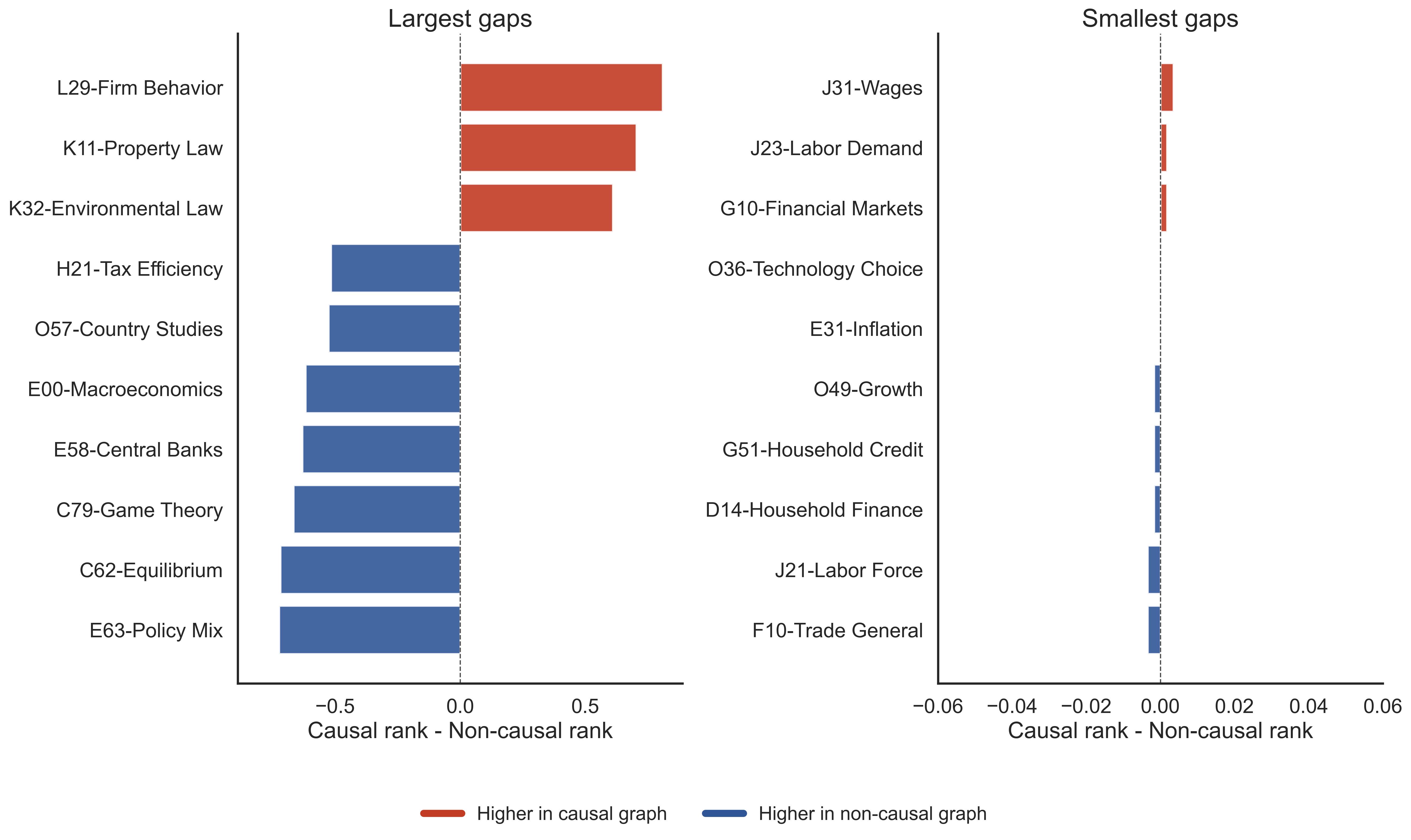}
\captionsetup{singlelinecheck=off,font=small,labelformat=empty,labelsep=none}
\caption*{\scriptsize\justifying \textbf{Note:} For each concept, the plotted statistic is rank in the causal graph minus rank in the non-causal graph, using mean eigenvector centrality over time. The left panel shows the ten largest absolute gaps; positive bars indicate concepts relatively more central in the causal graph, while negative bars indicate concepts relatively more central in the non-causal graph. Because selection is by absolute gap among high-centrality concepts, the sign mix is not forced to be symmetric (more negative bars can appear if downgrades dominate). The right panel shows ten concepts with the smallest gaps (near-zero differences), highlighting stable centrality across graph definitions. Labels are shown as JEL-short names for readability.}
\label{fig:top_20_jel_nodes_centrality_combined}
\end{figure}

Figure~\ref{fig:top_20_jel_nodes_centrality_combined} focuses on relative centrality shifts between the causal and non-causal graphs. For each concept, we compute the difference between its rank in the causal graph and its rank in the non-causal graph, then show the ten largest gaps and ten smallest gaps among sufficiently common concepts.\footnote{The distribution of rank differences is skewed toward negative values among high-centrality concepts, so the ``largest gaps'' panel is not forced to be symmetric: there are more concepts that drop in rank when we focus on causal edges than concepts that rise. This asymmetry is itself informative, suggesting that many highly central topics are widely discussed in non-causal narratives but less often supported by identified causal designs, whereas a smaller subset of topics rises in prominence under causal identification (e.g., firm behavior, property law, environmental regulation).}

The largest-gap panel highlights concepts whose prominence changes sharply once we restrict attention to causally identified claims (e.g., L29-Firm Behavior and K11-Property Law move up in the causal ranking, while E63-Policy Mix and C62-Equilibrium move down). The smallest-gap panel identifies concepts with very similar prominence in both graphs (e.g., E31-Inflation, J31-Wages, O49-Growth), indicating stable centrality regardless of identification framing.

\paragraph{Rise and Fall of Importance of Certain topics}

Beyond cross-sectional gaps, we also track dynamic movement in node centrality over time. An Appendix trend figure reports the top three rising and top three declining concepts (separately in non-causal and causal graphs), with explicit legends in JEL-short-name format. For example, I24-Education-Inequality and I12-Health-Production rise in the causal graph, while J31-Wages and E43-Interest Rates decline; in the non-causal graph, D72-Political Process and I31-Well-Being rise while F31-FX Markets and E22-Investment decline.

For reference, an Appendix level-comparison figure also reports the traditional top-20 centrality comparison.

\section{Claim-Graph Predictors of Publication and Influence} \label{sec:publication_citation_outcomes}

We now link claim-graph features to two distinct margins of research success. The first is editorial selection, captured by publication in increasingly selective journal tiers. The second is diffusion, captured by long-run citations. These outcomes need not move together: editorial screening may prioritize credibility and contribution as assessed ex ante, while citations also reflect topic salience and broad reuse. Claim graphs let us separate narrative structure from evidentiary support and ask which components are most closely associated with each margin.

To keep the main text focused, we summarize results for five pre-specified headline measure families in one consolidated figure. The Appendix reports (i) alternative operationalizations within each family and (ii) sensitivity to the edge-overlap (EO) threshold used to construct paper-level graphs.

\subsection{Empirical Framework and Variables}

For each paper-level measure \(M_p\), we estimate the following associational specification:
\begin{equation}
\label{eq:baseline_regression}
y_{p} = \alpha + \beta M_{p}^{(\text{type})} + \delta_{t(p)} + \varepsilon_{p},
\end{equation}
where \(y_{p}\) is one of four outcomes: publication in a top-five economics journal, publication in journals ranked 6--20, publication in journals ranked 21--100, or \(\log(\mathrm{Cites}_{p}+1)\). All displayed specifications include year fixed effects \(\delta_{t(p)}\). We interpret \(\beta\) as a conditional association, not a causal effect. The superscript \((\text{type})\) indicates whether the measure is computed on the causal subgraph or on the non-causal subgraph when both variants are defined.

\subsection{Headline predictors}

\begin{figure}[htp]
\centering
\caption{Headline Claim-Graph Predictors of Publication and Influence}
\includegraphics[width=\textwidth]{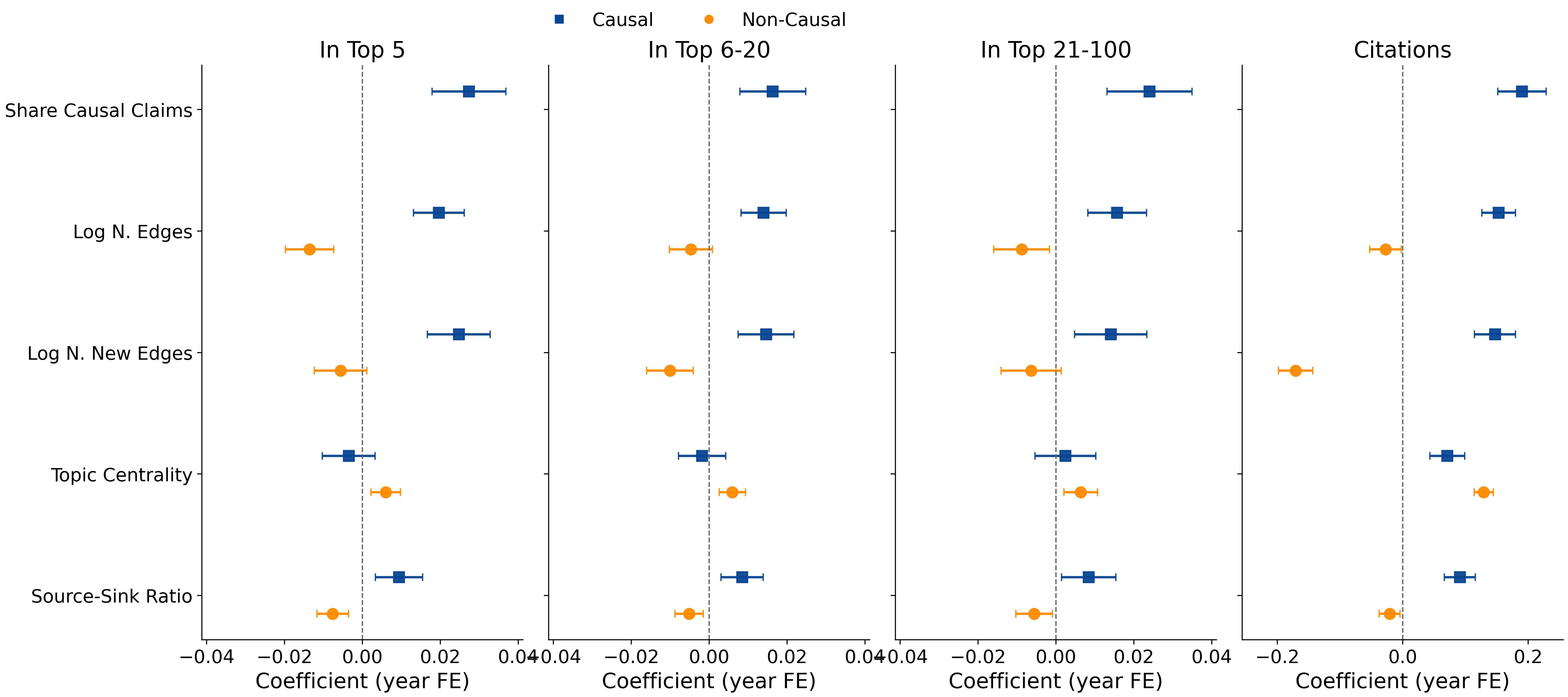}
\captionsetup{singlelinecheck=off,font=small,labelformat=empty,labelsep=none}
\caption*{\scriptsize\justifying \textbf{Note:} Each panel corresponds to one outcome (Top 5, Top 6--20, Top 21--100, citations). Rows report the five headline measure families used in the main text. Blue squares are coefficients for causal variants; orange circles are non-causal variants. For share causal claims, only the causal-series coefficient is defined. Horizontal bars are 95\% confidence intervals from year-fixed-effects regressions.}
\label{fig:publication_predictors_main}
\end{figure}

Figure~\ref{fig:publication_predictors_main} reports coefficients for five headline measure families:(i) the share of a paper's claims that are supported by canonical causal designs, (ii) the log number of edges (claim volume), (iii) the log number of new edges (edge-level novelty), (iv) topic centrality (conceptual positioning in the prior literature graph), and (v) the source--sink ratio (argument architecture).

These five families are designed to span distinct, interpretable dimensions: evidence basis (i), narrative breadth (ii), frontier expansion (iii), positioning in the cumulative literature (iv), and whether the paper's argument has a fan-in versus fan-out structure (v). The Appendix reports summary statistics, availability, and correlations, and shows that these families are not redundant in the baseline sample.

A central test in this section is whether evidentiary support matters beyond narrative size and topic choice. Many papers can be large or complex in purely descriptive or theoretical terms. By comparing causal and non-causal variants of the same construct (e.g., causal edges versus non-causal edges), we can assess whether associations with publication and citations are concentrated in relationships backed by identified designs, or whether they are similar for non-causal expansion of the narrative.

The headline pattern is that causal variants are consistently positive across publication tiers and citations, while non-causal counterparts are often weak or negative. For example, the share of causal claims is positive for top-five placement (\(\beta=0.027\), 95\% CI \([0.018, 0.037]\)) and citations (\(\beta=0.190\), 95\% CI \([0.151, 0.229]\)). Likewise, causal edge volume (\(\beta=0.020\) for top five; \(\beta=0.153\) for citations) and causal new-edge volume (\(\beta=0.025\) for top five; \(\beta=0.147\) for citations) are positive and precisely estimated, while the corresponding non-causal coefficients are negative for top-five placement and for citations.

Topic centrality behaves differently. Non-causal centrality is strongly positive for citations (\(\beta=0.129\), 95\% CI \([0.114, 0.144]\)) and is also positive across publication tiers, whereas causal centrality is mainly linked to citations (\(\beta=0.071\), 95\% CI \([0.043, 0.099]\)) and is imprecise for journal placement. This wedge is consistent with diffusion rewarding conceptual proximity to widely used topics even when editorial selection places relatively more weight on other dimensions of contribution.

Finally, the source--sink ratio is positive in its causal version (\(\beta=0.009\) for top five; \(\beta=0.091\) for citations) and negative in its non-causal version (\(\beta=-0.008\) for top five; \(\beta=-0.021\) for citations). Interpreted through the architecture lens, papers that use identified designs to speak to multiple determinants or channels for a focal outcome are more likely to place in top outlets and to be cited, while non-causal fan-in expansion is not similarly rewarded.

To compare magnitudes across families, the Appendix translates coefficients into common shifts: +10 percentage points for share causal claims, doubling for log-count predictors, and interquartile-range shifts for source--sink and centrality measures, alongside +1 SD shifts. The largest top-five association is for causal new-edge expansion: doubling is associated with +1.71 percentage points (about +15\% relative to the 11.35\% baseline rate, or about 17 additional top-five papers per 1,000). Causal edge volume is close behind at +1.36 percentage points (about +12\%). In contrast, a +10 percentage-point rise in the share of causal claims has a smaller top-five association (+0.27 percentage points), and the causal source--sink ratio has a modest association (+0.31 percentage points for an interquartile-range shift). For citations, the largest gains are linked to causal edge expansion (about +10.7\% to +11.2\%), while non-causal novelty is strongly negative (about -11.2\% for a doubling). Topic centrality is the main exception: citation gains are strongest for the non-causal variant, consistent with broader diffusion being tied to mainstream conceptual anchoring.

Taken together, these results suggest that journal placement and citation impact are most strongly associated with richer claim structure when it is supported by canonical identification designs, whereas expansion of non-causal claim structure is not similarly rewarded. At the same time, broad diffusion continues to reward central conceptual positioning, highlighting a separation between what appears most aligned with editorial selection and what travels most widely through the literature.

\subsection{Variants and robustness}

The Appendix reports alternative operationalizations for the same underlying constructs: narrative depth (unique paths, longest path), novelty (novel-path shares and gap filling), and conceptual positioning (topic diversity and PageRank-based centrality). These variants are useful because they probe whether the headline patterns reflect a particular measurement choice or a broader distinction between identified and non-identified structure.

The qualitative pattern remains aligned with the headline figure: causal variants are more often positive for top-tier placement and citations, while non-causal variants are weaker or negative in several outcomes. For example, causal longest-path coefficients remain positive for both top-five placement and citations, while non-causal longest-path coefficients are negative.

An Appendix EO-threshold grid figure shows that the headline coefficients are stable across EO thresholds in the region used for the main analysis (EO \(\geq 4\)), with uncertainty widening only at stricter thresholds where retained edge sets become thin.

\section{Discussion: Beyond Causal Claims}\label{sec:discussion}
Claim graphs provide a representation layer that makes claim content comparable across papers: they convert narrative prose into a standardized set of concepts and directed relationships, annotated by evidentiary basis. This representation enables claim-level meta-research that is difficult with bibliometrics alone, including comparisons of how identified and non-identified relationships combine into mechanisms, how novelty arises through new links or new chains, and how conceptual positioning shifts as the literature evolves.

The approach also has clear limitations. Claim graphs extract what papers state, not whether claims are true; mapping to JEL codes compresses nuance; and the fixed 30-page window can miss late-paper qualifications or robustness discussions. Extraction quality depends on document clarity and model stability. We address these issues by using repeated runs, overlap-based aggregation with an explicit precision--recall trade-off, and layered validation, but continued benchmarking and community feedback will remain important as claim-level infrastructure develops.

The same representation can support extensions that move beyond the causal/non-causal split. Natural next modules include richer mechanism and mediator classification, robustness and sensitivity language, data provenance and accessibility claims, explicit external-validity statements, and cross-paper synthesis of specific causal pathways. More broadly, the framework is not tied to economics; it can be adapted to other domains as suitable concept ontologies mature.

\section{Conclusion}
\label{sec:discussion_conclusion}

This paper introduces evidence-annotated claim graphs as a scalable representation of what economics papers claim and how those claims are supported. We map each paper to a directed graph of standardized economic concepts (JEL nodes) and stated relationships (edges), and we label edges by evidentiary basis, distinguishing links supported by canonical causal designs from links supported by non-causal evidence. Using a structured multi-stage LLM extraction pipeline with repeated runs and edge-overlap aggregation, we construct claim graphs for 44{,}852 NBER and CEPR working papers from 1980--2023.

Using these data, we document a large increase in the share of causally supported edges, from 7.7\% in 1990 to 31.7\% in 2020, with substantial heterogeneity across fields. We then relate paper-level graph measures to publication and citations. Measures of causal claim structure and causal novelty are positively associated with top-tier publication and long-run citations, while non-causal complexity and non-causal novelty are weakly related or negative. In contrast, citation diffusion remains strongly related to conceptual positioning: papers anchored in central concepts tend to receive more citations, even when editorial selection appears more closely aligned with causally documented structure and contribution.

More broadly, claim graphs offer a reusable representation for measuring how evidence, novelty, and narrative structure evolve, and for studying how incentives shape the construction and dissemination of claims. We view this paper as a first step toward claim-level infrastructure for cumulative science: a layer that makes the content of papers queryable, comparable, and aggregable at scale.

\clearpage
\appendix
\setcounter{table}{0}
\setcounter{figure}{0}
\renewcommand{\thetable}{A\arabic{table}}
\renewcommand{\thefigure}{A\arabic{figure}}

\clearpage
\thispagestyle{plain}
\begin{center}
{\Large\bfseries Appendix}
\end{center}
\vspace{0.75em}
\begin{center}
\begin{minipage}{0.92\textwidth}
\small
\noindent\textbf{Contents}
\vspace{0.35em}

\begin{tabular}{p{0.82\textwidth}r}
\hyperref[sec:appendix_retrieval_details]{A1. Details on LLM-based Information Retrieval} & \pageref{sec:appendix_retrieval_details} \\
\hyperref[subsec:appendix_jel_matching]{A2. Matching LLM Output to JEL Codes} & \pageref{subsec:appendix_jel_matching} \\
\hyperref[sec:edge_overlap_robustness]{A3. Edge-Overlap Robustness} & \pageref{sec:edge_overlap_robustness} \\
\hyperref[sec:validating_information_retrieval]{A4. Validating information retrieval} & \pageref{sec:validating_information_retrieval} \\
\hyperref[sec:appendix_graphical_measures]{A5. Paper-level Graphical Measures} & \pageref{sec:appendix_graphical_measures} \\
\end{tabular}
\end{minipage}
\end{center}
\clearpage

\section{Details on LLM-based Information Retrieval} \label{sec:appendix_retrieval_details}

\paragraph{Background on Large Language Models}\label{subsec:background_llms}
Large Language Models (LLMs) such as GPT-4o-mini have significantly advanced natural language processing by enabling machines to understand and generate human-like text. Pre-trained on extensive corpora (including academic papers, books, and web text), these models capture semantics and context and can be deployed for structured retrieval, summarization, and measurement tasks \citep{Ash2023-at, dell2024deep, aipnet, Korinek2023GenAI, ludwig2025economistsllm, humlum2025large, feyzollahi2025adoptionllm, asirvatham2026gptmeasurement}. The economics literature using LLMs has expanded rapidly through 2025 and 2026, including applications that explicitly use GPT-family models as measurement instruments.

In our study, we leverage the LLM's ability to extract structured information from unstructured text. By processing the first 30 pages of each economics working paper, the LLM identifies and extracts key metadata, methodological details, and claim-level relationships. This fixed input window captures the most policy-relevant and methodologically central sections while preserving comparability across papers and publication years.

\subsection{Prompt Design and Multi-Stage Extraction Process}

Our information retrieval process follows a prompt-chaining architecture, where Stage 1 produces structured summaries and Stage 2 transforms those summaries into claim-level graph edges with standardized attributes. We use repeated runs and edge-overlap aggregation to reduce reliance on any single stochastic response.

\subsubsection{Design Choices and Trade-offs}

We made several deliberate prompt-design choices to balance accuracy, coverage, and transparency:
\begin{itemize}
    \item \textbf{Fixed input window (first 30 pages).} We chose a fixed window to preserve comparability across papers and years while focusing on the abstract, introduction, and core empirical sections where identification and claims are typically stated. The trade-off is that some appendices or late-paper robustness checks may be missed.
    \item \textbf{Two-stage extraction rather than one-shot edges.} Stage 1 produces a curated summary plus verbatim snippets. Stage 2 turns those into edges. This improves recall on the full narrative but introduces a potential summary-to-edges translation step, which we later audit with snippet-only extraction.
    \item \textbf{Structured JSON schema.} We enforce schemas to avoid unconstrained natural-language drift and to ease downstream validation. The trade-off is occasional schema adherence errors, which we handle by re-running and aggregation.
    \item \textbf{Multiple iterations with aggregation.} Running each stage three times (nine edge lists per paper) enables stability checks and explicit EO thresholds. The trade-off is higher compute cost, but it provides a measurable precision--recall control.
    \item \textbf{Alternative prompting variants.} We also test a snippet-only Stage 2 that excludes summaries and uses only verbatim claim snippets. This is intentionally more conservative and forms a self-consistency validation rather than a human-labeled benchmark.
\end{itemize}

We emphasize that these choices reflect a cautious stance: they are designed to reduce hallucination risk, preserve traceability to text, and provide explicit robustness checks. Other reasonable designs include model-ensemble voting, page-by-page sliding windows, or direct full-text edge extraction with stricter filtering. Our current design reflects the empirical trade-offs we found most reliable for a large-scale corpus.

\subsubsection{Stability, Model Drift, and Truncation}

Our extraction is a measurement system rather than a deterministic parser, so we explicitly address stability and scope. First, we mitigate non-determinism by running three independent Stage 1 summaries and three Stage 2 edge extractions per paper, yielding nine runs. We aggregate edges by edge overlap (EO), and show stability of headline signs from EO \(\geq 1\) to EO \(\geq 4\) in the Appendix. Second, we fix prompts and model versions at the time of extraction and treat the resulting dataset as a snapshot; future model updates can shift outputs, so we archive prompts, schemas, and derived outputs alongside the analysis. Third, we define causal edges narrowly: DiD (including TWFE/event-study implementations), IV, RCTs, RDD, and synthetic control qualify; structural estimation and causal language without an identified design remain non-causal. This tight definition is aligned with the credibility-revolution canon and reduces ambiguity.

We also address truncation. We process the first 30 pages of each paper, which typically include the abstract, introduction, identification strategy, and main results where claims are stated most clearly. Our goal is not to recover precise effect sizes, but to capture whether a qualitative claim exists and which identification strategy is invoked. Late-page appendices and robustness checks can add nuance, but we find that the core claim structure is already present early in most economics papers. We further audit reliability using a snippet-only Stage 2 variant and external benchmarks (the Appendix), which together provide evidence that the key retrieval dimensions are stable and interpretable.

\subsubsection{Full Prompts and Schemas} \label{sec:appendix_full_prompt}

For submission readability, we summarize the prompt architecture in the main manuscript and provide full prompt text plus JSON schemas in the reproducibility package and linked project repository.
Direct archive URL: \url{https://github.com/prashgarg/CausalClaimsInEconomics}. Prompt and schema artifacts are versioned in the linked repository under dedicated \texttt{prompts/} and \texttt{schemas/} directories.

Stage 1 uses constrained system/user templates to extract structured paper summaries.
Stage 2 maps those summaries to claim-graph edges under a fixed schema.
A snippet-only Stage 2 variant is used for self-consistency validation.

\section{Matching LLM Output to JEL Codes} \label{subsec:appendix_jel_matching}

After the LLM extracted the causal claims and provided free-text descriptions of the source and sink variables, we needed to standardize these variables to facilitate aggregation and systematic analysis across the corpus. This standardization was achieved by mapping the variable descriptions to official Journal of Economic Literature (JEL) codes and OpenAlex Topics.

\subsubsection{Choice of Standardization Methods}

We considered several options for standardizing the terms used in the source and sink variables, including JEL codes. Each option had its advantages and limitations. JEL codes are well-understood within the economics community and facilitate interpretability but are relatively broad in classification. OpenAlex Topics offer more granularity with around 4,500 topics but are less familiar to economists. Concepts provide even more detail with approximately 60,000 terms but are being deprecated by OpenAlex and are not widely recognized in economics. The OECD Glossary contains about 6,700 economics-related terms but may be biased towards statistical concepts.

After careful consideration, we opted to focus on JEL codes for standardization. JEL codes were chosen as our primary method due to their familiarity and acceptance within the economics discipline, facilitating interpretation and communication of results.\footnote{OpenAlex Topics could also be used as a supplementary method, providing additional granularity and capturing interdisciplinary aspects not covered by JEL codes. OpenAlex Topics will be useful when scaling to research beyond economics.}

\subsubsection{Embedding-Based Matching Methodology}

To map the free-text variable descriptions to the standardized codes, we employed an embedding-based matching approach using vector representations of the texts. We utilized the OpenAI text embedding model \texttt{text-embedding-3 large}, which generates 3,072-dimensional vector embeddings that capture the semantic meaning of the text. Embeddings were generated for: (i) the free-text descriptions of the source and sink nodes extracted by the LLM, and (ii) the official descriptions of JEL codes, enhanced by concatenating descriptions from higher-level codes to provide richer context.

We calculated the cosine similarity between the embeddings of the variable descriptions and the embeddings of the JEL codes.\footnote{Cosine similarity measures the cosine of the angle between two vectors, providing a value between $-1$ and $1$, where higher values indicate greater similarity.} We assigned the variable to the codes with the highest similarity scores.

\subsubsection{Advantages and Limitations of the Embedding-Based Approach}

By leveraging embeddings, we moved beyond simple keyword matching, which can be limited by variations in terminology and phrasing. The embedding-based approach captures the semantic meaning of the texts, allowing us to match variable descriptions to standardized concepts even when different terms are used to describe similar ideas (e.g., unemployment rate'' vs.\ joblessness''). This method enhanced the robustness of our matching process, reducing the impact of typos, synonyms, and variations in language. It allowed us to systematically standardize a large number of variable descriptions across the corpus, facilitating cross-paper comparisons and aggregations.

While the embedding-based matching approach offers significant advantages, there are limitations to consider. The quality of the matches depends on the accuracy of the embeddings and the chosen similarity threshold, which in our case is the one with highest similarity. By focusing on only one matching concept, we are imposing a structure on the latent causal graph. It may well be that a source or a sink could be captured by multiple JEL codes. However, for simplicity and consistency across papers, we decided to capture only the best match. Future research can consider setting a threshold, e.g., 0.85, beyond which multiple JEL codes can be accepted. However, this comes with its own set of hyper-parameter selection: there is a trade-off between precision and recall; a higher threshold increases precision but may miss relevant matches, while a lower threshold increases recall but may include irrelevant matches.\footnote{One application of having one-to-many-mappings as a result of a similarity threshold is a study of growing interdisciplinary nature of economics research.}

% \subsection{Validation and Quality Assurance}

% To ensure the reliability of the extracted data, we implemented validation checks at each stage. The structured outputs were parsed and checked for compliance with the specified schemas. In cases where inconsistencies or missing data were detected, the prompts were refined, and the extraction was repeated.

% We also conducted manual reviews of a sample of the extracted data to assess accuracy. This included checking the correctness of the causal claims extracted, the appropriateness of the mapped JEL codes, and the consistency of metadata. The feedback from these reviews informed further refinements to the prompts and extraction process.

% % A large scale validation of our approach will follow in future drafts. This includes contacting corresponding authors to validate the causal graphs in their own papers. 

\subsection{Limitations and Considerations}

While the use of LLMs provides significant advantages in processing large volumes of complex text, there are limitations to consider. The LLM's extraction is dependent on the quality and clarity of the original text; ambiguities or omissions in the papers may lead to incomplete extraction. In theory, the LLM may occasionally misclassify or misinterpret information, particularly with nuanced methodological details (our multiple iterations of same prompt aims to minimize this limitation). While we collected additional attributes such as effect sizes and statistical significance, these features were experimental and not used in the main analysis due to variation in reporting standards, standardization of which is not yet a norm in economics.

Additionally, the embedding-based matching approach for standardizing variables may not capture all nuances of the economic concepts involved. There is a risk of misclassification if the variable descriptions are ambiguous or if the embeddings do not accurately represent the semantic content. Despite these limitations, we believe that the overall methodology provides a robust framework for large-scale analysis of economic research.

\subsection{Replication Package}

We provide a replication package with derived datasets (edge lists and paper-level measures), analysis code, prompts, and schemas. The package is available via \url{https://github.com/prashgarg/CausalClaimsInEconomics}. For legal reasons, we do not redistribute proprietary PDFs; researchers can re-download the source working papers from NBER/CEPR and rerun the retrieval using the included prompts and scripts. This openness level enables full regeneration of tables and figures from released analysis data, while retaining traceability to the original text extraction design.

\section{Edge-Overlap Robustness} \label{sec:edge_overlap_robustness}

Our baseline specification uses EO \(\geq 4\), meaning that an edge must appear in at least four of nine extraction runs to enter the paper-level graph. This threshold is intentionally conservative relative to EO \(\geq 1\) (union) while preserving coverage. The EO grid below shows that core patterns are stable over EO \(\geq 1\) to EO \(\geq 4\), with expected attenuation of sample size and increased noise at very strict thresholds (EO \(\geq 6\) and above).

\begin{figure}[htp]
\centering
\caption{Overlay Time Trends by EO Threshold}
\includegraphics[width=0.86\textwidth]{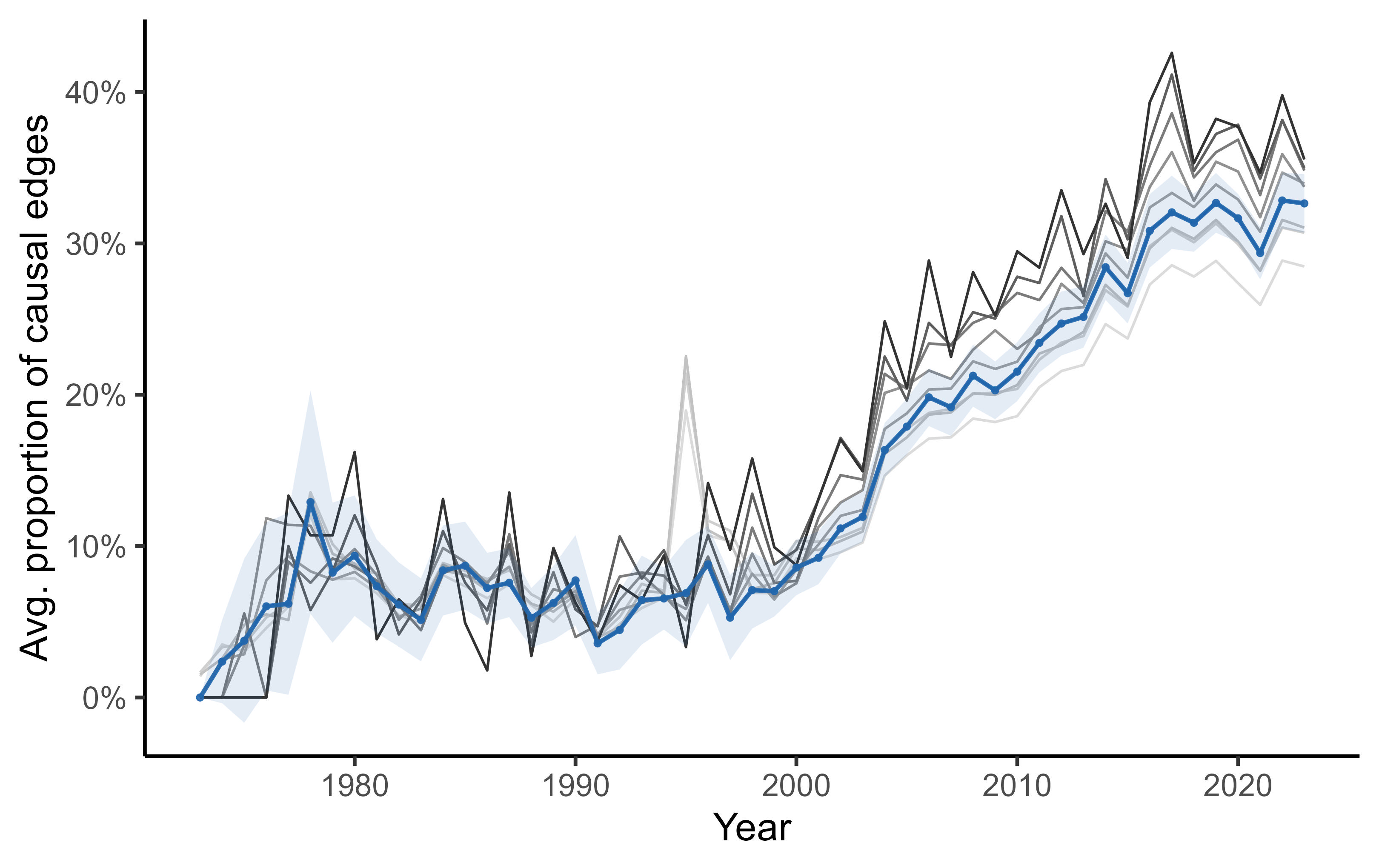}
\captionsetup{singlelinecheck=off,font=small,labelformat=empty,labelsep=none}
\caption*{\scriptsize\justifying \textbf{Note:} The x-axis is calendar year and the y-axis is the paper-level mean share of causal edges. Each line is one EO threshold from \(\geq 1\) to \(=9\). The highlighted EO \(\geq 4\) series is the baseline used in the main text. Tighter thresholds reduce retained edges, but the upward time trend is preserved, indicating that the rise in causal-claim share is not an artifact of a permissive overlap filter.}
\label{fig:prop_causal_edges_ts_overlay}
\end{figure}

\begin{figure}[htp]
\centering
\caption{Pre/Post-2000 Field Patterns by EO Threshold}
\includegraphics[width=0.88\textwidth]{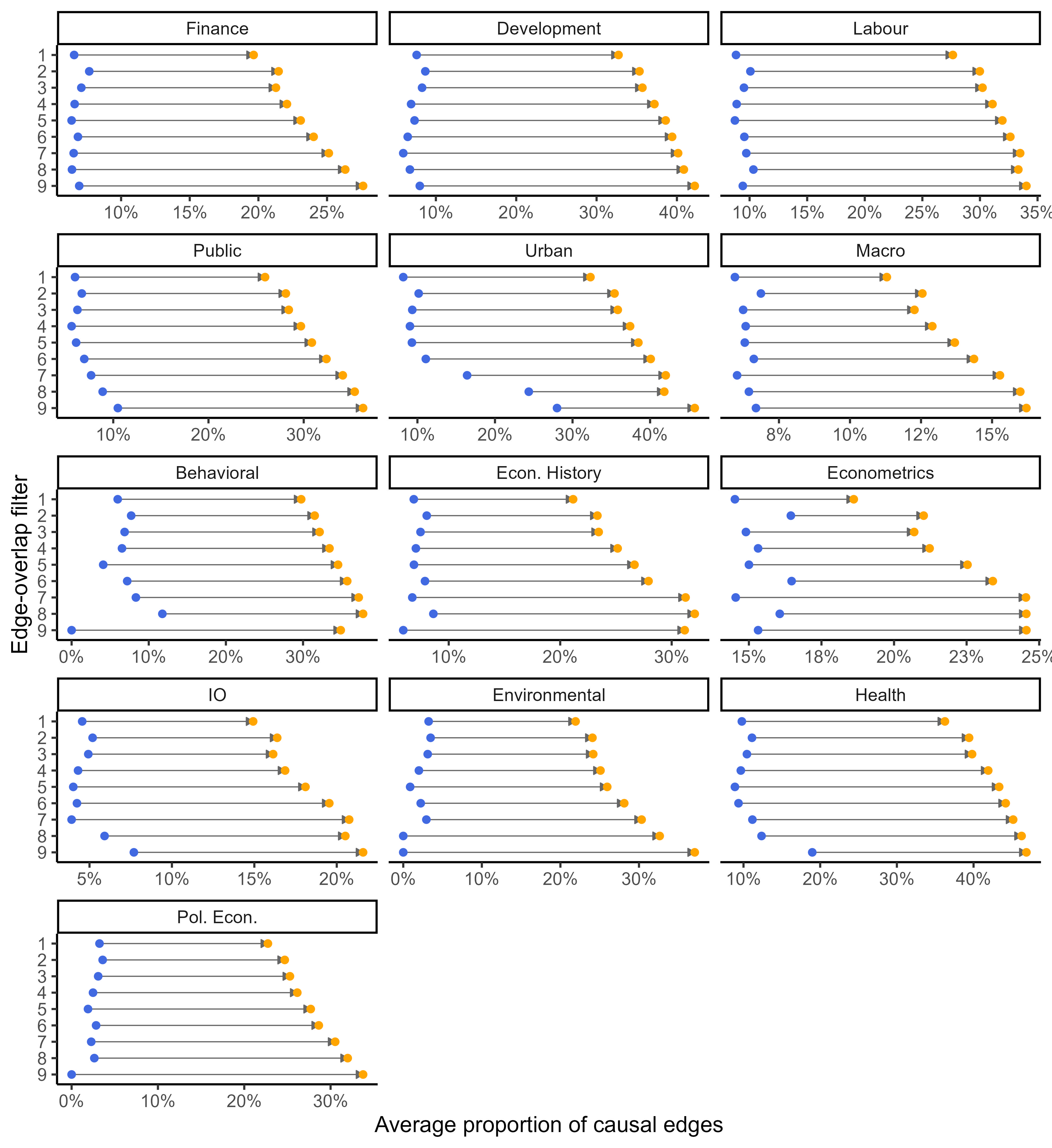}
\captionsetup{singlelinecheck=off,font=small,labelformat=empty,labelsep=none}
\caption*{\scriptsize\justifying \textbf{Note:} Each panel is a field; points show pre-2000 and post-2000 averages and arrows connect the two periods. The horizontal axis is the mean causal-edge share and the vertical axis indexes EO thresholds. EO \(\geq 4\) is the baseline in the main text. Directional increases remain visible for most fields across thresholds, supporting cross-field robustness of the credibility-revolution pattern.}
\label{fig:prop_causal_edges_by_field}
\end{figure}

\begin{figure}[htp]
\centering
\caption{Method Prevalence by EO Threshold}
\includegraphics[width=0.95\textwidth]{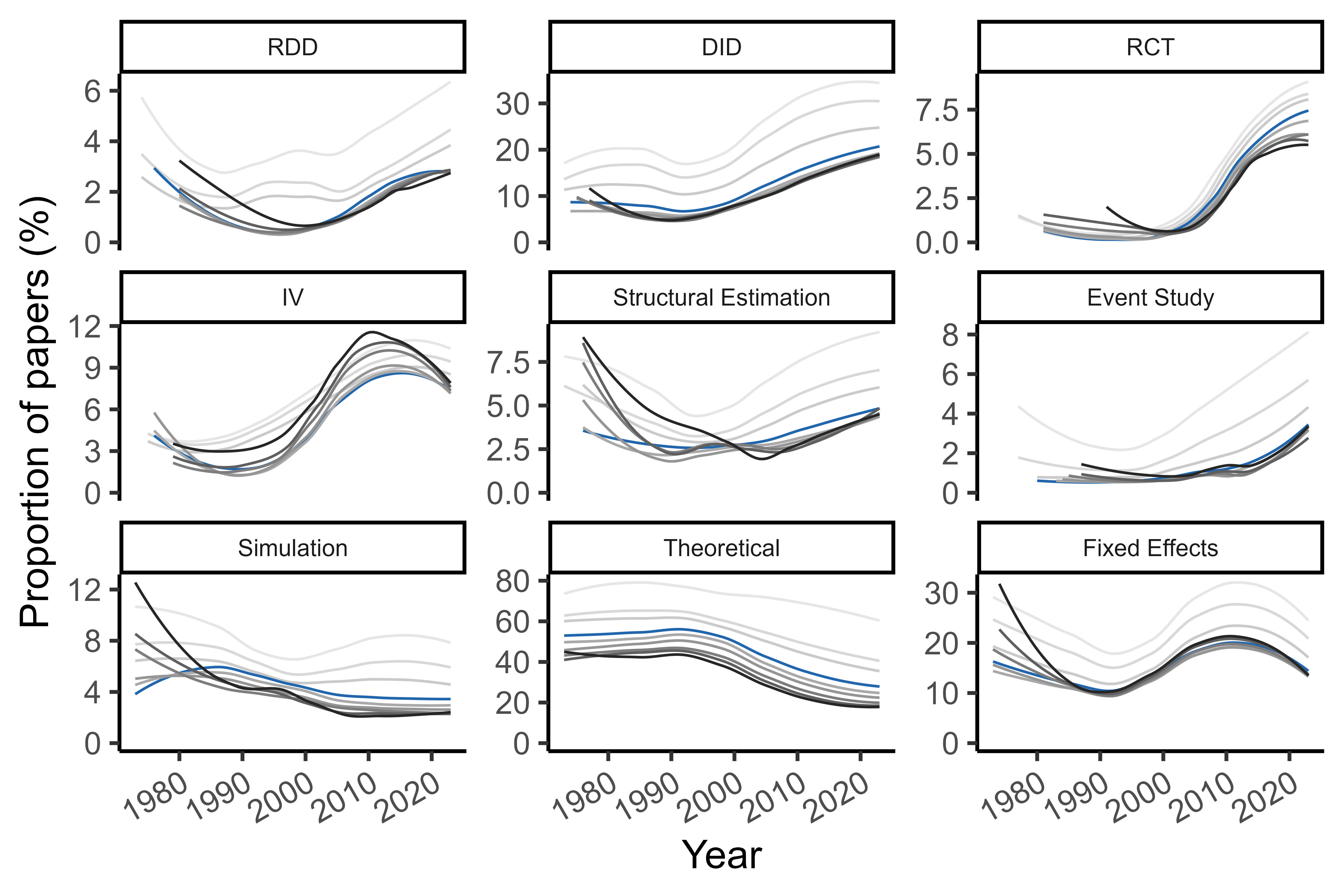}
\captionsetup{singlelinecheck=off,font=small,labelformat=empty,labelsep=none}
\caption*{\scriptsize\justifying \textbf{Note:} Each small panel tracks one method, with year on the x-axis and method prevalence among papers on the y-axis. Darker lines indicate stricter EO thresholds. The EO \(\geq 4\) trajectory is visually close to nearby thresholds for most methods, implying that method-proliferation patterns are not driven by single-run extraction noise.}
\label{fig:eo_methods}
\end{figure}

\begin{figure}[htp]
\centering
\caption{Method Prevalence Over Time: Single Stage-1 Iteration (Iter1)}
\includegraphics[width=0.95\textwidth]{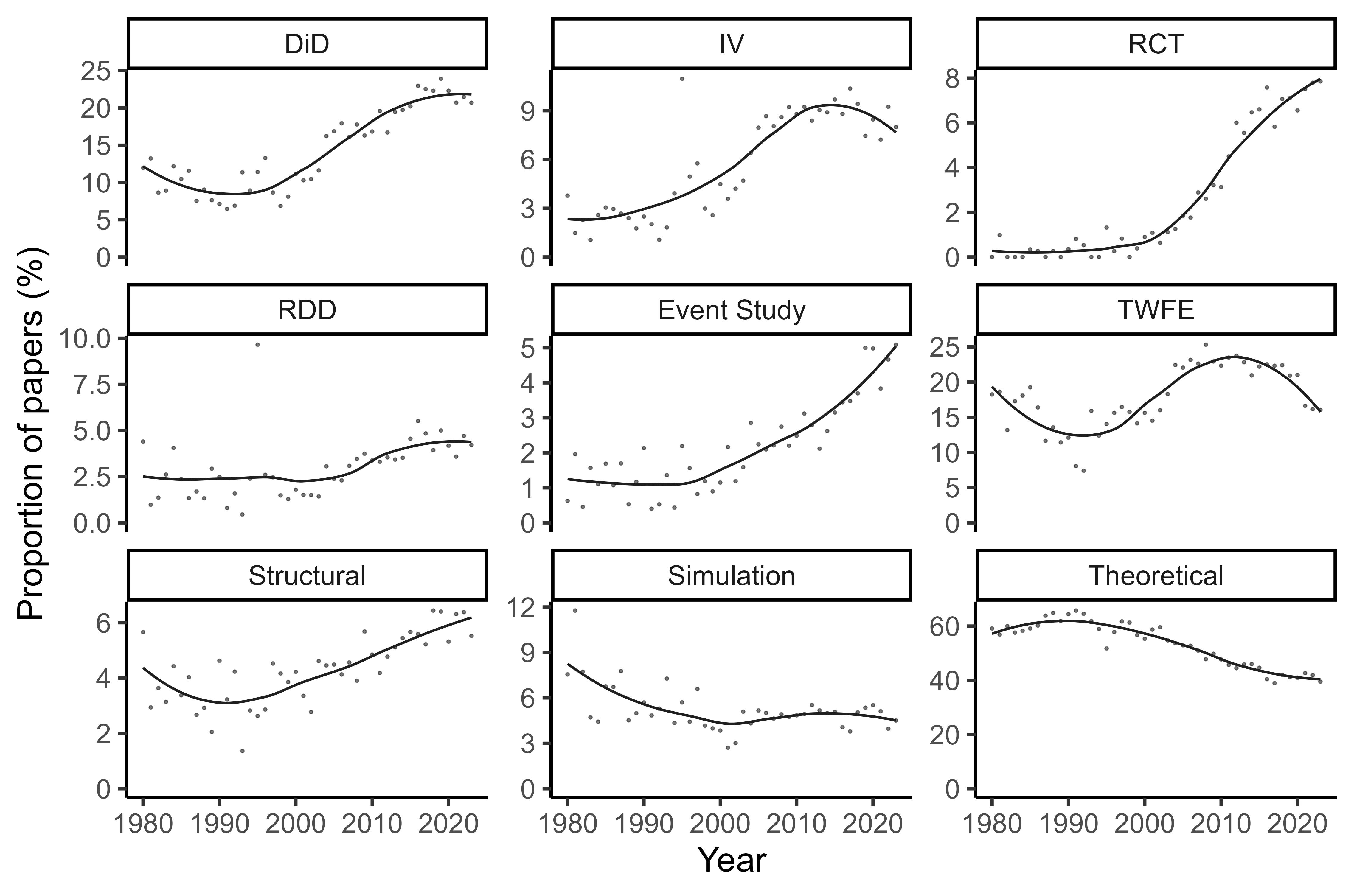}
\captionsetup{singlelinecheck=off,font=small,labelformat=empty,labelsep=none}
\caption*{\scriptsize\justifying \textbf{Note:} This figure mirrors the main-text method-proliferation figure using outputs constructed only from Stage-1 iteration 1 (with Stage-2 aggregation within that branch). Axes match the main specification: year on the x-axis and method prevalence on the y-axis. The directional method trends remain similar to the EO \(\geq 4\) baseline, indicating that conclusions do not depend on combining all three Stage-1 branches.}
\label{fig:eo_methods_iter1_time}
\end{figure}

\begin{figure}[htp]
\centering
\caption{Method Usage by Field: Single Stage-1 Iteration (Iter1)}
\includegraphics[width=0.95\textwidth]{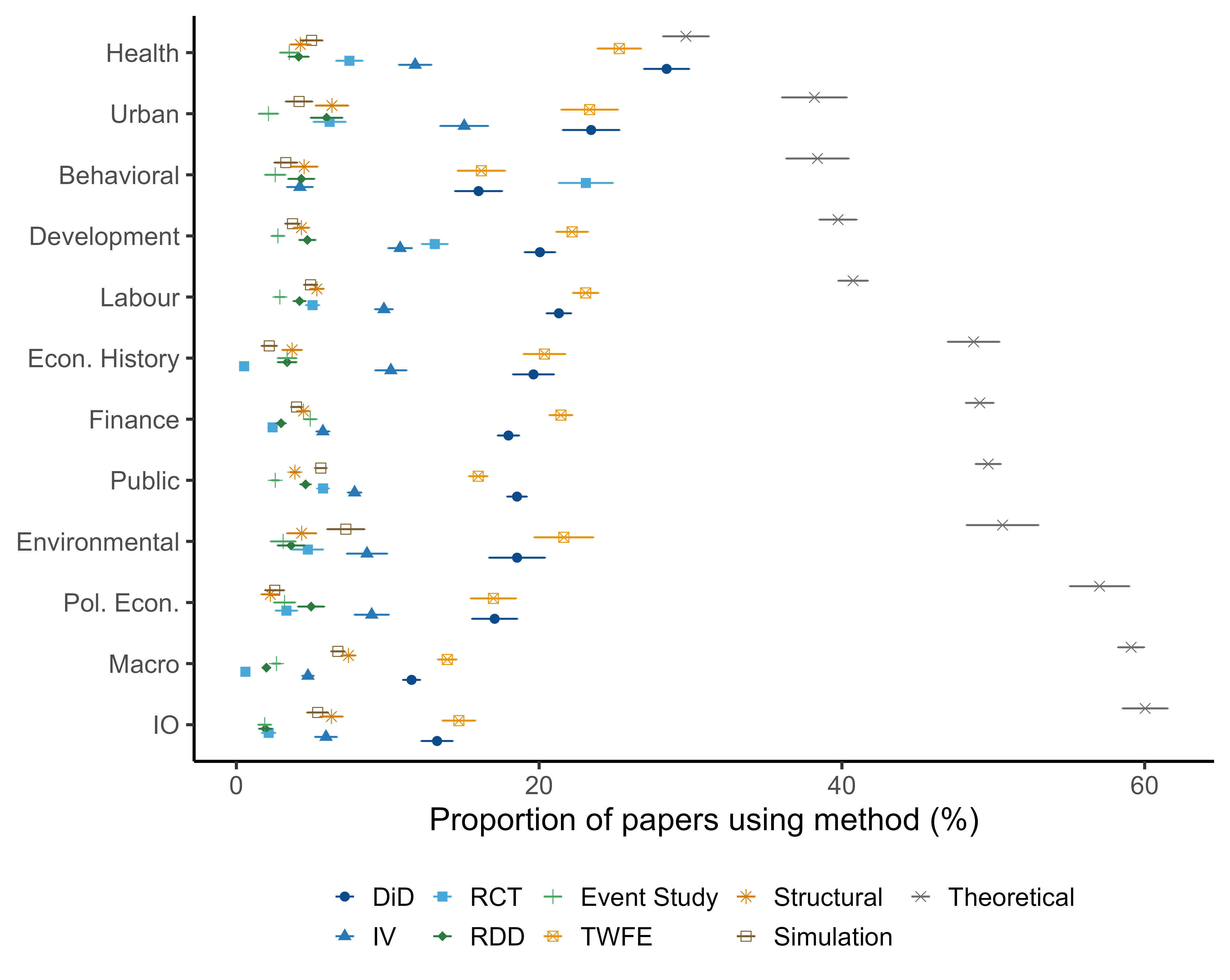}
\captionsetup{singlelinecheck=off,font=small,labelformat=empty,labelsep=none}
\caption*{\scriptsize\justifying \textbf{Note:} This figure mirrors the main-text cross-sectional method-by-field figure using outputs from Stage-1 iteration 1 only. The x-axis reports method prevalence by field and whiskers denote 95\% confidence intervals. Cross-field ranking patterns remain broadly aligned with the EO \(\geq 4\) baseline, supporting robustness to the specific Stage-1 branch used in the extraction pipeline.}
\label{fig:eo_methods_iter1_field}
\end{figure}

\begin{figure}[htp]
\centering
\caption{Headline Predictor Coefficients Across EO Thresholds}
\includegraphics[width=\textwidth]{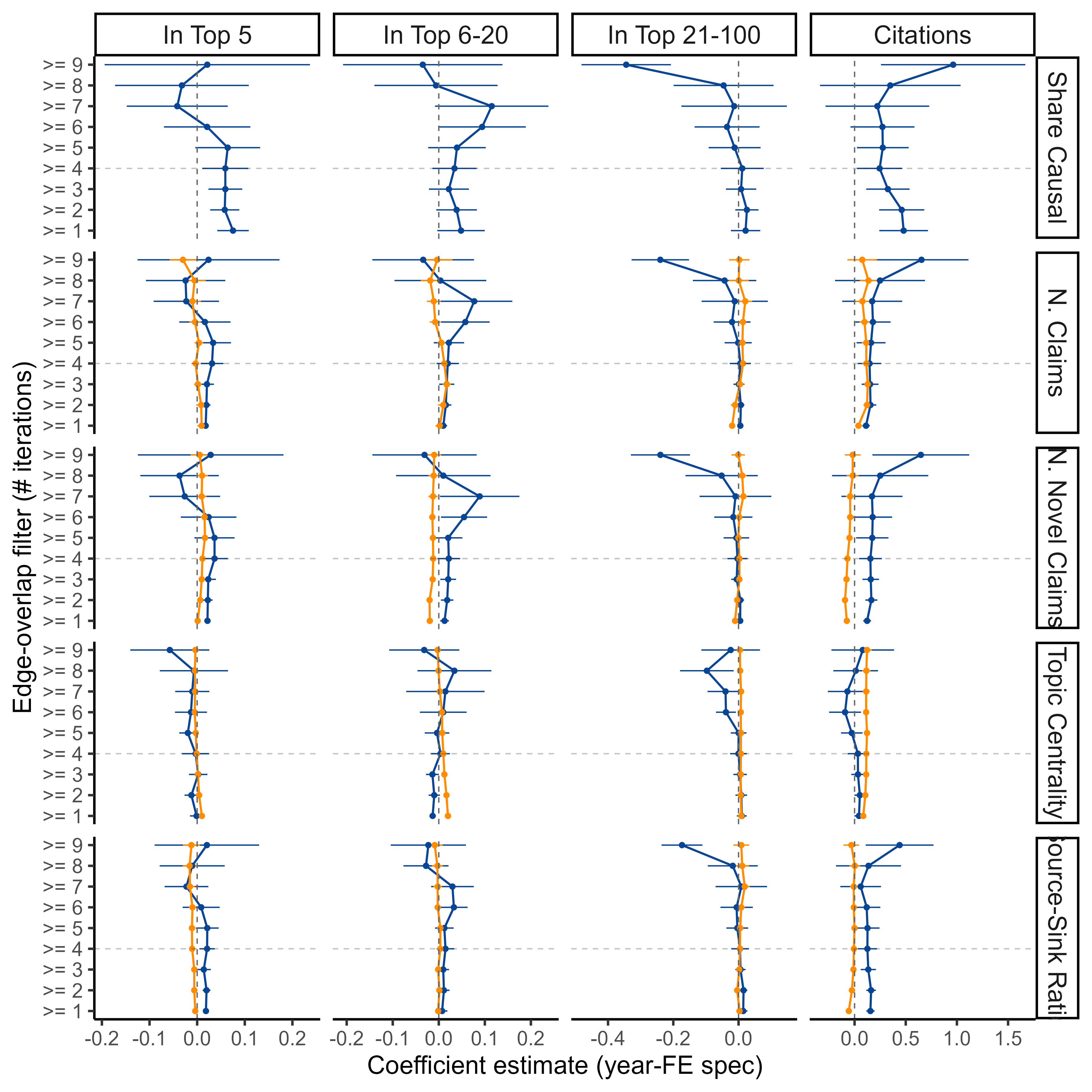}
\captionsetup{singlelinecheck=off,font=small,labelformat=empty,labelsep=none}
\caption*{\scriptsize\justifying \textbf{Note:} Rows are predictor families and columns are outcomes (Top 5, Top 6--20, Top 21--100, citations). Within each predictor family, the three journal-outcome panels use a common x-axis scale to support direct comparison; the citations panel is scaled separately because coefficient magnitudes differ. The y-axis is EO threshold (\(\geq 1\) to \(=9\)). Points are year-FE coefficients with 95\% confidence intervals. EO \(\geq 4\) is highlighted as the baseline; signs remain stable through this region and uncertainty expands only at very strict thresholds where support thins.}
\label{fig:eo_regression_grid}
\end{figure}

Figures~\ref{fig:prop_causal_edges_ts_overlay}--\ref{fig:eo_regression_grid} provide the EO-threshold robustness exhibits corresponding to the main-text trend, method, and predictor results. Figures~\ref{fig:eo_methods_iter1_time} and \ref{fig:eo_methods_iter1_field} then report single-iteration sensitivity checks.

\paragraph{Share of Causal Edges Across EO Thresholds}
We begin with the core quantity used in the paper, the share of causal edges. These exhibits test whether the headline time trend and field-level shifts are robust to stricter or looser overlap filtering.

\paragraph{Methods Over Time Across EO Thresholds}
We then repeat the EO-threshold exercise for method prevalence. The goal is to verify that method-composition trends are not driven by one specific overlap cutoff.

\paragraph{Single-Iteration Sensitivity (Stage-1 Iteration 1)}
These exhibits isolate a single Stage-1 branch and aggregate only within that branch. They provide a direct check that the main method patterns are not an artifact of pooling across Stage-1 summary branches.

\paragraph{Regression Stability Across EO Thresholds}
Finally, we examine whether publication and citation associations for headline predictors are stable across EO thresholds. This links threshold choice to the main inferential objects in the paper.

Overall, these checks support EO \(\geq 4\) as a pragmatic operating point: it is strict enough to suppress fragile one-off edges, but not so strict that it materially distorts coverage of paper-level claim structures.

\section{Validating information retrieval} \label{sec:validating_information_retrieval}

\subsection{Validation with the Brodeur et al. Dataset}

To validate method and field retrieval, we compare our extraction outputs against \citet{brodeur2024p}, which provides label annotations for economics articles. We run this validation directly on paper-level outputs, re-computed for each EO threshold, with title matching via exact cleaned-title matches plus Jaro--Winkler fuzzy fallback.

In the primary benchmark variant, matched sample sizes are 285 papers at EO \(\geq 4\), 169 at EO \(\geq 7\), and 90 at EO = 9. We compute one-vs-rest confusion metrics (accuracy, precision, recall, F1) for each threshold and each label (DiD, RCT, RDD, IV, Urban, Finance, Macroeconomics, Development), and display denominator-valid rows in the table.

Table \ref{tab:validation_results_brodeur2024p} reports EO \(\geq 4\), EO \(\geq 7\), and EO = 9 results with class prevalence, but displays only rows with defined precision and recall (\(tp+fp>0\) and \(tp+fn>0\)). Table \ref{tab:validation_results_brodeur_runlevel_summary} then summarizes the same metrics across the nine individual Stage-1$\times$Stage-2 runs (min/median/mean/max). Full EO-grid outputs (EO \(\geq 1\) through EO \(\geq 9\)), omitted rows, run-level metrics, and match logs are exported in the reproducibility package tables.

\subsection{Validation with the Plausibly Exogenous Galore Dataset}

To further validate extracted claim content, we compare our outputs to the Plausibly Exogenous Galore benchmark\footnote{The Plausibly Exogenous Galore dataset is a curated list of plausibly exogenous variations in the empirical economics literature, maintained by Sangmin S. Oh. Available at \url{https://www.notion.so/1a897b8106ca44eeaf31dcd5ae5a61b1?v=ff7dc75862c6427eb4243e91836e077e}.}. We aggregate edges to paper-level component text (causes, effects, and exogenous sources) and match papers using the same exact+fuzzy title protocol as above.

Matched counts are 491 papers at EO \(\geq 4\), 296 at EO \(\geq 7\), and 148 at EO = 9. For each component, we compute cosine similarity on deterministic text vectors and report max/mean/median/dispersion statistics. We also compute a shuffled-match random baseline (1,000 permutations per EO/component), with lift defined as observed mean minus baseline mean.

Table \ref{tab:validation_results_plausibly_exo} reports the EO \(\geq 4\), EO \(\geq 7\), and EO = 9 results, including baseline and lift. Full EO-grid outputs (EO \(\geq 1\) through EO \(\geq 9\)) and matching diagnostics are provided in the reproducibility package tables.

\subsection{Snippet-level Validation Across EO Candidate Sets}

The snippet validation exercise is a machine-assisted self-consistency check rather than a human-labeled benchmark. We construct an alternative edge list using the snippet-only Stage 2 prompt (which sees only verbatim claim snippets from Stage 1), then compare candidate edge sets (A, B, C, and pooled ABC) against those snippet-only edges. For each audited paper we flag matched edges (true positives), unsupported edges (false positives), and snippet-supported but missing edges (false negatives), and then aggregate micro-precision, micro-recall, and micro-F1. This isolates the effect of summary-based inputs versus snippet-only inputs on the extracted graph.

To align this audit with overlap filtering, we map candidate sets to EO proxies: set A to EO \(\geq 1\), set B to EO \(\geq 2\), and set C to EO = 3. Table \ref{tab:validation_snippet_eo} reports the resulting metrics. Precision rises monotonically as overlap is tightened (0.778 to 0.894 to 0.936), while recall falls (0.779 to 0.594 to 0.417), which is the expected precision--recall trade-off when moving from permissive to strict edge inclusion.

The table therefore provides a direct operational interpretation of threshold choice: lower EO retains more candidate edges and captures more true edges but accepts more noise; higher EO suppresses noise but drops a larger share of true edges.

\subsection{EO-threshold Perturbation Summary}

To connect robustness checks to threshold choice, we compute a perturbation table (Table \ref{tab:validation_perturbation_summary}) that benchmarks each EO cutoff against EO \(\geq 4\). The mean causal-edge share stays within a 10\% band for EO = 2 through EO = 6, but diverges in more permissive (EO = 1) and more restrictive (EO \(\geq 7\)) regions. This supports EO \(\geq 4\) as a practical baseline that balances stability and coverage.

\section{Paper-level Graphical Measures (Appendix)}\label{sec:appendix_graphical_measures}

%%%%%%%% Narrative complexity measures descirptive and illustrated
\subsection{Additional measures of Narratives Complexity}
We consider other measures of Narrative Complexity, including the number of unique paths in \( G_p \), denoted as \( P_p \), which is the total number of distinct directed paths between all pairs of nodes, excluding self-loops. This captures the interconnectedness of the narrative within the paper; a higher \( P_p \) indicates a more intertwined argument structure. The longest path length in \( G_p \), denoted as \( L_p \), represents the length of the longest directed path, indicating the depth of reasoning in the paper. We compute both \( P_p \) and \( L_p \) for the non-causal subgraph (denoted as \( P_p^{\text{non-causal}} \) and \( L_p^{\text{non-causal}} \)) and for the subgraph consisting only of causal edges (\( P_p^{\text{causal}} \) and \( L_p^{\text{causal}} \)).

\paragraph{Illustrative examples}

To concretely illustrate our graphical framework and the measures derived from it, we examine four landmark economic papers: \citet{chetty2014land}, \citet{banerjee2015miracle}, \citet{gabaix2011granular}, and \citet{goldberg2010imported}. These papers cover a diverse range of topics and methodologies, showcasing the versatility of our approach. The corresponding claim-graph visuals are shown in the main text.

In \citet{chetty2014land}, these claim graph measures for this paper are as follows: the number of unique paths $P_p^{\text{non-causal}} = 6$, and the longest path length $L_p^{\text{non-causal}} = 1$. The relatively high number of edges indicates a broad exploration of factors affecting upward mobility, while the longest path length of 1 reflects that the relationships are primarily direct associations rather than extended causal chains.

In \citet{banerjee2015miracle}, the measures are: the number of unique paths $P_p^{\text{causal}} = 12$, and the longest path length $L_p^{\text{causal}} = 3$. The high number of causal edges and unique paths indicates a complex causal narrative with multiple interconnected outcomes, while the longest path length reflects deeper causal chains.

\noindent See the main-text examples.

In \citet{gabaix2011granular}, the claim graph measures are: the number of edges $|E_p| = 6$ (all non-causal), the number of unique paths $P_p^{\text{non-causal}} = 11$, and the longest path length $L_p^{\text{non-causal}} = 3$. The relatively high number of unique paths and the longest path length indicate a complex theoretical narrative with multiple interconnected concepts and deeper reasoning chains.

Finally, \citet{goldberg2010imported}, the measures are: the number of edges $|E_p| = 5$, the number of causal edges $|E_p^{\text{causal}}| = 3$, the number of unique paths $P_p = 5$, and the longest path length $L_p = 2$. These measures reflect a focused exploration of specific causal relationships, with a moderate level of narrative complexity.

\noindent See the main-text examples.

These examples demonstrate the diversity in the structure and complexity of claim graphs across different types of economic research. They illustrate how our measures capture key aspects of the narratives, such as the breadth of topics covered, the depth of causal analysis, and the interconnectedness of concepts.
% \paragraph{Relation to Previous Findings}

\subsection{Additional measures of Novelty and Contribution to Literature}
\label{sec:novelty_contribution_app}

This appendix subsection summarizes the non-headline novelty measures used in robustness exercises. We keep definitions compact and interpretation-focused, because these statistics are variants of the same core constructs already discussed in the main text.

\subsubsection{Path-Based Novelty}
\label{sec:path_novelty_app}

Path-based variants include the number of unique directed paths, the longest directed path, and the share of paths that are novel relative to prior literature. Intuitively, these capture whether a paper contributes deeper or newly recombined mechanisms rather than only isolated links. We compute the same variants for the full graph and for the causal subgraph.

\subsubsection{Gap Filling (Co-occurrence Analysis)}
\label{sec:gap_filling_app}

Gap-filling variants track whether a paper connects JEL concept pairs that were historically rare in prior work. This is complementary to edge/path novelty: it focuses on bridging underexplored conceptual intersections even when individual edges are not fully new. In robustness tables and figures, higher values indicate more frontier-bridging combinations.

\subsubsection{Other Feasible Variants}

Other graph variants are possible but are not central for this paper: motif counts, betweenness-based brokerage, temporal persistence of newly introduced links, and cross-field bridge intensity. We do not foreground these to preserve focus on interpretable measures that map cleanly to journal placement and citation outcomes.

\paragraph{Which Journals Fill Gaps in Literature?}
\label{sec:gap_filling_journal_comparison_app}

\begin{figure}[htp]
\centering
\caption{Gap Filling Across Journal Quality Tiers and Within Top 5 Journals}
\begin{subfigure}[t]{0.78\textwidth}
    \centering
    \caption{Gap Filling by Journal Tier}
    \includegraphics[width=\textwidth]{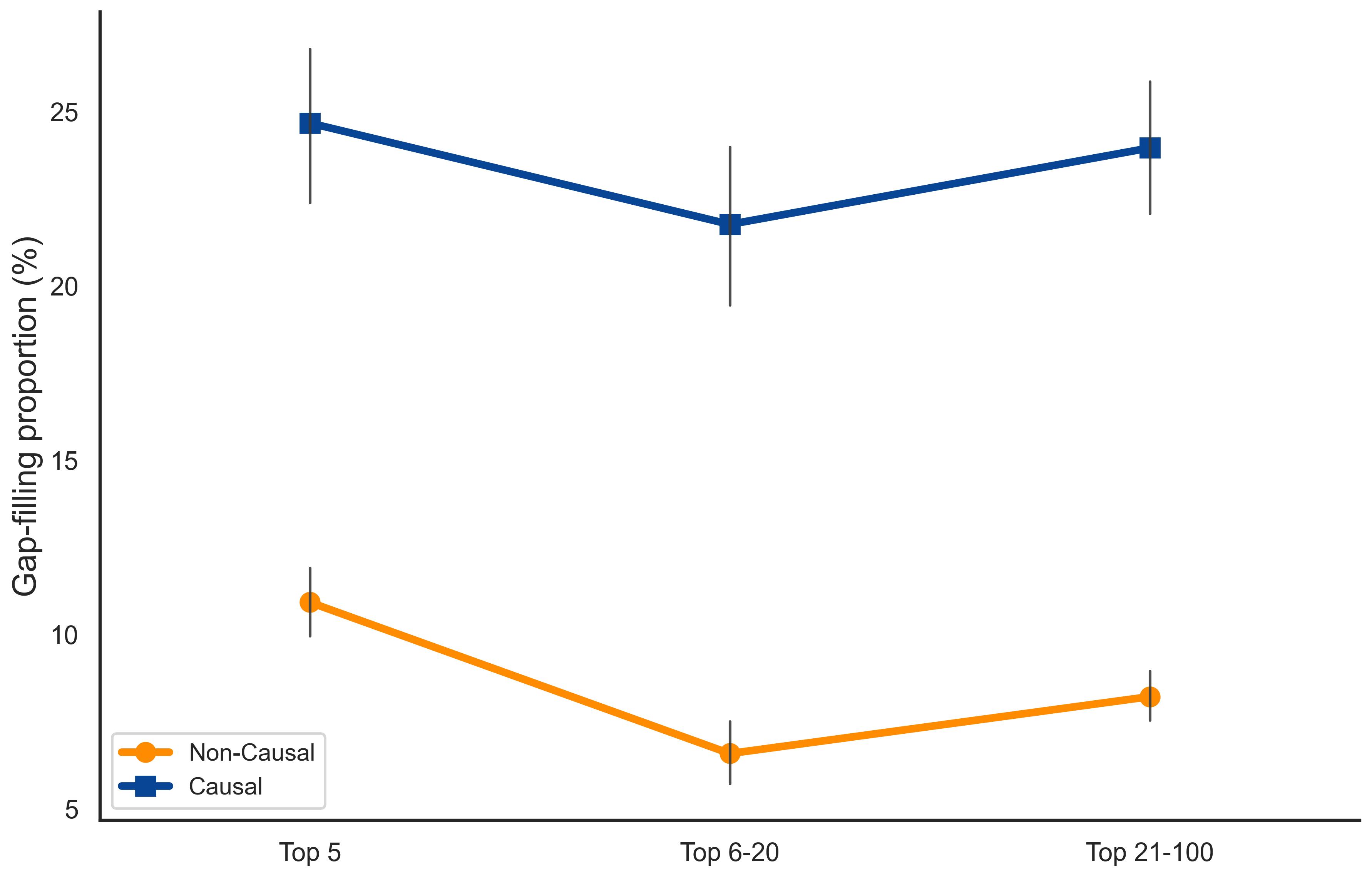}
    \label{fig:gap_filling_panel_a}
\end{subfigure}
\hfill
\begin{subfigure}[t]{0.78\textwidth}
    \centering
    \caption{Gap Filling Among Top 5 Journals}
    \includegraphics[width=\textwidth]{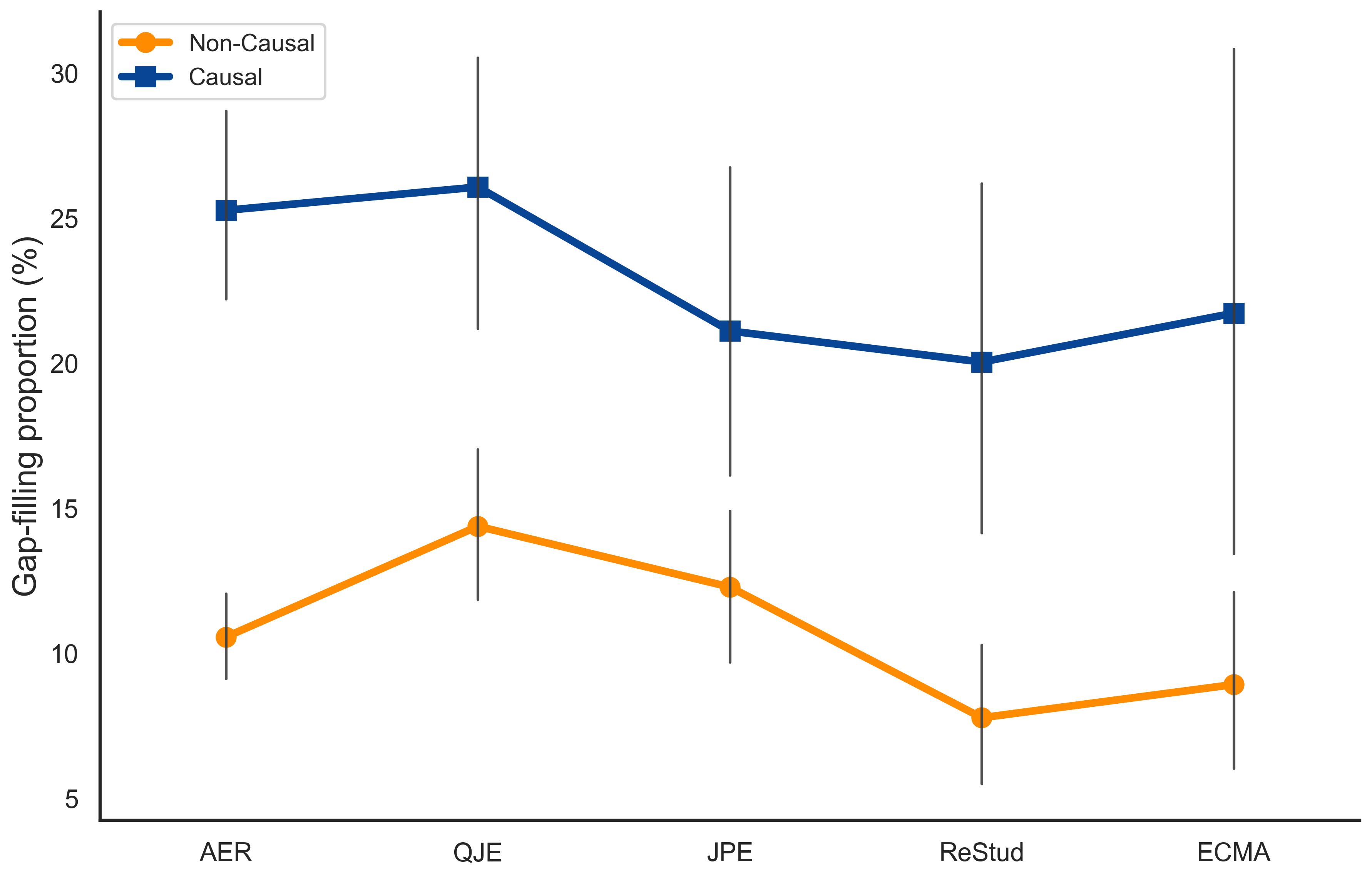}
    \label{fig:gap_filling_panel_b}
\end{subfigure}
\vspace{-0.2cm}
\captionsetup{singlelinecheck=off,font=small,labelformat=empty,labelsep=none}
\caption*{\scriptsize\justifying \textbf{Note:} Panel (a) reports average gap-filling rates (with 95\% confidence intervals) by journal tier; Panel (b) reports the same for the top-five journals using short labels (AER, QJE, JPE, ReStud, ECMA). Points show mean gap-filling shares and whiskers show 95\% confidence intervals aligned to the plotted means. Orange series use non-causal edges and blue series use causal edges. The figure shows higher average gap filling in top-tier journals, with notable variation across top-five outlets.}
\label{fig:gap_filling_journals_combined}
\end{figure}

To illustrate this family, Figure~\ref{fig:gap_filling_journals_combined} compares gap-filling rates across journal tiers and within top-5 outlets, split into non-causal and causal variants. The pattern is consistent with the main narrative: top-tier outlets reward frontier-bridging links, with especially strong values in selected top-5 journals.

\bigskip

Taken together, these measures of novelty and gap filling enrich our understanding of how each paper pushes the frontier of economic knowledge. By systematically tracking new edges, new paths, and underexplored concept pairs, we capture distinct facets of a paper's originality and the extent to which it addresses previously overlooked topics or mechanisms. As discussed in the main text, these indicators are significantly associated with both publication outcomes and the long-run citation impact of research.

% \paragraph{Gaps in political economy (or labour economics) over decade}
% - e.g. find the gaps within the JEL code (perhaps 2 digit) in 1990, 2000, 2010, 2020.

\subsection{Additional Measures of Topic Centrality and Diversity}
\label{sec:conceptual_importance_and_diversity_app}

We now turn to assessing how centrally each paper's concepts lie within the broader economic literature and whether the paper balances multiple sources and targets in its argumentation. These attributes may shape both the paper's perceived significance and its eventual scholarly influence. We consider two sets of measures: (i) the outdegree--indegree ratio, which captures the balance of causal flows in the paper, and (ii) centrality-based indicators (e.g., eigenvector or PageRank scores) computed for the nodes in a cumulative claim graph of prior literature. We then aggregate these centrality values (mean, variance) over the specific concepts a paper employs, providing measures of conceptual importance and conceptual diversity.

\subsubsection{Outdegree--Indegree Ratio}
The outdegree--indegree ratio summarizes whether a paper's claim graph emphasizes multiple drivers feeding into fewer outcomes, or the reverse. Values above one indicate a source-heavy structure (many distinct drivers relative to outcomes), while values below one indicate an outcome-heavy structure (fewer drivers linked to many downstream effects). We report the ratio for both the full graph and the causal-only graph.

This measure is included because it captures an interpretable feature of argument architecture that standard count metrics miss: whether papers are `many causes to fewer outcomes'' versus `few causes to many outcomes.'' the Appendix reports exact notation and implementation details.

To examine how the outdegree--indegree ratio evolves, the main-text source--sink figure panel (a) plots annual averages with 95\% confidence bands. The non-causal series is comparatively flat through 2000 and then edges down, while the causal series rises over time, indicating greater use of designs that test multiple identified drivers for a narrower set of outcomes.

The main-text source--sink figure panel (b) compares fields cross-sectionally. Economic History features relatively high non-causal ratios, while Behavioral and Health show higher causal ratios, consistent with broader use of designs that test multiple identified channels for policy-relevant outcomes.

\begin{figure}[htp]
\centering
\caption{Top 20 Concepts by Mean Eigenvector Centrality (Level Comparison)}
\includegraphics[width=0.92\textwidth]{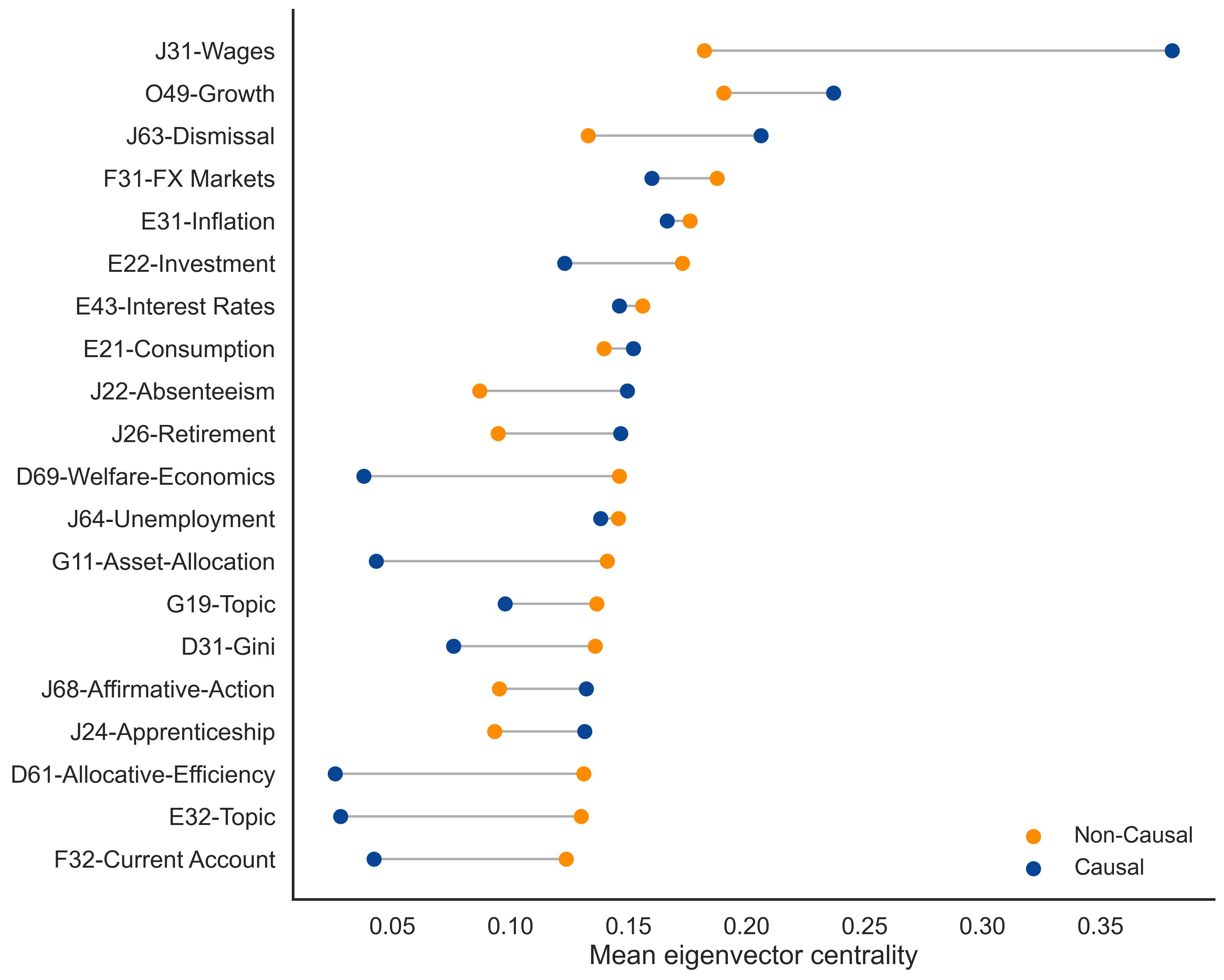}
\captionsetup{singlelinecheck=off,font=small,labelformat=empty,labelsep=none}
\caption*{\scriptsize\justifying \textbf{Note:} This appendix figure reports the level comparison for the top 20 concepts ranked by mean eigenvector centrality (using the maximum across the two graph definitions). Each row shows non-causal (orange) and causal (blue) centrality with a connecting segment. Labels are in \texttt{JEL-short name} format to improve readability.}
\label{fig:top_20_jel_nodes_centrality_levels_appendix}
\end{figure}

\begin{figure}[htp]
\centering
\caption{Top 3 Rising and Declining Concepts Over Time (Normalized)}
\vspace{-0.1cm}
\begin{subfigure}{0.95\textwidth}
    \centering
    \caption{Non-Causal Subgraph}
    \includegraphics[width=\textwidth]{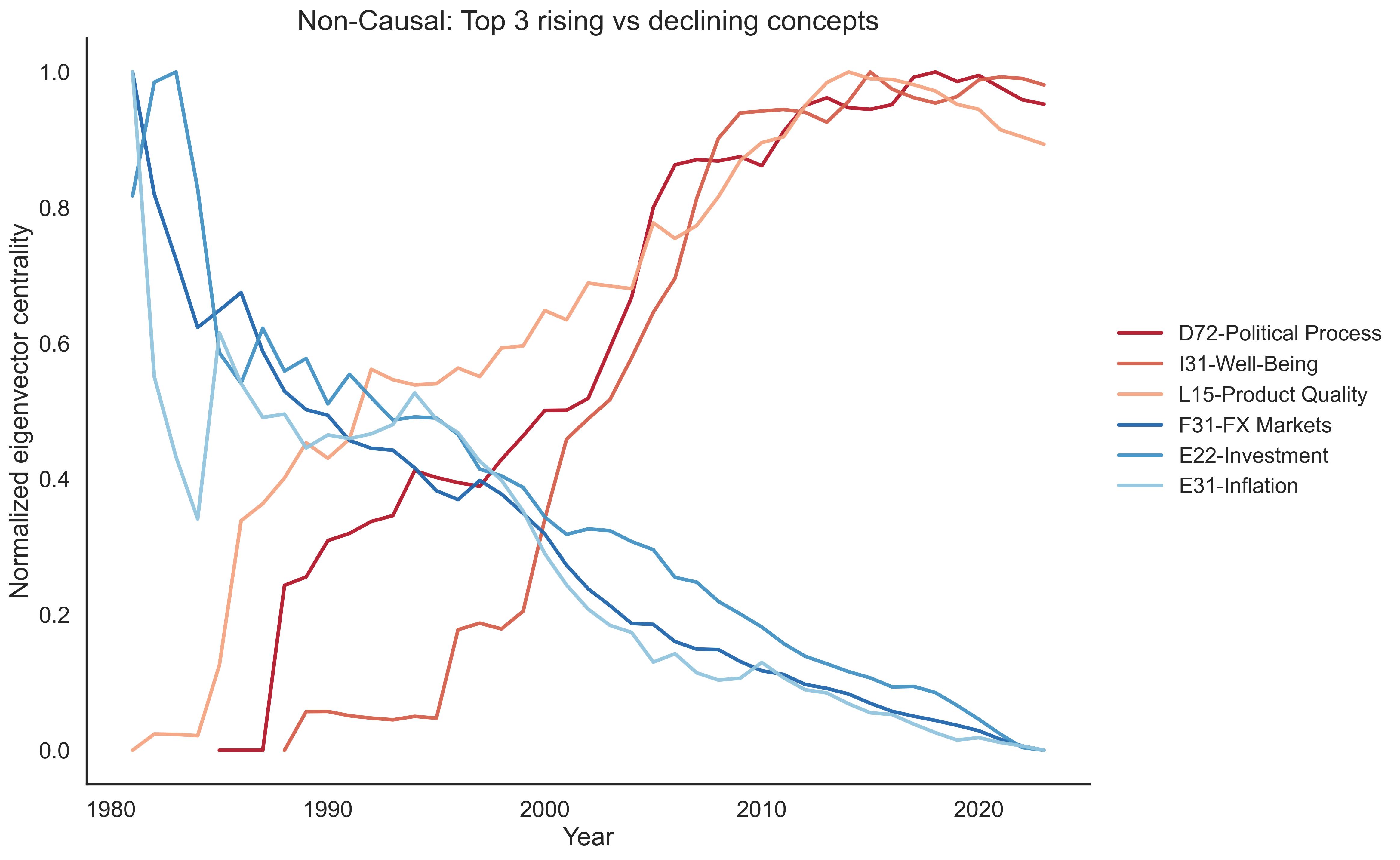}
    \label{fig:non_causal_node_centrality_selected_trends}
\end{subfigure}
\vspace{-0.25cm}
\begin{subfigure}{0.95\textwidth}
    \centering
    \caption{Causal Subgraph}
    \includegraphics[width=\textwidth]{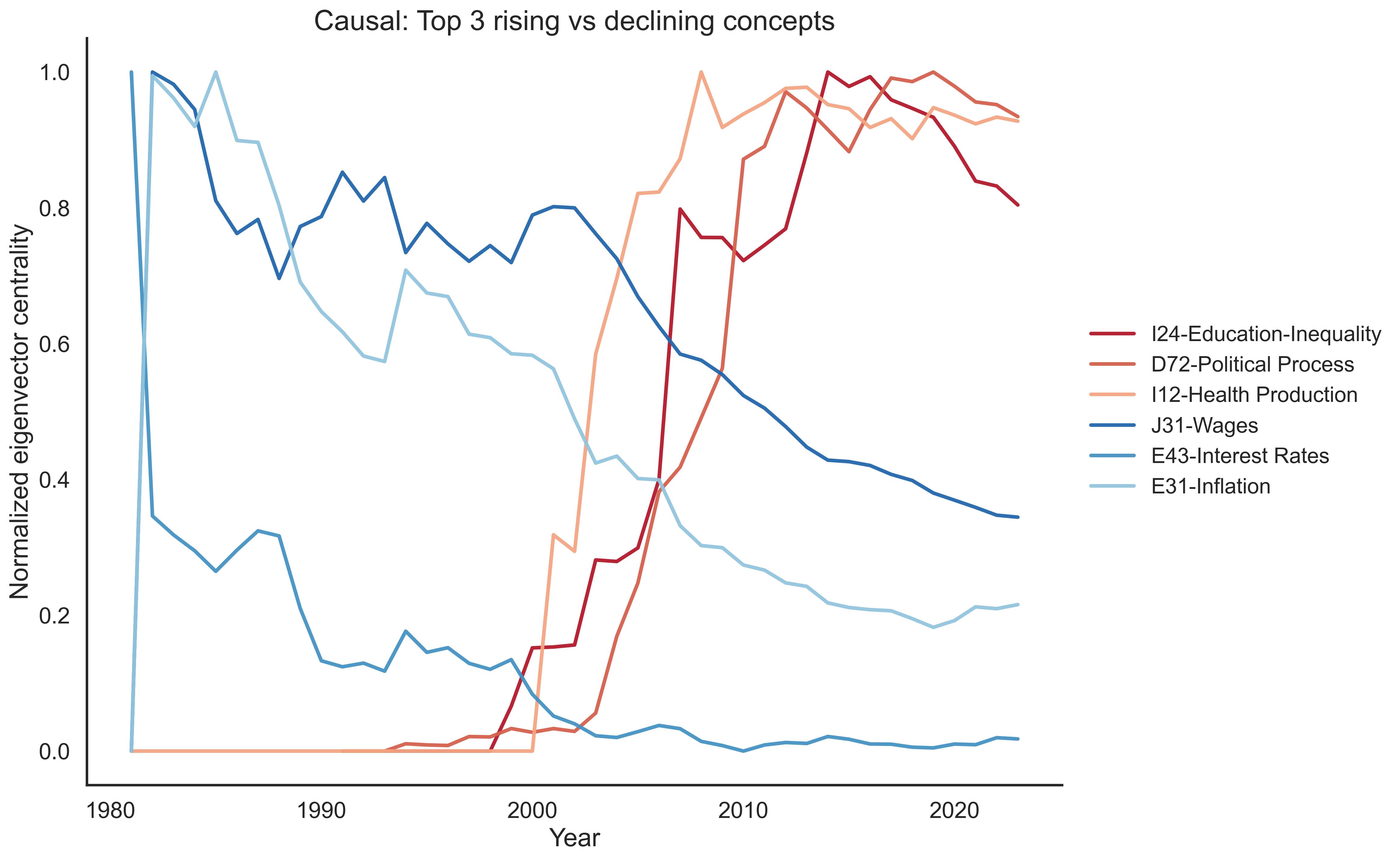}
    \label{fig:causal_node_centrality_selected_trends}
\end{subfigure}
\vspace{-0.2cm}
\captionsetup{singlelinecheck=off,font=small,labelformat=empty,labelsep=none}
\caption*{\scriptsize\justifying \textbf{Note:} Each panel plots six concepts: the three steepest risers (warm colors) and the three steepest decliners (cool colors) in normalized eigenvector centrality over time, selected from sufficiently common nodes to limit noise. Legends use \texttt{JEL-short name} labels to make trajectories interpretable without requiring memorization of JEL codes.}
\label{fig:node_centrality_selected_trends}
\end{figure}

\begin{figure}[htp]
    \centering
    \caption{Proportion of Papers Published in Top 5 Journals, Pre- and Post-2000.}
    \vspace{-0.2cm}    
    \begin{subfigure}[t]{0.6\textwidth}
        \centering
        \caption{By Field}
    \vspace{-0.2cm}    
        \includegraphics[width=\textwidth]{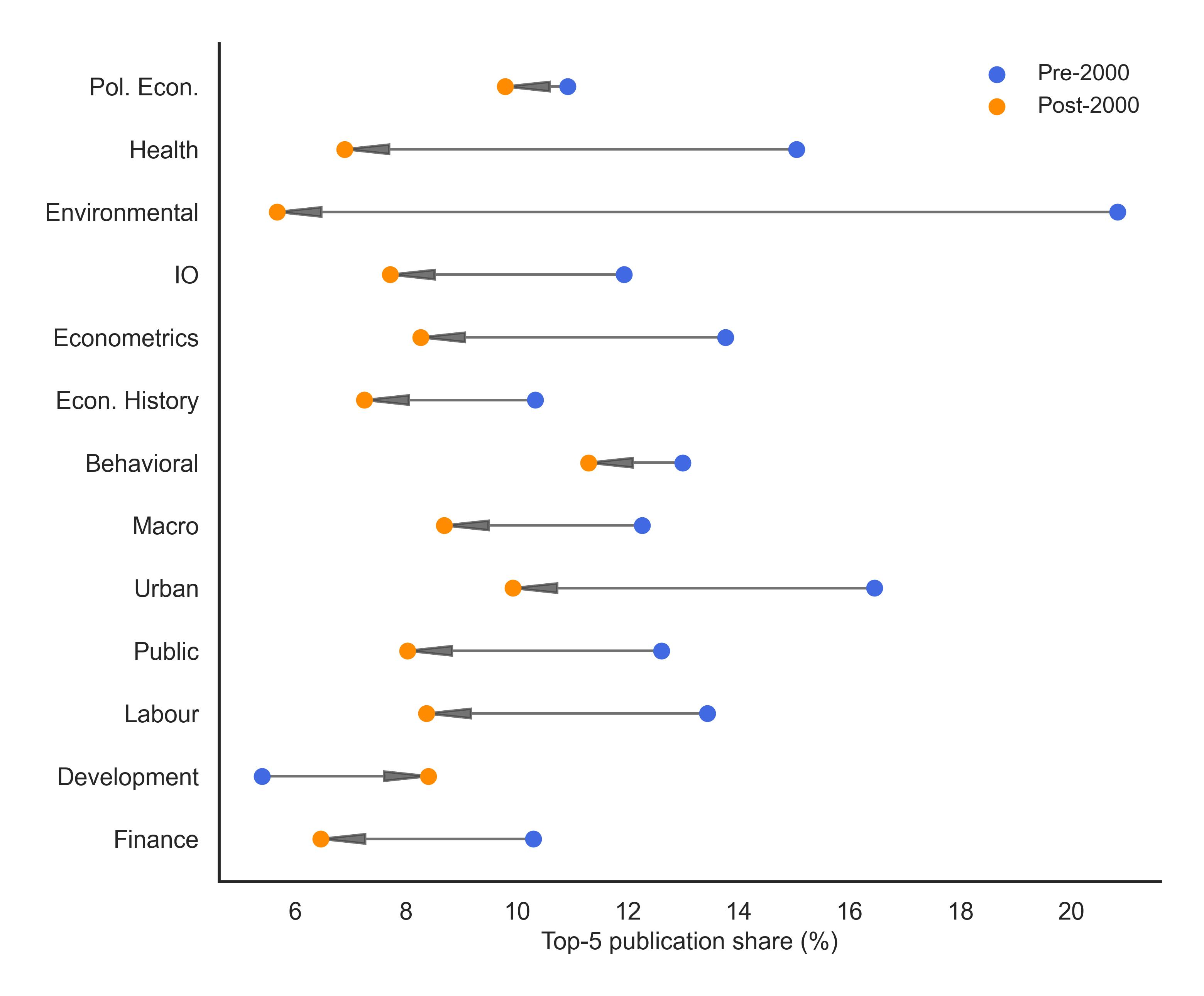}
        \label{fig:paper_is_top5_by_field_post_2000_arrow}
    \end{subfigure}
    \vspace{1em}    
    \begin{subfigure}[t]{0.6\textwidth}
        \centering
        \caption{By Field and Method}
    \vspace{-0.2cm}    
        \includegraphics[width=\textwidth]{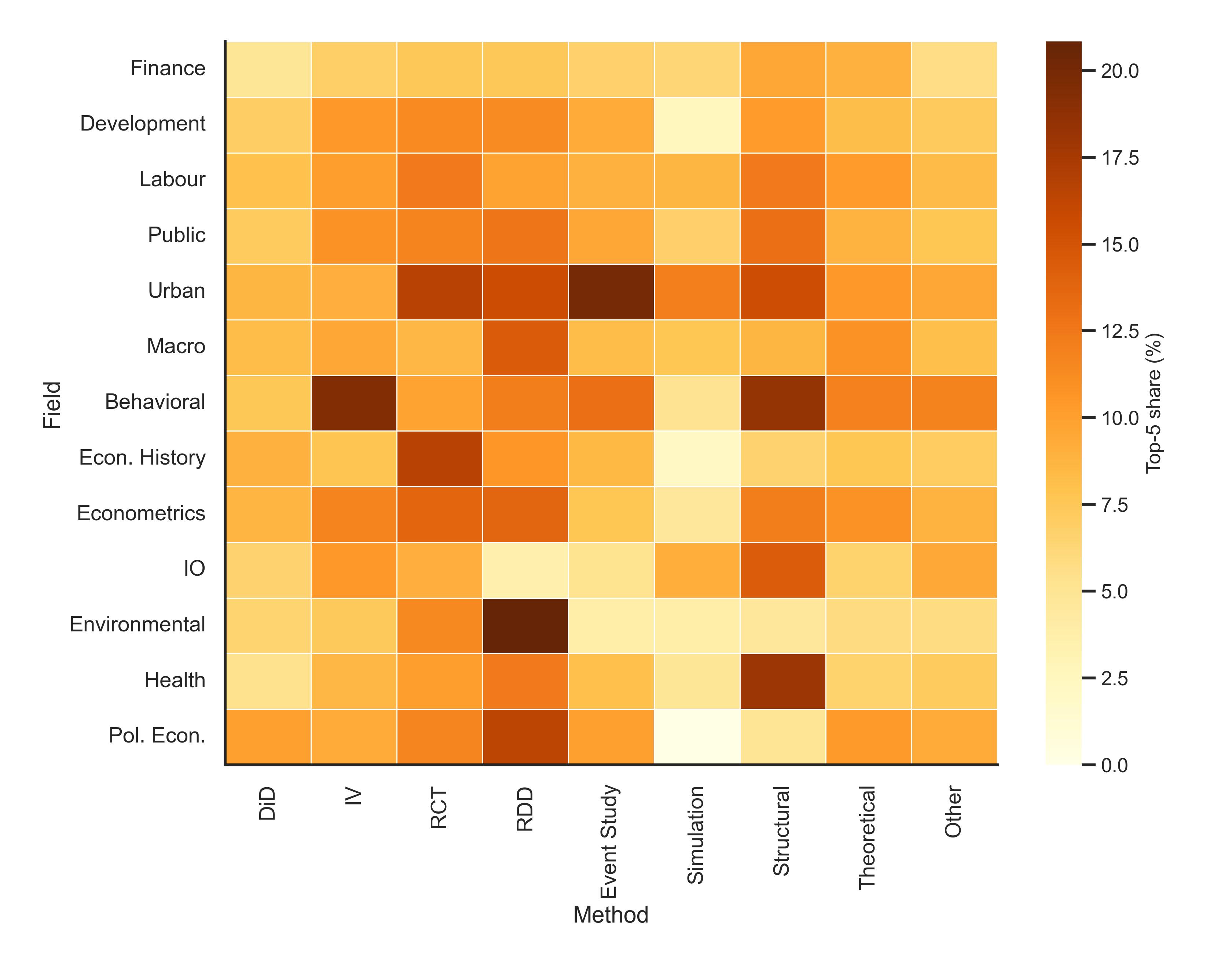}
        \label{fig:top5_papers_by_field_method_heatmap}
    \end{subfigure}
    \vspace{-1cm}    \captionsetup{singlelinecheck=off,font=small,labelformat=empty,labelsep=none}
    \caption*{\scriptsize\justifying \textbf{Note:} This figure displays the proportion of working papers that were eventually published in the top five economics journals, broken down by (a) field and (b) field by empirical method. Data are derived from our matched publication dataset. Arrows in panel (a) indicate changes between pre-2000 and post-2000 periods. The heatmap in panel (b) shows that certain fields and methods have higher publication rates in top journals, with field-method combinations such as Theoretical methods in Behavioural, Structural in IO or RCTs in Urban.}
    \label{fig:top_5_by_field_method}
\end{figure}

\begin{figure}[htp]
\centering
\caption{Distribution of Citation Percentiles by Journal Category}
\includegraphics[width=\textwidth]{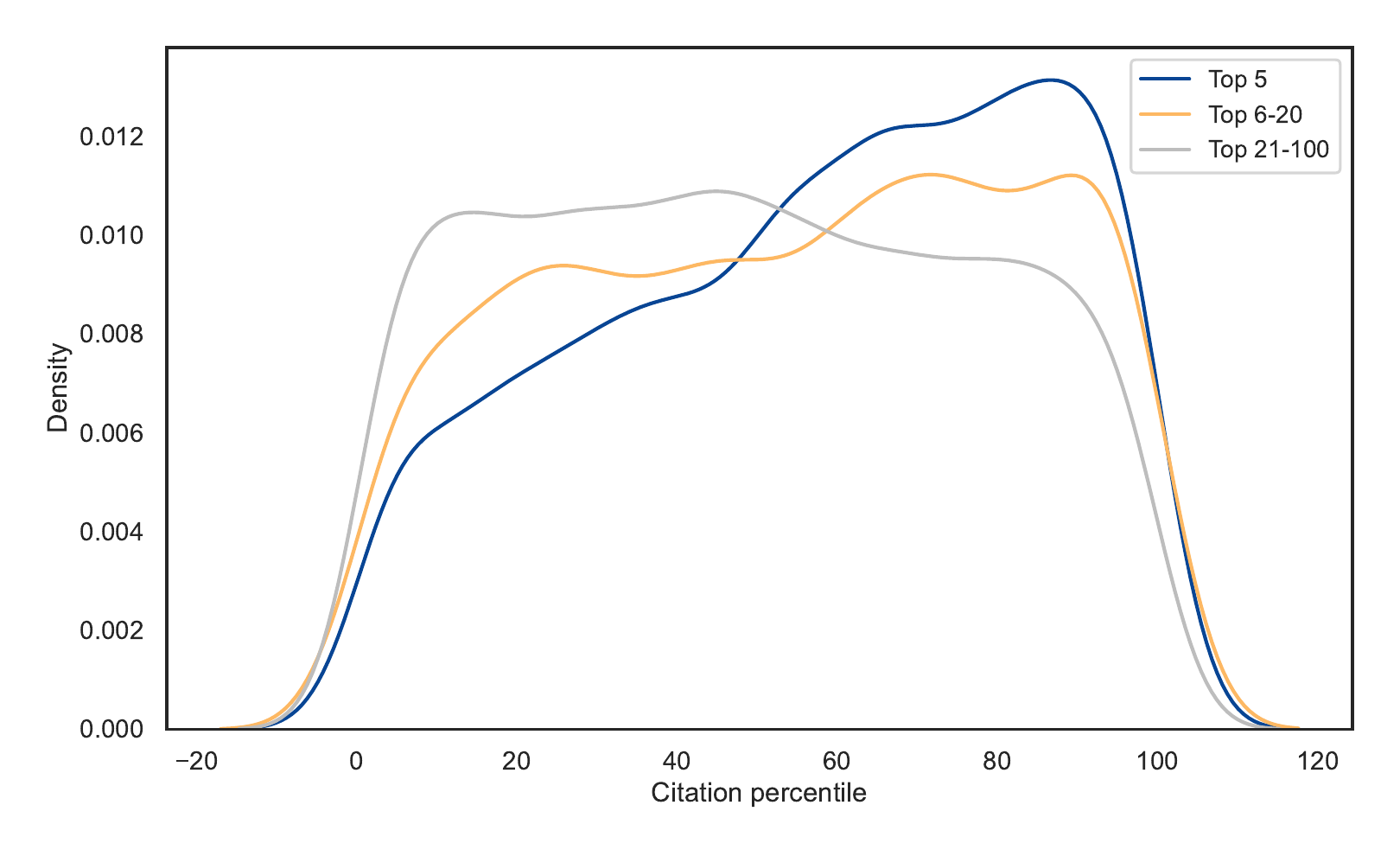}
\captionsetup{singlelinecheck=off,font=small,labelformat=empty,labelsep=none}
\caption*{\scriptsize\justifying \textbf{Note:} This figure displays kernel density plots of the citation percentiles for papers published in Top 5, Top 6--20, and Top 21--100 journals. The citation percentiles are calculated based on the entire sample, with higher values indicating higher citation counts relative to other papers. The plot shows that while papers published in higher-ranked journals tend to receive more citations on average, the most highly cited papers are more evenly distributed across journal categories. This suggests that exceptionally influential papers can emerge from a wide range of journals.}
\label{fig:citations_by_journal_cat_density}
\end{figure}

Table~\ref{tab:citation_percentiles_by_journal} reports the corresponding distribution moments by journal group.

\subsection{Additional Predictor Panels}

\subsubsection{Variant Predictor Definitions and Interpretation}
\label{sec:predictor_variants_appendix_details}

These panels report close variants of the five headline predictor families to show that the core results are not driven by a single operationalization. Narrative-complexity variants (log unique paths and log longest path) capture breadth versus depth of argument chains; novelty variants (novel-path shares) capture whether papers introduce previously unseen mechanism combinations; and positioning variants (topic-diversity variance and PageRank-based centrality) capture dispersion and prominence of concepts in the literature graph.

The broad pattern matches the headline specification: causal variants are more often positive for top-tier publication and citations, while non-causal variants are weaker or negative in several outcomes. The largest uncertainty appears in sparse novelty variants, where confidence intervals are wider; this is expected given thinner support. Additional graph statistics are possible (for example, brokerage or motif-based measures), but we prioritize interpretable variants that map directly to editorial selection and diffusion outcomes.

Tables~\ref{tab:measure_selection_framework}--\ref{tab:predictor_correlations} document why the headline predictors were selected, their empirical coverage, and their correlation structure. Table~\ref{tab:headline_effect_sizes} then translates headline coefficients into interpretable effect sizes.

\begin{table}[htp]
\centering
\caption{Measure Selection Framework and Alternatives}
\label{tab:measure_selection_framework}
\begin{tabularx}{\textwidth}{L{0.15\textwidth}L{0.22\textwidth}L{0.28\textwidth}L{0.27\textwidth}}
\hline
\textbf{Family} & \textbf{Headline Measure} & \textbf{Alternatives Considered} & \textbf{Selection Rationale} \\
\hline
Evidence share & Share causal claims & Method-count shares; paper-level causal dummy & Chosen for direct interpretability and broad availability; preserves continuous variation in evidentiary basis. \\
Complexity & Log number of edges & Unique paths; longest path; motif counts & Chosen as the most stable breadth measure; path metrics retained as appendix variants for depth checks. \\
Novelty & Log number of new edges & Novel-path shares; novelty-rate-only metrics & Chosen to capture frontier expansion at edge level with strong EO stability; path novelty reported as variant. \\
Positioning & Topic centrality (mean eigenvector) & PageRank; in-/out-degree centrality summaries & Chosen for comparability with literature-graph prominence; PageRank retained as variant robustness. \\
Architecture & Source--sink ratio & Reciprocity; clustering; brokerage indices & Chosen for transparent driver/outcome balance interpretation; higher-order topology deferred to future modules. \\
\hline
\end{tabularx}
\captionsetup{singlelinecheck=off,font=small,labelformat=empty,labelsep=none}
\caption*{\scriptsize\justifying \textbf{Note:} Selection follows five criteria: interpretability, construct coverage, EO-threshold sign stability, low redundancy, and broad availability. Appendix variants are retained to show that headline conclusions are not an artifact of one operationalization.}
\end{table}

\begin{table}[htp]
\centering
\caption{Summary Statistics for Headline Predictor Families (EO \(\geq 4\))}
\label{tab:predictor_summary_stats}
\begin{tabular}{lcccccc}
\hline
\textbf{Predictor} & \textbf{Mean} & \textbf{SD} & \textbf{P10} & \textbf{P50} & \textbf{P90} & \textbf{Undefined \%} \\
\hline
Share causal claims & 0.2249 & 0.3873 & 0.0000 & 0.0000 & 1.0000 & 0.00 \\
Log \(N\) edges (causal) & 0.3254 & 0.5582 & 0.0000 & 0.0000 & 1.3863 & 0.00 \\
Log \(N\) edges (non-causal) & 1.0125 & 0.5737 & 0.0000 & 1.0986 & 1.6094 & 0.00 \\
Log \(N\) new edges (causal) & 0.7537 & 0.5133 & 0.0000 & 0.6931 & 1.3863 & 71.76 \\
Log \(N\) new edges (non-causal) & 0.6018 & 0.5181 & 0.0000 & 0.6931 & 1.3863 & 16.97 \\
Topic centrality (causal) & 0.2539 & 0.1644 & 0.0604 & 0.2297 & 0.4855 & 71.89 \\
Topic centrality (non-causal) & 0.3404 & 0.1597 & 0.1345 & 0.3348 & 0.5464 & 17.34 \\
Source--sink ratio (causal) & 1.0335 & 0.6591 & 0.3333 & 1.0000 & 2.0000 & 71.76 \\
Source--sink ratio (non-causal) & 1.1792 & 0.7785 & 0.5000 & 1.0000 & 2.0000 & 16.97 \\
\hline
\end{tabular}
\captionsetup{singlelinecheck=off,font=small,labelformat=empty,labelsep=none}
\caption*{\scriptsize\justifying \textbf{Note:} Statistics are computed on the EO \(\geq 4\) baseline paper-level sample. ``Undefined \%'' is structural rather than a data-ingestion error: causal variants are undefined when a paper has no retained causal edge at the selected threshold, and analogous logic applies to non-causal variants in papers with no retained non-causal edge. This table is therefore a coverage diagnostic for measure availability, not evidence of missing raw records.}
\end{table}

\begin{table}[htp]
\centering
\caption{Pairwise Correlations: Causal Headline Predictors (EO \(\geq 4\))}
\label{tab:predictor_correlations}
\begin{tabular}{lccccc}
\hline
 & \textbf{Share} & \textbf{Log Edges} & \textbf{Log New} & \textbf{Centrality} & \textbf{S/S Ratio} \\
\hline
Share causal claims & 1.000 & 0.935 & 0.278 & 0.047 & -0.009 \\
Log \(N\) edges (causal) & 0.935 & 1.000 & 0.566 & 0.070 & -0.022 \\
Log \(N\) new edges (causal) & 0.278 & 0.566 & 1.000 & -0.329 & 0.000 \\
Topic centrality (causal) & 0.047 & 0.070 & -0.329 & 1.000 & -0.110 \\
Source--sink ratio (causal) & -0.009 & -0.022 & 0.000 & -0.110 & 1.000 \\
\hline
\end{tabular}
\captionsetup{singlelinecheck=off,font=small,labelformat=empty,labelsep=none}
\caption*{\scriptsize\justifying \textbf{Note:} Correlations use complete-pair observations in the EO \(\geq 4\) paper-level dataset. The matrix supports a low-redundancy interpretation: while share and causal edge volume are related, novelty, centrality, and source--sink balance add distinct variation.}
\end{table}

\begin{table}[htp]
\centering
\caption{Effect-Size Translation for Headline Predictors (EO \(\geq 4\))}
\label{tab:headline_effect_sizes}
\resizebox{\textwidth}{!}{%
\begin{tabular}{l l r r r r r r}
\hline
\textbf{Predictor (Variant)} & \textbf{Primary Shift} & \textbf{Top-5 \(\Delta\) pp} & \textbf{Top-5 per 1,000} & \textbf{Top-5 Rel. \%} & \textbf{Citation \(\%\Delta\)} & \textbf{Top-5 \(\Delta\) pp (+1 SD)} & \textbf{Citation \(\%\Delta\) (+1 SD)} \\
\hline
Share causal claims (Causal) & +10pp in share causal & +0.27 & +2.73 & +2.41 & +1.92 & +1.01 & +7.28 \\
Log \(N\) edges (Causal) & Doubling (\(\ln(2)\)) & +1.36 & +13.59 & +11.97 & +11.18 & +1.05 & +8.51 \\
Log \(N\) edges (Non-causal) & Doubling (\(\ln(2)\)) & -0.94 & -9.43 & -8.31 & -1.89 & -0.76 & -1.52 \\
Log \(N\) new edges (Causal) & Doubling (\(\ln(2)\)) & +1.71 & +17.13 & +15.10 & +10.74 & +1.06 & +6.49 \\
Log \(N\) new edges (Non-causal) & Doubling (\(\ln(2)\)) & -0.39 & -3.89 & -3.43 & -11.17 & -0.30 & -8.67 \\
Topic centrality (Causal) & IQR shift (P75--P25) & -0.41 & -4.10 & -3.61 & +8.60 & -0.18 & +3.64 \\
Topic centrality (Non-causal) & IQR shift (P75--P25) & +0.77 & +7.66 & +6.75 & +18.01 & +0.55 & +12.61 \\
Source--sink ratio (Causal) & IQR shift (P75--P25) & +0.31 & +3.12 & +2.75 & +3.08 & +0.53 & +5.32 \\
Source--sink ratio (Non-causal) & IQR shift (P75--P25) & -0.51 & -5.08 & -4.48 & -1.39 & -0.64 & -1.75 \\
\hline
\end{tabular}
}
\captionsetup{singlelinecheck=off,font=small,labelformat=empty,labelsep=none}
\caption*{\scriptsize\justifying \textbf{Note:} This table translates year-fixed-effects coefficients from the main-text headline predictor figure into interpretable magnitudes. For journal outcomes, \(\Delta\)pp \(=100\times \beta \times \Delta X\), ``per 1,000'' \(=1000\times \beta \times \Delta X\), and ``Top-5 Rel.\%'' scales \(\beta\Delta X\) by the EO \(\geq 4\) Top-5 baseline rate (11.35\%). For citations, the dependent variable is \(\log(\textit{Cites}+1)\), so \(\%\Delta=100\left[\exp(\beta \Delta X)-1\right]\). Primary shifts are: +10 percentage points for share causal claims, doubling for log-count predictors, and interquartile-range shifts for source--sink and centrality predictors. ``+1 SD'' columns report standardized magnitude comparisons across predictors.}
\end{table}

\begin{figure}[htp]
\centering
\caption{Variant Predictor Specifications for Publication and Citations}
\includegraphics[width=\textwidth]{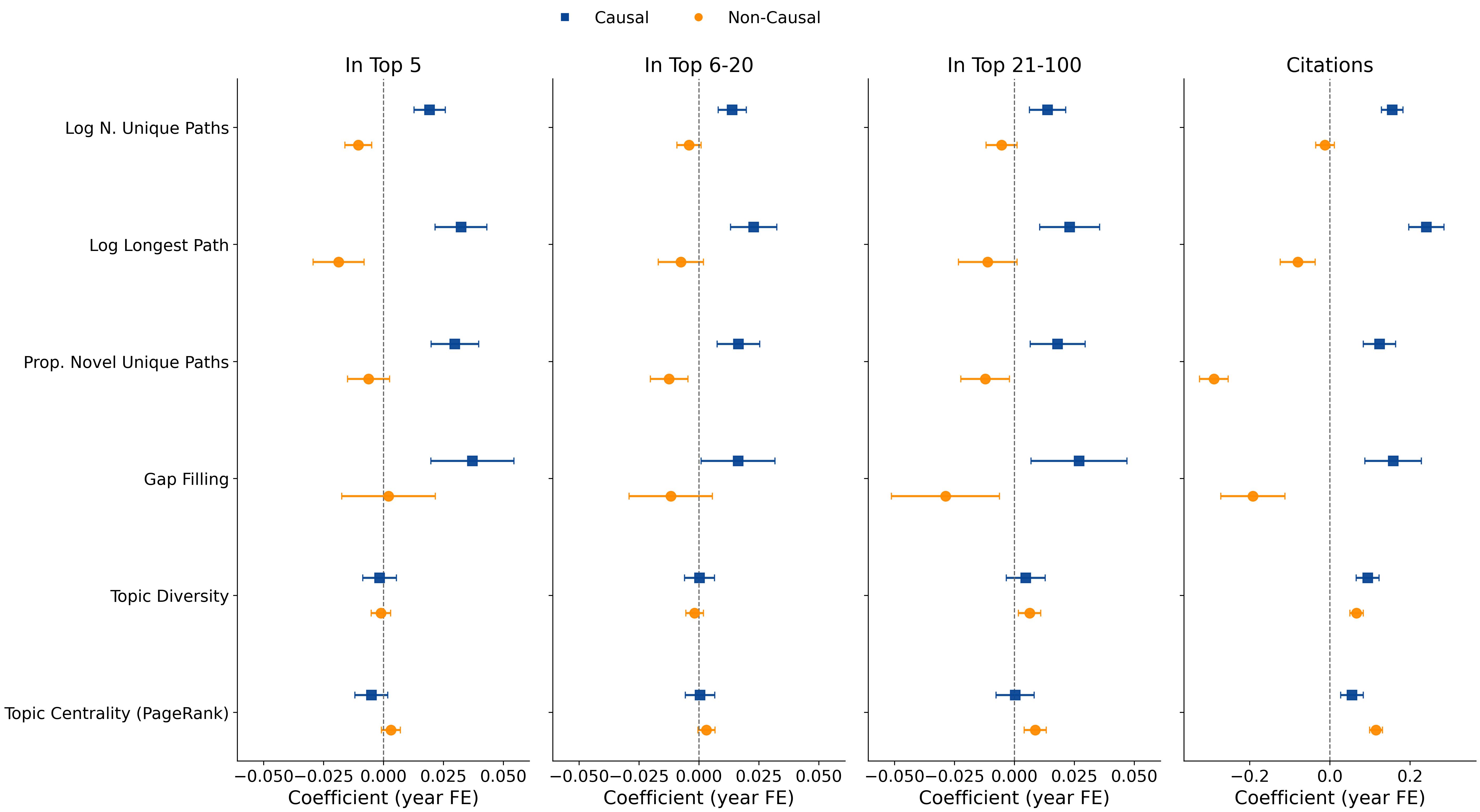}
\captionsetup{singlelinecheck=off,font=small,labelformat=empty,labelsep=none}
\caption*{\scriptsize\justifying \textbf{Note:} This appendix figure reports year-fixed-effects coefficient estimates for variant measures corresponding to the main predictor families: log unique paths, log longest path, novel-path shares, gap filling, topic diversity (variance in centrality), and PageRank-based topic centrality. Blue squares denote causal variants and orange circles denote non-causal variants. Horizontal bars are 95\% confidence intervals. The directional pattern broadly mirrors the headline figure: causal variants are more often positive for top-tier publication and citations, while non-causal variants are weaker or negative in several specifications.}
\label{fig:publication_predictors_variants_appendix}
\end{figure}

\begin{table}[htp]
\centering
\caption{Summary Statistics of Citation Percentiles by Journal Category}
\begin{tabular}{lccc}
\hline
\textbf{Journal Category} & \textbf{Mean Percentile} & \textbf{Median Percentile} & \textbf{Standard Deviation} \\
\hline
Top 5 & 62.18 & 68 & 25.86 \\
Top 6--20 & 57.06 & 61 & 27.97 \\
Top 21--100 & 52.12 & 53 & 27.37 \\
\hline
\end{tabular}
\captionsetup{singlelinecheck=off,font=small,labelformat=empty,labelsep=none}
\caption*{\scriptsize\justifying \textbf{Note:} This table summarizes the citation percentiles for papers published in different journal categories. The mean and median percentiles indicate that papers published in Top 5 journals have higher citation impact on average compared to those in lower-ranked journals. However, the overlap in distributions suggests that highly cited papers can also be found in lower-ranked journals.}
\label{tab:citation_percentiles_by_journal}
\end{table}

\begin{table}[htp]
\centering
\caption{Validation Results of Information Retrieval (\citet{brodeur2024p})}
\label{tab:validation_results_brodeur2024p}
\begin{tabular}{llccccccc}
\hline
\textbf{EO Threshold} & \textbf{Variable} & \textbf{Accuracy} & \textbf{Precision} & \textbf{Recall} & \textbf{F1} & \textbf{Prevalence} & \textbf{Support} \\
\hline
EO \(\geq 4\) & Method: DiD & 0.8281 & 0.7375 & 0.6782 & 0.7066 & 0.3053 & 285 \\
EO \(\geq 4\) & Method: RDD & 0.9263 & 0.7317 & 0.7500 & 0.7407 & 0.1404 & 285 \\
EO \(\geq 4\) & Method: IV & 0.7895 & 0.9265 & 0.5339 & 0.6774 & 0.4140 & 285 \\
EO \(\geq 7\) & Method: DiD & 0.8639 & 0.8372 & 0.6923 & 0.7579 & 0.3077 & 169 \\
EO \(\geq 7\) & Method: RDD & 0.9231 & 0.7407 & 0.7692 & 0.7547 & 0.1538 & 169 \\
EO \(\geq 7\) & Method: IV & 0.7751 & 0.9744 & 0.5067 & 0.6667 & 0.4438 & 169 \\
EO = 9 & Method: DiD & 0.8667 & 0.8696 & 0.6897 & 0.7692 & 0.3222 & 90 \\
EO = 9 & Method: RDD & 0.9444 & 0.6667 & 0.8889 & 0.7619 & 0.1000 & 90 \\
EO = 9 & Method: IV & 0.7444 & 0.9615 & 0.5319 & 0.6849 & 0.5222 & 90 \\
\hline
\end{tabular}
\captionsetup{singlelinecheck=off,font=small,labelformat=empty,labelsep=none}

    \caption*{\scriptsize\justifying \textbf{Note:} This table reports Brodeur validation for EO \(\geq 4\), EO \(\geq 7\), and EO \(=9\). We omit rows where precision or recall is undefined (i.e., \(tp+fp=0\) or \(tp+fn=0\)), which occurs primarily for very low-prevalence labels in the matched sample. Prevalence is the benchmark positive-class share in the matched sample. Full EO-grid outputs, including these undefined cases, and match logs are provided in the reproducibility package.}
\end{table}

\begin{table}[htp]
\centering
\caption{Brodeur Validation Dispersion Across Nine Individual Stage-1$\times$Stage-2 Runs}
\label{tab:validation_results_brodeur_runlevel_summary}
\resizebox{\textwidth}{!}{%
\begin{tabular}{llccccc}
\hline
\textbf{Variable} & \textbf{Metric} & \textbf{Runs} & \textbf{Min} & \textbf{Median} & \textbf{Mean} & \textbf{Max} \\
\hline
Method: DiD & Accuracy & 9 & 0.8076 & 0.8127 & 0.8152 & 0.8369 \\
Method: DiD & Precision & 9 & 0.6701 & 0.6860 & 0.6947 & 0.7500 \\
Method: DiD & Recall & 9 & 0.6705 & 0.6818 & 0.6950 & 0.7386 \\
Method: DiD & F1 & 9 & 0.6782 & 0.6982 & 0.6943 & 0.7125 \\
Method: DiD & Prevalence & 9 & 0.2979 & 0.3024 & 0.3020 & 0.3039 \\
Method: RCT & Accuracy & 9 & 0.7544 & 0.7732 & 0.7701 & 0.7872 \\
Method: RCT & Precision & 9 & 0.0143 & 0.0156 & 0.0216 & 0.0323 \\
Method: RCT & Recall & 9 & 0.5000 & 0.5000 & 0.7222 & 1.0000 \\
Method: RCT & F1 & 9 & 0.0278 & 0.0303 & 0.0420 & 0.0625 \\
Method: RCT & Prevalence & 9 & 0.0069 & 0.0069 & 0.0069 & 0.0071 \\
Method: RDD & Accuracy & 9 & 0.9329 & 0.9381 & 0.9391 & 0.9485 \\
Method: RDD & Precision & 9 & 0.7250 & 0.7500 & 0.7570 & 0.7949 \\
Method: RDD & Recall & 9 & 0.7750 & 0.8205 & 0.8164 & 0.8684 \\
Method: RDD & F1 & 9 & 0.7532 & 0.7848 & 0.7850 & 0.8193 \\
Method: RDD & Prevalence & 9 & 0.1307 & 0.1375 & 0.1363 & 0.1375 \\
Method: IV & Accuracy & 9 & 0.7774 & 0.7869 & 0.7915 & 0.8110 \\
Method: IV & Precision & 9 & 0.8873 & 0.9125 & 0.9087 & 0.9254 \\
Method: IV & Recall & 9 & 0.5167 & 0.5455 & 0.5582 & 0.6066 \\
Method: IV & F1 & 9 & 0.6631 & 0.6804 & 0.6910 & 0.7291 \\
Method: IV & Prevalence & 9 & 0.4140 & 0.4192 & 0.4187 & 0.4255 \\
Field: Urban Economics & Accuracy & 9 & 0.8935 & 0.9053 & 0.9056 & 0.9175 \\
Field: Urban Economics & Precision & 9 & 0.0909 & 0.1000 & 0.1025 & 0.1154 \\
Field: Urban Economics & Recall & 9 & 0.7500 & 0.7500 & 0.8333 & 1.0000 \\
Field: Urban Economics & F1 & 9 & 0.1622 & 0.1818 & 0.1820 & 0.2000 \\
Field: Urban Economics & Prevalence & 9 & 0.0105 & 0.0137 & 0.0127 & 0.0137 \\
Field: Finance & Accuracy & 9 & 0.8728 & 0.8772 & 0.8779 & 0.8830 \\
Field: Finance & Precision & 9 & 0.4211 & 0.4333 & 0.4315 & 0.4561 \\
Field: Finance & Recall & 9 & 0.8889 & 0.9286 & 0.9265 & 0.9630 \\
Field: Finance & F1 & 9 & 0.5714 & 0.5977 & 0.5887 & 0.6118 \\
Field: Finance & Prevalence & 9 & 0.0928 & 0.0928 & 0.0944 & 0.0993 \\
Field: Macroeconomics & Accuracy & 9 & 0.9450 & 0.9485 & 0.9480 & 0.9509 \\
Field: Macroeconomics & Precision & 9 & 0.0625 & 0.0667 & 0.0669 & 0.0714 \\
Field: Macroeconomics & Recall & 9 & 0.5000 & 0.5000 & 0.5000 & 0.5000 \\
Field: Macroeconomics & F1 & 9 & 0.1111 & 0.1176 & 0.1179 & 0.1250 \\
Field: Macroeconomics & Prevalence & 9 & 0.0069 & 0.0069 & 0.0069 & 0.0071 \\
Field: Development & Accuracy & 9 & 0.7509 & 0.7527 & 0.7550 & 0.7595 \\
Field: Development & Precision & 9 & 0.0533 & 0.0548 & 0.0543 & 0.0556 \\
Field: Development & Recall & 9 & 0.8000 & 0.8000 & 0.8000 & 0.8000 \\
Field: Development & F1 & 9 & 0.1000 & 0.1026 & 0.1017 & 0.1039 \\
Field: Development & Prevalence & 9 & 0.0172 & 0.0172 & 0.0173 & 0.0177 \\
\hline
\end{tabular}
}
\captionsetup{singlelinecheck=off,font=small,labelformat=empty,labelsep=none}
    \caption*{\scriptsize\justifying \textbf{Note:} Each row summarizes one metric across the nine individual extraction runs (3 Stage-1 runs $\times$ 3 Stage-2 runs), using the same Brodeur title-matching protocol as the EO table. This isolates non-deterministic run-to-run variation before EO aggregation. Medians and means are stable for method labels; field metrics are sensitive to low prevalence in this benchmark, which is visible in the prevalence rows.}
\end{table}

\begin{table}[htp]
\centering
\caption{Validation Results with Plausibly Exogenous Galore Dataset}
\label{tab:validation_results_plausibly_exo}
\resizebox{\textwidth}{!}{%
\begin{tabular}{l l c c c c c c c c}
\hline
\textbf{EO Threshold} & \textbf{Variable} & \textbf{N} & \textbf{Max Similarity} & \textbf{Mean Similarity} & \textbf{Median} & \textbf{Std Dev} & \textbf{Min} & \textbf{Random Baseline} & \textbf{Lift vs Baseline} \\
\hline
EO \(\geq 4\) & Cause Similarity & 491 & 0.8112 & 0.1747 & 0.1031 & 0.2037 & 0.0006 & 0.0060 & 0.1687 \\
EO \(\geq 4\) & Effect Similarity & 491 & 0.7602 & 0.1439 & 0.0827 & 0.1658 & 0.0010 & 0.0072 & 0.1366 \\
EO \(\geq 4\) & Exogenous Variation Similarity & 491 & 1.0000 & 0.2682 & 0.1832 & 0.2377 & 0.0097 & 0.0451 & 0.2231 \\
EO \(\geq 7\) & Cause Similarity & 296 & 0.8485 & 0.1729 & 0.0884 & 0.2238 & 0.0014 & 0.0073 & 0.1656 \\
EO \(\geq 7\) & Effect Similarity & 296 & 0.7690 & 0.1318 & 0.0551 & 0.1831 & 0.0012 & 0.0084 & 0.1235 \\
EO \(\geq 7\) & Exogenous Variation Similarity & 296 & 1.0000 & 0.2579 & 0.1780 & 0.2314 & 0.0107 & 0.0530 & 0.2049 \\
EO = 9 & Cause Similarity & 148 & 0.8428 & 0.1971 & 0.1011 & 0.2445 & 0.0029 & 0.0095 & 0.1876 \\
EO = 9 & Effect Similarity & 148 & 0.7695 & 0.1320 & 0.0605 & 0.1847 & 0.0020 & 0.0094 & 0.1226 \\
EO = 9 & Exogenous Variation Similarity & 148 & 1.0000 & 0.2604 & 0.1795 & 0.2275 & 0.0158 & 0.0655 & 0.1950 \\
\hline
\end{tabular}
}
\captionsetup{singlelinecheck=off,font=small,labelformat=empty,labelsep=none}
    \caption*{\scriptsize\justifying \textbf{Note:} This table reports Plausibly validation for EO \(\geq 4\), EO \(\geq 7\), and EO = 9. Similarity is computed as cosine on deterministic text vectors, and the random baseline is a shuffled-match permutation benchmark (1,000 permutations per EO/component). The permutation p-values are \(0.001\) in all nine rows. Full EO-grid outputs and match logs are included in the replication outputs linked from the project repository.}
\end{table}

\begin{table}[htp]
\centering
\caption{Snippet Validation by EO Proxy}
\label{tab:validation_snippet_eo}
\begin{tabular}{lccccc}
\hline
\textbf{EO Proxy} & \textbf{Validation Cases} & \textbf{Support Edges} & \textbf{Micro Precision} & \textbf{Micro Recall} & \textbf{Micro F1} \\
\hline
EO \(\geq 1\) & 77,070 & 864,085 & 0.7784 & 0.7792 & 0.7788 \\
EO \(\geq 2\) & 59,463 & 378,255 & 0.8942 & 0.5941 & 0.7139 \\
EO = 3 & 33,471 & 170,644 & 0.9361 & 0.4172 & 0.5772 \\
\hline
\end{tabular}
\captionsetup{singlelinecheck=off,font=small,labelformat=empty,labelsep=none}
    \caption*{\scriptsize\justifying \textbf{Note:} This table reports snippet-audit performance by EO proxy. Moving from EO \(\geq 1\) to EO \(\geq 2\) and EO = 3 increases precision (0.778 to 0.894 to 0.936) while reducing recall (0.779 to 0.594 to 0.417), quantifying the expected precision--recall trade-off in overlap filtering.}
\end{table}

\begin{table}[htp]
\centering
\caption{EO-threshold Perturbation Summary Relative to EO \(\geq 4\)}
\label{tab:validation_perturbation_summary}
\begin{tabular}{lccc}
\hline
\textbf{EO Threshold} & \textbf{Papers} & \textbf{Mean Share Causal Edges} & \textbf{\% Change vs EO \(\geq 4\)} \\
\hline
EO = 1 & 34,723 & 0.1518 & -32.50 \\
EO = 2 & 45,656 & 0.2176 & -3.25 \\
EO = 3 & 45,540 & 0.2175 & -3.33 \\
EO = 4 & 42,404 & 0.2249 & 0.00 \\
EO = 5 & 37,005 & 0.2343 & 4.17 \\
EO = 6 & 29,781 & 0.2442 & 8.58 \\
EO = 7 & 20,638 & 0.2569 & 14.19 \\
EO = 8 & 14,410 & 0.2647 & 17.67 \\
EO = 9 & 8,792 & 0.2702 & 20.14 \\
\hline
\end{tabular}
\captionsetup{singlelinecheck=off,font=small,labelformat=empty,labelsep=none}
    \caption*{\scriptsize\justifying \textbf{Note:} This table compares descriptive graph quantities across EO thresholds relative to EO \(\geq 4\). EO = 2 through EO = 6 remain close to the baseline, while EO = 1 and EO \(\geq 7\) show larger departures, supporting EO \(\geq 4\) as a balanced operating threshold.}
\end{table}

\subsection{Corpus Audits and Data Processing Checks}

\begin{table}[htp]
\centering
\caption{Corpus Deduplication Audit Summary}
\label{tab:corpus_dedup_summary}
\begin{tabular}{lc}
\hline
\textbf{Audit Statistic} & \textbf{Value} \\
\hline
Unique paper IDs in metadata union snapshot & 46,014 \\
Unique normalized title-year keys & 43,352 \\
Title-year keys with any duplicates & 2,574 \\
Title-year keys overlapping across NBER/CEPR & 2,520 \\
Cross-repo overlap share of unique title-year keys & 5.81\% \\
\hline
\end{tabular}
\captionsetup{singlelinecheck=off,font=small,labelformat=empty,labelsep=none}
\caption*{\scriptsize\justifying \textbf{Note:} Counts are computed from the metadata union using normalized title-year keys. Cross-repository overlaps are collapsed to canonical IDs before paper-level aggregation, so publication-outcome regressions and trend statistics are not mechanically inflated by duplicate NBER/CEPR versions of the same paper.}
\end{table}

\clearpage
\bibliographystyle{plainnat}
\bibliography{bib}

@article{garg2024political,
  author    = {Garg, Prashant and Fetzer, Thiemo},
  title     = {Political expression of academics on Twitter},
  journal   = {Nature Human Behaviour},
  year      = {2025},
  volume    = {9},
  pages     = {1815--1832},
  doi       = {10.1038/s41562-025-02199-1},
  url       = {https://doi.org/10.1038/s41562-025-02199-1}
}

@article{besley2024mediamultiplier,
  author    = {Besley, Timothy and Fetzer, Thiemo and Mueller, Hannes},
  title     = {How Big Is the Media Multiplier? Evidence from Dyadic News Data},
  journal   = {Review of Economics and Statistics},
  year      = {2024},
  pages     = {1--45},
  doi       = {10.1162/rest_a_01415},
  url       = {https://doi.org/10.1162/rest_a_01415},
  publisher = {MIT Press}
}

@article{fetzer2025losinghomefront,
  author    = {Fetzer, Thiemo and Souza, Pedro C. L. and Vanden Eynde, Oliver and Wright, Austin L.},
  title     = {Losing on the Home Front? Battlefield Casualties, Media, and Public Support for Foreign Interventions},
  journal   = {American Journal of Political Science},
  year      = {2025},
  volume    = {69},
  number    = {4},
  pages     = {1300--1316},
  doi       = {10.1111/ajps.12907},
  url       = {https://doi.org/10.1111/ajps.12907}
}

@article{pearson2024can,
  title={Can AI review the scientific literature—and figure out what it all means?},
  author={Pearson, Helen},
  journal={Nature},
  volume={635},
  number={8038},
  pages={276--278},
  year={2024},
  publisher={Nature}
}

@article{Ash2023-at,
  title     = "Text Algorithms in Economics",
  author    = "Ash, Elliott and Hansen, Stephen",
  journal   = "Annu. Rev. Econom.",
  publisher = "Annual Reviews",
  month     =  jul,
  year      =  2023,
  language  = "en"
}

@article{Korinek2023GenAI,
  author = {Anton Korinek},
  title = {Generative AI for Economic Research: Use Cases and Implications for Economists},
  journal = {Journal of Economic Literature},
  volume = {61},
  number = {4},
  pages = {1281--1317},
  year = {2023}
}

@article{dell2024deep,
  author    = {Melissa Dell},
  title     = {Deep Learning for Economists},
  journal   = {National Bureau of Economic Research Working Paper Series},
  number    = {32768},
  year      = {2024},
  month     = {August},
  url       = {http://www.nber.org/papers/w32768},
  address   = {1050 Massachusetts Avenue, Cambridge, MA 02138}
}

@article{pugliese2024conduct,
  title={How to conduct efficient and objective literature reviews using natural language processing: A step-by-step guide for marketing researchers},
  author={Pugliese, Serena and Giannetti, Verdiana and Banerjee, Sourindra},
  journal={Psychology \& Marketing},
  volume={41},
  number={2},
  pages={427--441},
  year={2024},
  publisher={Wiley Online Library}
}

@article{backhouse2017age,
  title={The age of the applied economist: the transformation of economics since the 1970s},
  author={Backhouse, Roger E and Cherrier, B{\'e}atrice},
  journal={History of Political Economy},
  volume={49},
  number={Supplement},
  pages={1--33},
  year={2017},
  publisher={Duke University Press}
}

@article{brodeur2024p,
  title={P-hacking, data type and data-sharing policy},
  author={Brodeur, Abel and Cook, Nikolai and Neisser, Carina},
  journal={The Economic Journal},
  volume={134},
  number={659},
  pages={985--1018},
  year={2024},
  publisher={Oxford University Press}
}

@article{deaton2010instruments,
  title={Instruments, randomization, and learning about development},
  author={Deaton, Angus},
  journal={Journal of Economic Literature},
  volume={48},
  number={2},
  pages={424--455},
  year={2010},
  publisher={American Economic Association}
}

@book{imbens2015causal,
  title={Causal Inference for Statistics, Social, and Biomedical Sciences: An Introduction},
  author={Imbens, Guido W and Rubin, Donald B},
  year={2015},
  publisher={Cambridge University Press}
}

@article{waltman2012new,
  title={A new methodology for constructing a publication-level classification system of science},
  author={Waltman, Ludo and Van Eck, Nees Jan},
  journal={Journal of the American Society for Information Science and Technology},
  volume={63},
  number={12},
  pages={2378--2392},
  year={2012},
  publisher={Wiley Online Library}
}

@article{hidalgo2009building,
author = {César A. Hidalgo  and Ricardo Hausmann },
title = {The building blocks of economic complexity},
journal = {Proceedings of the National Academy of Sciences},
volume = {106},
number = {26},
pages = {10570-10575},
year = {2009},
doi = {10.1073/pnas.0900943106},
URL = {https://www.pnas.org/doi/abs/10.1073/pnas.0900943106},
eprint = {https://www.pnas.org/doi/pdf/10.1073/pnas.0900943106}}

@article{buehler2024accelerating,
  title = "{Accelerating Scientific Discovery with Generative Knowledge Extraction, Graph-Based Representation, and Multimodal Intelligent Graph Reasoning}",
  author = {Buehler, Markus J.},
  journal = {Machine Learning: Science and Technology},
  volume = {5},
  number = {4},
  pages = {045005},
  year = {2024},
  doi = {10.1088/2632-2153/ad7228},
  url = {https://iopscience.iop.org/article/10.1088/2632-2153/ad7228}
}

@article{chan2024steps,
  title={Steps Towards an Infrastructure for Scholarly Synthesis},
  author={Chan, Joel and Akamatsu, Matthew and Vargas, David and Kawerau, Lukas and Gartner, Michael},
  journal={arXiv preprint arXiv:2407.20666},
  year={2024}
}

@article{goldsmith2024tracking,
  title={Tracking the Credibility Revolution across Fields},
  author={Goldsmith-Pinkham, Paul},
  journal={arXiv preprint arXiv:2405.20604},
  year={2024}
}

@article{rawat2014publish,
  title={Publish or perish: Where are we heading?},
  author={Rawat, Seema and Meena, Sanjay},
  journal={Journal of research in medical sciences: the official journal of Isfahan University of Medical Sciences},
  volume={19},
  number={2},
  pages={87},
  year={2014},
  publisher={Wolters Kluwer--Medknow Publications}
}

@techreport{aipnet,
  title={AI-Generated Production Networks: Measurement and Applications to Global Trade},
  author={Fetzer, Thiemo and Lambert, John Peter and Garg, Prashant and Feld, Bennet},
  year={2024},
  institution={Working Paper}
}

@article{banerjee2015miracle,
  title={The miracle of microfinance? Evidence from a randomized evaluation},
  author={Banerjee, Abhijit and Duflo, Esther and Glennerster, Rachel and Kinnan, Cynthia},
  journal={American economic journal: Applied economics},
  volume={7},
  number={1},
  pages={22--53},
  year={2015},
  publisher={American Economic Association 2014 Broadway, Suite 305, Nashville, TN 37203-2425}
}

@article{chetty2014land,
  title={Where is the land of opportunity? The geography of intergenerational mobility in the United States},
  author={Chetty, Raj and Hendren, Nathaniel and Kline, Patrick and Saez, Emmanuel},
  journal={The quarterly journal of economics},
  volume={129},
  number={4},
  pages={1553--1623},
  year={2014},
  publisher={MIT Press}
}

@article{stromberg2004radio,
  title={Radio's impact on public spending},
  author={Str{\"o}mberg, David},
  journal={The Quarterly Journal of Economics},
  volume={119},
  number={1},
  pages={189--221},
  year={2004},
  publisher={MIT Press}
}

@article{mccombs1972agenda,
  title={The agenda-setting function of mass media},
  author={McCombs, Maxwell E and Shaw, Donald L},
  journal={Public opinion quarterly},
  volume={36},
  number={2},
  pages={176--187},
  year={1972},
  publisher={Oxford University Press}
}

@article{entman1993framing,
  title={Framing: Towards clarification of a fractured paradigm},
  author={Entman, Robert M and others},
  journal={McQuail's reader in mass communication theory},
  volume={390},
  pages={397},
  year={1993}
}

@article{dellavigna2007fox,
  title={The Fox News effect: Media bias and voting},
  author={DellaVigna, Stefano and Kaplan, Ethan},
  journal={The Quarterly Journal of Economics},
  volume={122},
  number={3},
  pages={1187--1234},
  year={2007},
  publisher={MIT Press}
}

@article{gentzkow2010drives,
  title={What drives media slant? Evidence from US daily newspapers},
  author={Gentzkow, Matthew and Shapiro, Jesse M},
  journal={Econometrica},
  volume={78},
  number={1},
  pages={35--71},
  year={2010},
  publisher={Wiley Online Library}
}

@article{snyder2010press,
  title={Press coverage and political accountability},
  author={Snyder Jr, James M and Str{\"o}mberg, David},
  journal={Journal of political Economy},
  volume={118},
  number={2},
  pages={355--408},
  year={2010},
  publisher={The University of Chicago Press}
}

@article{enikolopov2011media,
  title={Media and political persuasion: Evidence from Russia},
  author={Enikolopov, Ruben and Petrova, Maria and Zhuravskaya, Ekaterina},
  journal={American economic review},
  volume={101},
  number={7},
  pages={3253--3285},
  year={2011},
  publisher={American Economic Association}
}

@article{gabaix2011granular,
  title={The granular origins of aggregate fluctuations},
  author={Gabaix, Xavier},
  journal={Econometrica},
  volume={79},
  number={3},
  pages={733--772},
  year={2011},
  publisher={Wiley Online Library}
}

@article{goldberg2010imported,
  title={Imported intermediate inputs and domestic product growth: Evidence from India},
  author={Goldberg, Pinelopi Koujianou and Khandelwal, Amit Kumar and Pavcnik, Nina and Topalova, Petia},
  journal={The Quarterly journal of economics},
  volume={125},
  number={4},
  pages={1727--1767},
  year={2010},
  publisher={MIT Press}
}

@article{deaton2018understanding,
  title={Understanding and misunderstanding randomized controlled trials},
  author={Deaton, Angus and Cartwright, Nancy},
  journal={Social Science \& Medicine},
  volume={210},
  pages={2--21},
  year={2018},
  publisher={Elsevier}
}

@inproceedings{currie2020technology,
  title={Technology and big data are changing economics: Mining text to track methods},
  author={Currie, Janet and Kleven, Henrik and Zwiers, Esm{\'e}e},
  booktitle={AEA Papers and Proceedings},
  volume={110},
  pages={42--48},
  year={2020},
  organization={American Economic Association 2014 Broadway, Suite 305, Nashville, TN 37203}
}

@article{cartwright2010rcts,
  title={Are RCTs the gold standard?},
  author={Cartwright, Nancy},
  journal={BioSocieties},
  volume={2},
  number={1},
  pages={11--20},
  year={2007},
  publisher={Springer}
}

@article{angrist2017economic,
  author       = {Angrist, Joshua and Azoulay, Pierre and Ellison, Glenn and Hill, Ryan and Lu, Susan Feng},
  title        = {Economic Research Evolves: Fields and Styles},
  journal      = {American Economic Review: Papers and Proceedings},
  volume       = {107},
  number       = {5},
  pages        = {293--297},
  year         = {2017}
}

@unpublished{falk2021whats,
  author       = {Andre, Peter and Falk, Armin},
  title        = {What's Worth Knowing in Economics? A Global Survey among Economists},
  year         = {2021},
  note         = {Working Paper}
}

@article{park2023papers,
  title={Papers and patents are becoming less disruptive over time},
  author={Park, Michael and Leahey, Erin and Funk, Russell J},
  journal={Nature},
  volume={613},
  number={7942},
  pages={138--144},
  year={2023},
  publisher={Nature Publishing Group UK London}
}

@article{bloom2020ideas,
  title={Are ideas getting harder to find?},
  author={Bloom, Nicholas and Jones, Charles I and Van Reenen, John and Webb, Michael},
  journal={American Economic Review},
  volume={110},
  number={4},
  pages={1104--1144},
  year={2020},
  publisher={American Economic Association 2014 Broadway, Suite 305, Nashville, TN 37203}
}

@article{baumann2020have,
  title={Where have all the working papers gone? Evidence from four major economics working paper series},
  author={Baumann, Alexandra and Wohlrabe, Klaus},
  journal={Scientometrics},
  volume={124},
  number={3},
  pages={2433--2441},
  year={2020},
  publisher={Springer}
}

@article{sims2010but,
  title={But economics is not an experimental science},
  author={Sims, Christopher A},
  journal={Journal of Economic Perspectives},
  volume={24},
  number={2},
  pages={59--68},
  year={2010},
  publisher={American Economic Association}
}

@article{heckman2001micro,
  title={Micro data, heterogeneity, and the evaluation of public policy: Nobel lecture},
  author={Heckman, James J},
  journal={Journal of Political Economy},
  volume={109},
  number={4},
  pages={673--748},
  year={2001},
  publisher={University of Chicago Press}
}

@article{angrist2010credibility,
  title={The credibility revolution in empirical economics: How better research design is taking the con out of econometrics},
  author={Angrist, Joshua D and Pischke, J{\"o}rn-Steffen},
  journal={Journal of Economic Perspectives},
  volume={24},
  number={2},
  pages={3--30},
  year={2010},
  publisher={American Economic Association}
}

@misc{angrist1994identification,
  title={Identification and estimation of local average treatment effects},
  author={Angrist, Joshua and Imbens, Guido},
  year={1994},
  publisher={National Bureau of Economic Research Cambridge, Mass., USA}
}

@incollection{van2014visualizing,
  title={Visualizing bibliometric networks},
  author={Van Eck, Nees Jan and Waltman, Ludo},
  booktitle={Measuring scholarly impact: Methods and practice},
  pages={285--320},
  year={2014},
  publisher={Springer}
}

@article{small1973co,
  title={Co-citation in the scientific literature: A new measure of the relationship between two documents},
  author={Small, Henry},
  journal={Journal of the American Society for information Science},
  volume={24},
  number={4},
  pages={265--269},
  year={1973},
  publisher={Wiley Online Library}
}

@book{angrist2008mostly,
  title={Mostly Harmless Econometrics: An Empiricist's Companion},
  author={Angrist, Joshua D. and Pischke, J{\"o}rn-Steffen},
  year={2008},
  publisher={Princeton University Press},
  address={Princeton, NJ}
}

@article{hamermesh2013six,
  title={Six decades of top economics publishing: Who and how?},
  author={Hamermesh, Daniel S},
  journal={Journal of Economic Literature},
  volume={51},
  number={1},
  pages={162--172},
  year={2013},
  publisher={American Economic Association}
}

@article{card2013nine,
  title={Nine facts about top journals in economics},
  author={Card, David and DellaVigna, Stefano},
  journal={Journal of Economic Literature},
  volume={51},
  number={1},
  pages={144--161},
  year={2013},
  publisher={American Economic Association}
}

@techreport{ludwig2025economistsllm,
  title={Large language models: An applied econometric framework},
  author={Ludwig, Jens and Mullainathan, Sendhil and Rambachan, Ashesh},
  year={2025},
  institution={National Bureau of Economic Research}
}

@techreport{humlum2025large,
  title={Large language models, small labor market effects},
  author={Humlum, Anders and Vestergaard, Emilie},
  year={2025},
  institution={National Bureau of Economic Research}
}

@techreport{asirvatham2026gptmeasurement,
  title={GPT as a Measurement Tool},
  author={Asirvatham, Juhar and Inbar, Levi and Lu, Felix and Wang, Yiqing},
  institution={National Bureau of Economic Research},
  type={NBER Working Paper},
  number={34834},
  year={2026},
  month={feb},
  doi={10.3386/w34834},
  url={https://www.nber.org/system/files/working_papers/w34834/w34834.pdf}
}

@article{feyzollahi2025adoptionllm,
  title={The adoption of large language models in economics research},
  author={Feyzollahi, Ali and Rafizadeh, Merhdad},
  journal={Economics Letters},
  volume={251},
  pages={112265},
  year={2025},
  doi={10.1016/j.econlet.2025.112265},
  url={https://doi.org/10.1016/j.econlet.2025.112265}
}

@article{carnehl2025quest,
  title={A Quest for Knowledge},
  author={Carnehl, Matthias and Schneider, Martin},
  journal={Econometrica},
  volume={93},
  number={2},
  pages={565--595},
  year={2025},
  doi={10.3982/ECTA22144},
  url={https://doi.org/10.3982/ECTA22144}
}
\end{document}